\documentclass[twocolumn,showpacs,superscriptaddress,preprintnumbers,amsmath,amssymb,aps,prc,floatfix]{revtex4}
\usepackage{graphicx}
\usepackage{dcolumn}
\usepackage{bm}

\begin{document}

\title{Applications of quark-hadron duality in $F_{2}$ structure function}

\author{
S.P. ~Malace}
\affiliation{Hampton University, Hampton, Virginia 23668}
\affiliation{University of South Carolina, Columbia, South Carolina 29208}
\author{
G.S.~Adams}
\affiliation{Rensselaer Polytechnic Institute, Troy, New York 12180}
\author{
A.~Ahmidouch}
\affiliation{North Carolina A \& T State University, Greensboro, North Carolina 27411}
\author{
T.~Angelescu}
\affiliation{Bucharest University, Bucharest, Romania}
\author{
J.~Arrington}
\affiliation{Physics Division, Argonne National Laboratory, Argonne, Illinois 60565}
\author{
R.~Asaturyan}
\thanks{Deceased}
\affiliation{Yerevan Physics Institute, Yerevan, Armenia}
\author{
O.K.~Baker}
\affiliation{Hampton University, Hampton, Virginia 23668}
\affiliation{Thomas Jefferson National Accelerator Facility, Newport News, Virginia 23606}
\author{
N.~Benmouna}
\affiliation{The George Washington University, Washington, D.C. 20052}
\author{
H.P.~Blok}
\affiliation{Department of Physics, VU-university, 1081 HV, Amsterdam, The Netherlands}
\author{
W.U.~Boeglin}
\affiliation{Florida International University, University Park, Florida 33199}
\author{
P.E.~Bosted}
\affiliation{Thomas Jefferson National Accelerator Facility, Newport News, Virginia 23606}
\author{
H.~Breuer}
\affiliation{University of Maryland, College Park, Maryland 20742}
\author{
M.E.~Christy}
\affiliation{Hampton University, Hampton, Virginia 23668}
\author{
Y.~Cui}
\affiliation{University of Houston, Houston, TX 77204}
\author{
M.M.~Dalton}
\affiliation{University of the Witwatersrand, Johannesburg, South Africa}
\author{ 
S.~Danagoulian}
\affiliation{North Carolina A \& T State University, Greensboro, North Carolina 27411}
\author{
D.~Day}
\affiliation{University of Virginia, Charlottesville, Virginia 22901}
\author{
J.A.~Dunne}
\affiliation{Mississippi State University, Mississippi State, Mississippi 39762}
\author{
D.~Dutta}
\affiliation{Mississippi State University, Mississippi State, Mississippi 39762}
\author{
R.~Ent}
\affiliation{Thomas Jefferson National Accelerator Facility, Newport News, Virginia 23606}
\affiliation{Hampton University, Hampton, Virginia 23668}
\author{
H.C.~Fenker}
\affiliation{Thomas Jefferson National Accelerator Facility, Newport News, Virginia 23606}
\author{
L.~Gan}
\affiliation{University of North Carolina Wilmington, Wilmington, North Carolina 28403}
\author{
D.~Gaskell}
\affiliation{Thomas Jefferson National Accelerator Facility, Newport News, Virginia 23606}
\author{
K.~Hafidi}
\affiliation{Physics Division, Argonne National Laboratory, Argonne, Illinois 60565}
\author{
W.~Hinton}
\affiliation{Hampton University, Hampton, Virginia 23668}
\author{
R.J.~Holt}
\affiliation{Physics Division, Argonne National Laboratory, Argonne, Illinois 60565}
\author{
T.~Horn}
\affiliation{University of Maryland, College Park, Maryland 20742}
\author{
G.M.~Huber}
\affiliation{University of Regina, Regina, Saskatchewan, Canada S4S 0A2}
\author{
E.~Hungerford}
\affiliation{University of Houston, Houston, TX 77204}
\author{
X.~Jiang}
\affiliation{Rutgers, The State University of New Jersey, Piscataway, New Jersey 08855}
\author{
M.~Jones}
\affiliation{Thomas Jefferson National Accelerator Facility, Newport News, Virginia 23606}
\author{
K.~Joo}
\affiliation{University of Connecticut, Storrs, Connecticut 06269}
\author{ 
N.~Kalantarians}
\affiliation{University of Houston, Houston, TX 77204}
\author{
J.J.~Kelly}
\thanks{Deceased}
\affiliation{University of Maryland, College Park, Maryland 20742}
\author{
C.E.~Keppel}
\affiliation{Hampton University, Hampton, Virginia 23668}
\affiliation{Thomas Jefferson National Accelerator Facility, Newport News, Virginia 23606}
\author{
E.R.~Kinney}
\affiliation{University of Colorado, Boulder, Co 80309}
\author{
V.~Kubarovsky}
\affiliation{Thomas Jefferson National Accelerator Facility, Newport News, Virginia 23606}
\author{ 
Y.~Li}
\affiliation{University of Houston, Houston, TX 77204}
\author{
Y.~Liang}
\affiliation{Ohio University, Athens, Ohio 45071}
\author{ 
P.~Markowitz}
\affiliation{Florida International University, University Park, Florida 33199}
\author{
E.~McGrath}
\affiliation{James Madison University, Harrisonburg, Virginia 22807}
\author{ 
P.~McKee}
\affiliation{University of Virginia, Charlottesville, Virginia 22901}
\author{ 
D.G.~Meekins}
\affiliation{Thomas Jefferson National Accelerator Facility, Newport News, Virginia 23606}
\author{ 
H.~Mkrtchyan}
\affiliation{Yerevan Physics Institute, Yerevan, Armenia}
\author{ 
T.~Navasardyan}
\affiliation{Yerevan Physics Institute, Yerevan, Armenia}
\author{
G.~Niculescu}
\affiliation{Ohio University, Athens, Ohio 45071}
\affiliation{James Madison University, Harrisonburg, Virginia 22807}
\author{ 
I.~Niculescu}
\affiliation{James Madison University, Harrisonburg, Virginia 22807}
\author{ 
P.E.~Reimer}
\affiliation{Physics Division, Argonne National Laboratory, Argonne, Illinois 60565}
\author{ 
J.~Reinhold}
\affiliation{Florida International University, University Park, Florida 33199}
\author{
J.~Roche}
\affiliation{Thomas Jefferson National Accelerator Facility, Newport News, Virginia 23606}
\author{ 
E.~Schulte}
\affiliation{Physics Division, Argonne National Laboratory, Argonne, Illinois 60565}
\author{ 
E.~Segbefia}
\affiliation{Hampton University, Hampton, Virginia 23668}
\author{ 
C.~Smith}
\affiliation{University of Virginia, Charlottesville, Virginia 22901}
\author{ 
G.R.~Smith}
\affiliation{Thomas Jefferson National Accelerator Facility, Newport News, Virginia 23606}
\author{ 
V.~Tadevosyan}
\affiliation{Yerevan Physics Institute, Yerevan, Armenia}
\author{ 
L.~Tang}
\affiliation{Hampton University, Hampton, Virginia 23668}
\affiliation{Thomas Jefferson National Accelerator Facility, Newport News, Virginia 23606}
\author{
M.~Ungaro}
\affiliation{University of Connecticut, Storrs, Connecticut 06269}
\author{
A.~Uzzle}
\affiliation{Hampton University, Hampton, Virginia 23668}
\author{ 
S.~Vidakovic}
\affiliation{University of Regina, Regina, Saskatchewan, Canada S4S 0A2}
\author{
A.N.~Villano}
\affiliation{Rensselaer Polytechnic Institute, Troy, New York 12180}
\author{ 
W.F.~Vulcan}
\affiliation{Thomas Jefferson National Accelerator Facility, Newport News, Virginia 23606}
\author{
F.R.~Wesselmann}
\affiliation{University of Virginia, Charlottesville, Virginia 22901}
\author{ 
B.~Wojtsekhowski}
\affiliation{Thomas Jefferson National Accelerator Facility, Newport News, Virginia 23606}
\author{
S.A.~Wood}
\affiliation{Thomas Jefferson National Accelerator Facility, Newport News, Virginia 23606}
\author{ 
L.~Yuan}
\affiliation{Hampton University, Hampton, Virginia 23668}
\author{ 
X.~Zheng}
\affiliation{Physics Division, Argonne National Laboratory, Argonne, Illinois 60565}

\date{\today}

\begin{abstract}
Inclusive electron-proton and electron-deuteron inelastic cross sections 
have been measured at Jefferson Lab (JLab) 
in the resonance region, at large Bjorken $x$, up to 0.92, and four-momentum 
transfer squared $Q^{2}$ up to 7.5 GeV$^{2}$ in the experiment E00-116. 
These measurements are used to extend to larger $x$ and $Q^{2}$ 
precision, quantitative, studies of the phenomenon of quark-hadron duality. 
Our analysis confirms, both globally and locally, the apparent ``violation'' of quark-hadron duality 
previously observed at a $Q^{2}$ of 3.5 GeV$^{2}$ when resonance data are compared to 
structure function data created from CTEQ6M and MRST2004 parton distribution functions (PDFs). 
More importantly, our new data show that 
this discrepancy saturates by $Q^{2}$ $\sim$ 4 Gev$^{2}$, becoming $Q^{2}$ independent. 
This suggests only small violations of $Q^{2}$ evolution 
by contributions from the higher-twist terms in the resonance region which 
is confirmed by our comparisons to ALEKHIN and ALLM97.
We conclude that the unconstrained strength of the CTEQ6M and MRST2004 PDFs at 
large $x$ is the major source of the disagreement between data and these parameterizations 
in the kinematic regime we study and that, in view of quark-hadron duality, 
properly averaged resonance region data could be used in global QCD fits to reduce PDF 
uncertainties at large $x$.

\end{abstract}


\maketitle

\section{Introduction}

To understand how Quantum ChromoDynamics (QCD) works remains one of the great
challenges in nuclear physics today. The challenge arises from the fact that
the degrees of freedom observed in nature, hadrons and nuclei, are not the same
as the ones appearing in the QCD Lagrangian, quarks and gluons. The challenge
is then to formulate a connection between the description of hard, or
short-distance, scattering processes which can be calculated perturbatively
in terms of quark and gluon degrees of freedom and their weak couplings, and
soft, or long-distance, scattering processes, where the physical asymptotic
states are prominent and the quarks and gluons interact strongly.

Given these strong quark-gluon interactions, or the large value of the strong
coupling constant $\alpha_{s}$, the spectra of the asymptotic hadron states are
not calculable within a perturbative QCD (pQCD) framework, and are difficult to directly
connect to the underlying quark-gluon or parton dynamics. Yet, several
instances exist in nature where the behavior of low-energy scattering cross
sections, averaged over appropriate energy intervals, closely coincide with 
asymptotically high-energy scattering cross sections, calculated in
terms of quark-gluon degrees of freedom. This phenomenon is referred to as
{\sl quark-hadron duality}, and may be a general property of quantum field
theories with inherent weak and strong coupling limits, with QCD as a prime
example.

The observation of a non-trivial relationship between inclusive
electron--nucleon scattering cross sections at low energy, in the region
dominated by the nucleon resonances, and that in the deep inelastic scaling
regime at high energy predates QCD itself. While analyzing the data from the
early deep inelastic scattering experiments at SLAC, Bloom and Gilman observed
\cite{BG1,BG2} that the inclusive structure function at low hadronic final
state mass, $W$, generally follows a global scaling curve which describes
high-$W$ data, and to which the resonance structure function averages. Following 
the development of QCD in the early 1970s, Bloom--Gilman duality was reformulated 
in terms of an operator product ({\sl twist}) expansion of moments 
of the structure functions \cite{gp1,gp2}. This allowed a systematic classification of terms 
responsible for duality and its violations in terms of so-called {\sl higher-twist} 
operators which describe long-range interactions between quarks and gluons. 
However, this description could not explain {\sl why} particular multi-parton 
correlations were suppressed, and {\sl how} the physics of resonances gave way 
to scaling \cite{wally_duality}.

Since then, with the development of high luminosity beams at modern accelerator facilities
such as JLab, a wealth of new information on structure
functions, with unprecedented accuracy and over a wide range of kinematics,
has become available. One of the striking findings of the new JLab data
\cite{F2JL1,F2JL2,wally_duality} is that Bloom-Gilman duality appears to work
exceedingly well, down to $Q^2$ values as low as 1~GeV$^2$ or even below. This 
is considerably lower than previously believed, and well into the
region where $\alpha_{s}$ is relatively large. Furthermore, 
the equivalence of the averaged resonance and scaling structure functions appeared
to hold for each resonance, over restricted regions in $W$, so that the 
resonance--scaling duality holds also locally. It was also found that 
quark-hadron duality manifests itself in the separated proton 
transverse ($F_{1}^{p}$) and longitudinal ($F_{L}^{p}$) structure functions. 

The more recent JLab resonance structure function studies have revealed an
important application of duality: if the workings of the resonance--deep
inelastic interplay are sufficiently well understood, the region of
high Bjorken-$x$ ($x \agt 0.7$, where $x$ is the longitudinal momentum
fraction of the hadron carried by the parton in the infinite momentum
frame) would become accessible to quantitative studies. This region remains
largely unexplored experimentally due to the requirement of high-energy
beams with sufficiently high luminosity.

The $x \to 1$ region is an important testing ground for
nonperturbative and perturbative mechanisms underpinning valence
quark dynamics, and is vital to map out if we hope to achieve a
complete description of nucleon structure.
Data from the nucleon resonance region, where quark-hadron duality
has been established, could be used to better constrain QCD
parameterizations of parton distribution functions (PDFs), from which also
the hadronic backgrounds in high-energy collisions are computed \cite{St03}.
The large-$x$ region also constitutes an appreciable amount of the
moments of polarized and unpolarized structure functions, especially
for the higher moments.
It is precisely these moments that can be calculated from first
principles in QCD on the lattice \cite{LATTICE_MOM}, in terms of
matrix elements of local operators.

Note that, since the $x$ dependence of structure functions
cannot be calculated on the lattice directly, one cannot easily use
the lattice to learn about the degree to which duality holds locally.
Indeed, the ability to calculate a leading-twist moment on the
lattice implicitly uses quark-hadron duality to average the
resonance contributions to a smooth, scaling function.

In this article, we quantitatively study the application of quark-hadron
duality to access parton dynamics in the region of large $x$, up to $x$ $\sim$ 0.9.
For this, we accumulated a series of inclusive electron-proton and
electron-deuteron scattering data in the nucleon resonance region
($W^{2}$ $<$ 4 GeV$^{2}$), at the highest momentum transfers accessible at JLab. 
These data are at values of $Q^2$ far above where duality was quantitatively
found to be valid in previous JLab experiments. The extracted $F_2$ structure
function data are also compared with various state-of-the-art parameterizations
of $F_2$ world data to improve our understanding of parton dynamics at
large values of Bjorken $x$. 

The article is structured in five sections. Section 2 
summarizes techniques of modeling the dynamics of the nucleon in terms of structure functions 
computed from PDFs and 
examines in detail few representative parameterizations of the nucleon $F_{2}$ structure function 
focusing on the large $x$ region. Section 3 is an overview of the experimental apparatus utilized 
to collect these experimental data and of the analysis steps taken to extract the cross section and 
the $F_{2}$ structure function. In Section 4 we present our studies of the application of 
quark-hadron duality to gain insight in the parton dynamics at large $x$. 
In Section 5 we draw conclusions.


\section{$F_2$ parameterizations at large Bjorken $x$}
The purpose of this Section is to give an overview of the techniques typically employed to map out the 
dynamics of the nucleon via structure functions. This discussion points out the importance, but also 
the difficulty, of obtaining a parameterization of the $F_{2}$ 
structure function for the entire kinematic range. In particular, the exclusion of data in regions where 
the perturbative QCD mechanisms are not the only ones expected to contribute, greatly limits the 
applicability of these parameterizations and also our knowledge of the nucleon structure. 
In this context, quark-hadron duality might be the tool which could open kinematic regions not 
easily accessible otherwise to detailed studies. 
Four representative parameterizations will be examined in detail with an emphasis on the large $x$ region: 
ALLM97 \cite{allm97}, CTEQ6M \cite{cteq}, MRST2004 \cite{mrst} and ALEKHIN \cite{alekhin_05,alekhin_03}. 
These parameterizations were used in our duality studies which will be presented in Sect. 4. 

Lastly, the parameterization of the structure function $F_{2}^{p}$ from Bourrely {\it et al.} 
\cite{bourrely} will also be considered. This parameterization is obtained from parton distribution functions constructed in 
a statistical physical picture of the nucleon, where the nucleon is viewed as a gas of massless partons 
(quarks, antiquarks and gluons) in equilibrium at a given temperature in a finite size volume. The 
$x$ dependence of the parton distributions is chosen to correspond to a Fermi-Dirac distribution for 
quarks and antiquarks, and to a Bose-Einstein distribution for gluons. The parameterization involves 
a total of eight free parameters which are constrained by fitting high $W^{2}$ data from various 
experiments: NMC, BCDMS, E665, ZEUS, CCFR. A comparison of $F_{2}^{p}$ from Bourrely {\it et al.} to 
results from CTEQ6M and ALLM97 will be shown in Sect. 4. 

{\bf Empirical Parameterization of $F_{2}^{p}$: ALLM97}. ALLM97, proposed as an update of ALLM \cite{allm} 
published in 1991, is a Regge motivated parameterization extended to the large $Q^2$ regime in a way 
compatible with QCD expectations. The data set used to obtain the ALLM97 fit coefficients included 
all $\gamma^*$p measurements published up to 1997, with $W^2$ $>$ 3 GeV$^2$, and covering a wide range in 
$Q^2$, 0 $\leq$ $Q^{2}$ $\leq$ 5000 GeV$^{2}$ . The ALLM97 fit function has a total of 23 parameters, half of 
which are needed for the description of the low $W^2$ (high $x$) region where higher-twist terms are expected 
to be important. There are two important aspects to be noted in relation to the behavior of the ALLM97 
parameterization at large $x$ (see Figs. 1 and 2).
\begin{figure}
\centering 
\includegraphics[width=9cm]{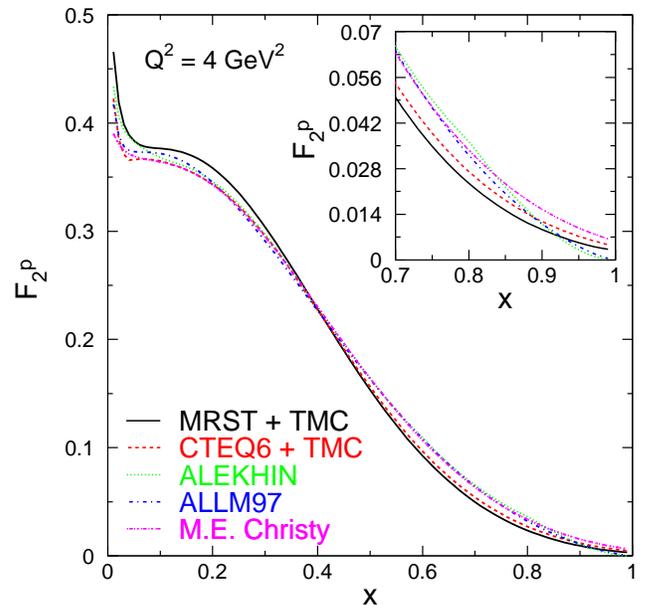}
\caption{(Color online) Existing $F_{2}^{p}$ parameterizations at a $Q^2$ value of 4 GeV$^2$. CTEQ6M \cite{cteq}, 
MRST2004 \cite{mrst}, ALLM97 \cite{allm97} 
and ALEKHIN \cite{alekhin_05,alekhin_03} were used for the quark-hadron duality studies 
presented in Sect. 4. Target mass corrections were added to 
CTEQ6M and MRST2004 (see text). The M.E. Christy \cite{eric_param} parameterization is shown 
for comparison.}
\end{figure}
\begin{figure}
\centering 
\includegraphics[width=9cm]{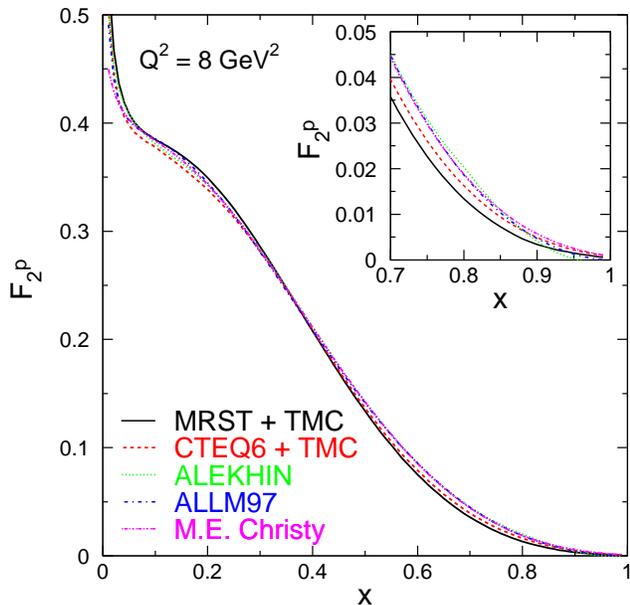}
\caption{(Color online) Existing $F_{2}^{p}$ parameterizations at a $Q^2$ value of 8 GeV$^2$. CTEQ6M \cite{cteq}, 
MRST2004 \cite{mrst}, ALLM97 \cite{allm97} 
and ALEKHIN \cite{alekhin_05,alekhin_03} were used for the quark-hadron duality studies presented 
in Sect. 4. Target mass corrections were added to 
CTEQ6M and MRST2004 (see text). The M.E. Christy \cite{eric_param} parameterization is shown 
for comparison.}
\end{figure}
On one hand, the data set used to obtain the fit coefficients is selected with a rather low $W^2$ cut. 
Thus, it is expected that, if duality holds {\sl globally}, the extrapolation of ALLM97 below $W^2$ of 
3 GeV$^2$ in the resonance region will work reasonably well, {\sl on average}. On the other hand, 
ALLM97 being 
an empirical fit, some of its shortcomings like unconstrained $x$ and $Q^{2}$ dependence or inability 
to fully 
account for target mass effects will become obvious as we probe kinematic regimes outside its domain of 
applicability. This will most likely be revealed in a clear manner when extrapolating to 
low $W^{2}$ regions.

{\bf QCD Parameterization of $F_{2}^{p}$}. 
Starting from two basic ideas of pQCD, {\sl factorization} and {\sl evolution}, the $F_{2}^{p}$ 
structure function can be calculated from PDFs 
extracted from hard-scattering data \cite{handbook_QCD}. The theorem of factorization 
of long-distance from short-distance dependence in deep inelastic scattering (DIS) allows 
for the structure function to be expressed 
as a generalization of the parton model results: 
\begin{equation}
F_{2}^{\gamma p}(x,Q^{2}) = \sum_{i=f,\bar{f},G}{\int_{0}^{1}{d\xi C_{2}^{\gamma i}(\frac{x}{\xi},\alpha_{s}(Q^{2})) \times \phi_{i/p}(\xi,Q^{2})}},
\end{equation}
where {\sl i} denotes a sum over all partons (quarks, antiquarks and gluons) inside the proton, 
$C_{2}^{\gamma i}$ are coefficient functions {\sl independent} of the long-distance effects while 
$\phi_{i/p}$ are parton distributions sensitive to the nonperturbative, long-distance 
effects inside the proton \cite{f_recon}. The evolution, on the 
other hand, enables the systematic, perturbative computation of logarithmic scale-braking effects and 
ensures that measuring $F_{2}^{\gamma p}$($x$,$Q^{2}$) is enough to predict not only 
$F_{2}^{\gamma p}$($x$,$Q^{2}$) 
but also $F_{2}^{\gamma p}$($x$,$Q^{'2}$) for all $Q^{'2}$, assuming that both $Q^{2}$ and $Q^{'2}$ are 
large enough that a perturbative expansion in $\alpha_{s}$ is still appropriate. This is typically done 
by using the DGLAP evolution equations to evolve the parton distributions: 
\begin{multline}
Q^{2}\frac{d}{dQ^{2}}\phi_{i/p}(x,Q^{2}) = \\ 
\sum_{j=f,\bar{f},G}{\int_{x}^{1}{\frac{d\xi}{\xi} P_{ij}(\frac{x}{\xi},\alpha_{s}(Q^{2})) \times \phi_{j/p}(\xi,Q^{2})}},
\end{multline}
where $P_{ij}$ are the evolution kernels (splitting functions) given by a 
perturbative expansion in $\alpha_{s}$, 
beginning with the leading order (LO) $O(\alpha_{s})$ but also calculable to higher orders, 
next-to-leading order (NLO) or next-to-next-leading order (NNLO). 
The kernels have the physical interpretation as probability densities 
of obtaining a parton of type {\sl i} from one of type {\sl j} carrying a fraction of the parent 
parton's momentum. 

Thus, the three basic quantities are the coefficient functions $C_{2}^{(\gamma i)}$, 
the evolution kernels $P_{ij}$ and the PDFs $\phi_{i/p}$. 
Of these, the first two are 
computed perturbatively as power series in $\alpha_{s}$. The physical non-perturbative 
parton distributions are extracted by combining theory and experiment and performing 
QCD fits. In a typical QCD fitting 
procedure the $x$ dependence of the parton distributions is parameterized at some low scale, $Q_{0}^{2}$, 
where higher-order corrections in $\alpha_{s}$ are expected to be negligible, and then a fixed order (
either LO or NLO or NNLO) DGLAP evolution is performed to specify the distributions at higher scales 
where data exist. A global fit to the data then determines the parameters of the input distributions. 
There is considerable freedom in choosing the parametric form of the parton distributions at scale 
$Q_{0}^{2}$ \cite{handbook_QCD}. The parameterization should be general enough to accommodate all 
possible $x$. A typical choice is:
\begin{equation}
\phi(x,Q_{0}^{2}) = A_{0} x^{A_{1}} (1-x)^{A_{2}} P(x),
\end{equation}
where $P(x)$ is a smooth function of $x$, $x^{A_{1}}$ and $(1-x)^{A_{2}}$ determines the small and 
large $x$ behavior, respectively, and $A_{0,1,2}$ are coefficients to be determined from fits to data.

When performing QCD fits, there are several conceptual difficulties to take into account. First of all, 
a QCD analysis of $F_{2}$ measurements involves the use of the gluon distribution which is, 
a priori, unknown. In fact, the gluon distribution is the most uncertain of the PDFs and is 
particularly ill-determined for $x$ $>$ 0.3, with uncertainties reaching 200\% by $x$ = 0.5 \cite{cteq}. 
This in turn translates in an uncertainty in the $\alpha_{s}$ determination from QCD analysis of PDFs 
since there is a correlation between the {\sl hardness} of the gluon and the magnitude 
of $\Lambda_{QCD}$, 
the quantity which sets the scale for $\alpha_{s}$ \cite{roberts}. The $xF_{3}$ measurements should be 
able to provide a precise value of $\Lambda_{QCD}$ since the gluon distribution does not enter into 
the evolution. However, the experimental uncertainties of $xF_{3}$ are still larger than those of $F_{2}$ 
and discrepancies in extracting $xF_{3}$ were observed between different experiments in 
the region of $x$ $>$ 0.4 \cite{tzanov_nutev}.  

Then, it should be pointed out that, a standard QCD analysis of PDFs does not take into account all 
residual $Q^{2}$ effects arising, for example, from higher-order radiative corrections in $\alpha_{s}$ 
or/and non-perturbative higher-twist corrections. In particular, the higher-twist terms 
are non-factorisable and process dependent and QCD has no rigorous prescription to account for it. As 
a result, most PDFs extractions are performed using {\sl safe} kinematic cuts for data selection in order 
to exclude regions where higher-twist or/and higher-order corrections in $\alpha_{s}$ play an important 
role. A typical set of cuts employed to select data for fitting is $Q^{2}$ $>$ 2 GeV$^{2}$ and 
$W^{2}$ $>$ 12 GeV$^{2}$ (this type of $W^{2}$ cut rejects the whole resonance region). 
Limiting the data coverage to a 
particular range in $Q^{2}$ and $W^{2}$ will result in a limitation in the $x$ coverage. For example, 
for a fixed $Q^{2}$, a $W^{2}$ cut of $W^{2}$ $>$ $W^{2}_{lim}$ will limit the $x$ range to 
$x$ $<$ $Q^{2}$/($W^{2}_{lim}$ - $M^{2}$ + $Q^{2}$). Considering that the $x$ dependence of the PDFs 
is parameterized empirically, as exemplified in Eq. 3, and that the parameterization coefficients are 
extracted from fits to data, these data selection cuts, though they make possible the extraction of 
PDFs without the complications specified above, yield to unconstrained strengths of the PDFs at large 
$x$ \cite{cteq}. 
 
To date, the large experimental uncertainties in the large $x$ regime, when excluding 
the low $W^{2}$ data, 
prevent answers to basic questions as to why the $d$ quark distribution 
appears to be {\sl softer} than the $u$ quark distribution. 
For the same reason, the $d$/$u$ behavior at large $x$, a critical test of the mechanism of 
spin-flavor symmetry breaking, is highly unconstrained. Furthermore, knowledge of PDFs at large $x$ 
is essential for determining high-energy cross sections at collider energies such as in search of 
new physics beyond the Standard Model, where structure information at large $x$ feeds down through 
perturbative $Q^{2}$ evolution to lower $x$ and higher values of $Q^{2}$ \cite{wally_duality}.
Thus, it is of paramount importance to decrease the uncertainties in the extraction of PDFs by 
deriving the parameterizations directly from data without resorting to theoretical assumptions 
alone for the extrapolation to $x$ $\sim$ 1. 

Extending to larger $x$ at a finite $Q^{2}$ means encountering the resonance region. 
An important consequence 
of duality is that the resonance and deep inelastic regions are deeply connected 
and properly averaged resonance region data could facilitate our understanding of 
the deep inelastic region. 
In some of the QCD analysis performed, the higher-twist terms have been extracted from data selected with 
a kinematic cut of $W^{2}$ $>$ 10 GeV$^2$ \cite{virchaux,yang,alekhin_03}. However, it was 
shown in several 
analyses \cite{simonetta,liuti_ioana} that only a relatively small higher-twist contribution consistent 
with the one obtained in Ref. \cite{virchaux,yang,alekhin_03} would be necessary to describe the 
entire $F_{2}$ structure function spectrum. Indeed, S. Liuti {\it et al.} analyzed resonance region data 
within a fixed-$W^{2}$ framework. This study found that the higher-twist 
contributions in the resonance region are similar to those from $W^{2}$ $>$ 10 GeV$^2$, 
with the exception 
of $\Delta$ region where the effects seem to be larger. This is in no way surprising if one thinks of it 
as a consequence of quark-hadron duality which ensures that, on average, higher-twists are small or 
cancel. As a consequence of duality, the wealth of resonance region data could be utilized to access the 
large $x$ region and constrain the PDFs in this regime. This approach, however, requires a 
very good understanding of 
the $Q^{2}$ dependence of the data in these kinematic regions of $x$ and $Q^{2}$ where the 
perturbative evolution is no more the only mechanism responsible for the $Q^{2}$ behavior. 

Figures 1 and 2 depict three pQCD parameterizations of the $F_{2}$ structure function at two $Q^{2}$ values, 
4 and 8 GeV$^{2}$, with a zoom-in of the large-$x$ region. 
The CTEQ6M parameterization shown is a QCD fit to hard scattering and DIS data (BCDMS, NMC, CCFR, E605, 
CDF, H1, ZEUS, D0) with $Q^{2}$ $>$ 4 GeV$^{2}$ and $W^{2}$ $>$ 12.25 GeV$^{2}$. 
The $x$ dependence of the PDFs 
is parameterized at a $Q^{2}$ of 1.3 GeV$^{2}$ and then the QCD 
evolution equations are utilized to evolve the distributions at higher Q$^{2}$ in the NLO (and LO). 
The authors employed the twist-2 pQCD formalism 
so the kinematic cuts used for data selection were tested to ensure that the introduction of 
simple phenomenological higher twist terms would not improve significantly the quality of the fit. 
The CTEQ6M fit shown in Figs. 1 and 2 is obtained in the $\overline{MS}$ (modified minimal subtraction) 
factorization scheme.  
One of the main improvements over earlier CTEQ fits is the addition to the global set of data of 
new measurements (H1, ZEUS, D0) which provide better constraints on the PDFs, in general, and on 
the gluon distribution at large $x$, the result being a harder gluon distribution in this region. 
The other noteworthy improvement is the full treatment of uncertainties of the PDFs and their 
physical predictions, using an eigenvector-basis approach. 

The MRST2004 parameterization is a QCD fit to a wide set of deep inelastic and related hard scattering 
data (BCDMS, SLAC, NMC, CCFR, CDF, H1, ZEUS, HERA, D0) with 
$Q^{2}$ $>$ 2 GeV$^{2}$ and $W^{2}$ $>$ 12.5 GeV$^{2}$. The $x$ dependence of the PDFs 
is parameterized at $Q^{2}$ of 1 GeV$^{2}$ and a fixed order, LO or NLO or NNLO, QCD evolution 
is performed to specify the distributions at larger $Q^{2}$ where data exist. A global fit to the data 
then determines the parameters of the input distributions. 
Though the fits are performed in the standard $\overline{MS}$ scheme, the gluon distribution is 
parameterized in the DIS (deep-inelastic scattering) factorization scheme and then transformed 
to the $\overline{MS}$ scheme. Together with more precise calculations of the 
splitting functions up to NNLO, this is actually the main improvement over earlier MRST fits 
(MRST2001 \cite{mrst_2001}). Indeed, the NLO 
global analysis with this new gluon parameterization appears to work extremely well when compared to 
Tevatron jet data and is even better for the NNLO fit. This objective 
couldn't be accomplished by previous MRST parameterizations.

Both CTEQ6M and MRST2004 are shown here with target mass corrections (TMC) included according to Ref. 
\cite{georgi_politzer_tmc}. 
CTEQ6M has more strength at large $x$ than MRST2004. For most part, 
this discrepancy originates from the fact that 
the two groups use different functional forms for the parameterization 
of the non-perturbative input parton distributions and neither parameterization is constrained by 
measurements in the large $x$ regime.

The ALEKHIN parameterization shown in Figs. 1 and 2 is an update of an earlier parameterization 
\cite{alekhin_03}, the significant improvement being the use of recent calculations of the exact 
NNLO evolution kernel. The data used were 
from SLAC, BCDMS, NMC, HERA, H1, ZEUS with kinematic cuts of $Q^{2}$ $>$ 2.5 GeV$^{2}$, $W^{2}$ $>$ 3.24 
GeV$^{2}$ and x $<$ 0.75. 
The model for the data description was based on pQCD with phenomenological parameterization of the 
twist-2 and 
higher-twist contributions to the structure functions. The analysis was performed in the $\overline{MS}$ 
scheme with the number of 
flavors fixed  at 3. The twist-2 PDFs were parameterized at $Q^{2}$ = 9 GeV$^{2}$. The pQCD analysis 
was done up to NNLO. 
Given the rather low $W^{2}$ cut used to select the 
data set, ALEKHIN parameterization includes, besides the typical parameters of pQCD, 
parameters to account for the target mass and dynamical higher-twist effects. This is a novelty, 
considering the standard procedure 
of performing QCD fits. The higher-twist contributions to the structure function were 
parameterized in additive form:
\begin{equation}
F_{2} = F_{2}^{LT,TMC} + \frac{H_{2}(x)}{Q^{2}},
\end{equation}
where $F_{2}^{LT,TMC}$ has contributions from the twist-2 terms with target mass corrections 
included according to Ref. \cite{georgi_politzer_tmc} and the dynamical twist-4 term $H_{2}(x)$ 
is parameterized in a model-independent way as a piece-linear function of $x$. The use of 
the exact NNLO corrections 
made possible an improvement in the positivity of the gluon distributions extrapolated to 
small $x$ and $Q$: in this 
parameterization the gluon distributions are positive up to $Q^{2}$ = 1 GeV$^{2}$, i.e. throughout the 
kinematic region where the parton model proved to be applicable. Since the ALEKHIN parameterization 
is based on fits to data with lower $W^{2}$ than CTEQ6M and MRST2004, its PDFs are expected to be 
better constrained at large $x$. 

{\bf Parameterizations of F$_{2}^{d}$}. The parameterizations discussed above provide parton 
distribution functions from which the nucleon structure function can be constructed in QCD frame-work. 
ALLM97 is a fit to just the nucleon (proton) structure function. In order to construct the structure 
function for a nucleon inside a nucleus substantial additional challenges need to be overcome. 
There are a host of well-documented issues in extracting nucleon
structure functions from nuclear data, even from deuterium data (see,
for instance \cite{Bodek:1983qn,Melnitchouk:1995fc,Kulagin:2004ie,Bosted:2007xd,Arrington:2008zh}). 
At large $x$ in particular, the effects of
Fermi motion, nuclear binding, the EMC effect, off-shell corrections,
and the like are quite large, and must be taken into account. Since
there is no consensus on how best to accomplish this, we have here
chosen to compare the measured deuterium resonance region data
directly to deep inelastic deuterium structure functions.
Specifically, we have chosen to multiply the array of structure
functions previously discussed by the following parameterization of 
d/p (deuteron over proton) \cite{d_p_antje}:
\begin{equation}
\frac{d}{2p} = 0.9851 - 0.5648 x - 0.0904 x^{2} + 0.7183 x^{3} - 0.3428 x^{4}.
\end{equation}
This equation is the result of a data fit up to $x$ = 0.8, and may not be constrained
correctly at the highest $x$. Moreover, it assumes no $Q^{2}$-dependence, a
need for which has been indicated in other works \cite{vladas_thesis}. A more thorough
approach might in the future consider specifically structure functions
formed from nuclear PDFs, such as those found in \cite{Hirai:2007sx,Schienbein:2007fs,
deFlorian:2003qf,Eskola:2007my}.

{\bf Target Mass Corrections}. At large enough values of $Q^{2}$ and $W^{2}$, QCD 
provides a rather clear and rigorous perturbative description of 
the physics that generates the $Q^{2}$ behavior of the structure function. 
When $W$ $\rightarrow$ $M$, where $M$ is the proton mass, both the 
nonperturbative kinematical power corrections (target mass corrections) and the dynamical higher 
twist have to be taken into account. 
Since these characterize the long-range non-perturbative interactions between quarks and gluons, 
the dynamical higher-twist terms contain information about the dynamics of confinement. 
The target mass corrections, on the other hand, arise from purely kinematic effects 
associated with finite values of $Q^2$/$\nu$ = $4M^{2}x^{2}$/$Q^{2}$. The target mass terms are related 
to the twist-2 operators and 
contain no additional information on the non-perturbative multi-parton correlations. In consequence, 
target mass effects should either be corrected for in the data or the effect should be included 
in the QCD fits if one aims for a consistent comparison of data to QCD fits. The target 
mass effects were taken into account in the CTEQ6M, MRST2004 and ALEKHIN parameterizations of the 
structure function according to the prescription of Georgi and Politzer  
\cite{georgi_politzer_tmc}. It is non-trivial to note that there is not an universally agreed-upon 
prescription to account for target mass \cite{accardi_tmc1,accardi_tmc,ingo_tm}, and so the choice of 
approach inherently introduces some uncertainty to this analysis.


\section{Experiment and Data Analysis}

 The experiment E00-116 was carried out in Summer 2003 in Hall C, at JLab. 
A fixed electron beam of energy 5.5 GeV came 
incident on cryogenic targets. The target system consisted of 4 cm long liquid hydrogen and 
deuterium, contained in circular aluminum cans. Scattered electrons were detected 
in the High Momentum 
Spectrometer (HMS). The Short Orbit Spectrometer (SOS) was used for detection of positrons, 
which was used to estimate possible electron background originating from charge-symmetric 
processes such as $\pi^0$ production and subsequent decay in the target. The data were taken 
at various scattering angles and momenta as follows: for each fixed spectrometer angle, 
the central momentum was varied in order to cover a region in $W^{2}$ from about 
1.2 to 4.5 GeV$^{2}$. The kinematics covered by this experiment are shown 
in Table 1. These data extend the existing Hall C resonance region measurements at 
larger $x$ and $Q^{2}$ \cite{F2JL1,F2JL2}.

\begin{table}
  \squeezetable
  \label{kinem}
  \caption{The kinematic regime covered by E00-116 at a beam energy of 5.5 GeV.}
  \begin{tabular}{cccc}
    \hline
    \hline
    Angle(deg) & Momentum(GeV/$c$) & $x$ & $Q^{2}$((GeV)$^{2}$)\\
    \hline
      & 2.26 & & \\
	{37.93} & 1.94 & 0.48-0.92 & 3.58-5.48 \\
	& 1.67 & & \\
	\hline
	& 2.17 & & \\
	    {41} & 1.86 & 0.53-0.94 & 3.99-5.86 \\
	    & 1.60 & & \\    
	    \hline
	    & 1.94 & & \\
		{45} & 1.67 & 0.55-0.95 & 4.28-6.29 \\
		& 1.44 & & \\     
		\hline
		& 1.34 & & \\
		    {55} & 1.16 & 0.60-0.94 & 5.01-7.07 \\
		    & 1.47 & & \\   
		    \hline
		    & 1.31 & & \\
			{60} & 1.19 & 0.52-0.95 & 4.52-7.38 \\
			& 1.04 & & \\
			& 0.89 & & \\
			\hline
			    {70}& 0.91 & 0.60-0.83 &5.38-7.11 \\
			    & 0.80 & & \\
			    \hline
			    \hline
  \end{tabular}
\end{table}
             
\subsection{Experimental Setup}

\subsubsection{Beam Line}

During E00-116, the Continuous Electron Beam Accelerator Facility (CEBAF) at JLab 
provided an unpolarized, 
Continuous Wave (CW) electron beam of 5.5 GeV, with currents up to 100 $\mu$A. 
The beam was steered from the Beam Switch Yard to the 
experimental hall through the beam line. Hall C beam line is equipped with magnets used to focus 
and steer the beam, as well as several monitors needed to measure the energy, current, position 
and profile of the beam.
The profile and the absolute position of the beam is monitored utilizing superharps. A superharp 
consists of a frame and three tungsten wires (two horizontal and one vertical) which are moved back 
and forth through the beam to determine the centroid position to about 10 $\mu$m. However, the 
superharp cannot be used during the data taking because it has a destructive interaction with the 
beam. Therefore, Beam Position Monitors (BPMs) \cite{bpm_ref} are used to continuously 
monitor the relative beam 
position during data taking. The BPMs are nondestructive to the beam and are calibrated with 
superharp scans. During this experiment, the typical relative variation of the beam 
position at the target was found to be less than 0.2 mm.

The beam energy is measured using the superharps and the dipole magnets in the beam line. 
Due to the fact that the dipole fields are accurately mapped and that the beam path is 
determined with high precision by the superharps, the accuracy of the absolute beam 
energy measurement is at the level of 5 $\times$ 10$^{-4}$ GeV.

The beam current and charge in Hall C is measured by a system of beam current monitors (BCMs) 
together with a parametric current transformer (Unser) \cite{armstr_thesis}. All these 
monitors are placed in the 
beam line before the target in the following order: BCM1, Unser, BCM2 and BCM3. Although the 
BCMs have a very stable offset, the gain drifts with time and the Unser is used for BCMs gain 
calibrations. Dedicated calibration runs are typically performed to minimize the effects 
of drifts in the BCMs gains. For this experiment, BCM2 was used for monitoring 
but due to time constraints, no BCM calibration runs were taken. 
However, the experiment that ran just before E00-116 had the same set-point 
for the BCM2 gain, such that it was possible to use their calibration runs taken 5 weeks and 
1 week before this experiment \cite{m_dalton_baryon}. 
\begin{figure}
\centering 
\includegraphics[width=8.8cm]{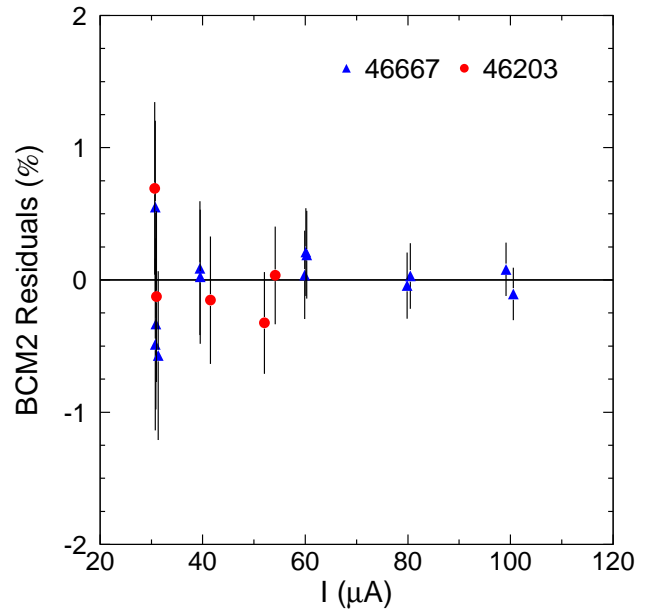}
\caption{(Color online) The difference between the current as given by the BCM2 after calibration and the 
current as given by the Unser. The two calibration runs, 46203 and 46667 were taken 
one month apart and yielded very similar results for the BCM2 gain and offset (see text).}
\end{figure}
The difference in BCM2 gain and offset when using each run individually or 
the combined runs was below 0.1\%, indicating that there were no significant drifts 
in the BCM2 gain over a month period. Figure 3 shows that the difference between the current as given 
by the BCM2 after calibration and the current as given by the Unser (the residuals) are within 
0.15\% above 50 $\mu$A.
The calibration result from the combined runs was used to calculate the current and the charge 
for this experiment.
For this experiment current regime, the normalization uncertainty in the current measurement 
was estimated to be $\sim$0.3\% at 100 $\mu$A and originated from possible small drifts of the BCM gain, 
from the precision 
of the BCM calibration and from the accuracy of the Unser in measuring the current, the latter bringing 
the largest contribution. The point-to-point uncertainty was estimated to be 0.05\% by taking the 
difference in the normalization uncertainties propagated at 80 $\mu$A and 100 $\mu$A, this being roughly 
the range in the beam current used in this experiment.

The electron beam generated by CEBAF is a high current beam with a very small transverse size 
(100-200 $\mu$m FWHM). To prevent damage to the targets and to minimize the changes in the 
cryotarget densities due to localized boiling, a rastering system is used to distribute the 
deposited energy of the beam in an uniform manner over the target volume. The raster consists 
of two sets of steering magnets: the first set of magnets rasters the beam vertically and the 
second horizontally. For this experiment the raster consisted of a 2$\times$2 mm uniform 
structure. A detailed description of the Hall C raster system is given in Ref. \cite{raster_guy}.

\subsubsection{Target}
This experiment used the standard liquid hydrogen and liquid deuterium cryogenic target system 
in Hall C. The liquid targets were contained in aluminum cans. Data were taken on aluminum 
foils (dummy target) for background measurement and 
subtraction. Also several runs were taken using a carbon foil target in order to determine the 
beam offsets relative to the pivot of the target. The cryogenic target cells were mounted on a 
cryostack together with the combination of carbon and aluminum target sled. The cryogenic 
system ensured that the temperature and density of the liquid targets were maintained during 
data taking at optimum values of 19 K and 0.0723 g/cm$^3$ for hydrogen and 22 K and 0.1674 g/cm$^3$ 
for deuterium.

To accurately determine the experimental 
luminosity it is necessary to have a precise knowledge, among others, of the 
targets' density and thickness. Though the cryogenic system is designed to ensure that 
the liquid hydrogen and deuterium targets are maintained at a fixed nominal 
temperature in all conditions, in reality, when the beam passes 
through the targets and deposits heat there are local changes in the temperature 
and density of the cryogen (boiling effect). Dedicated data (luminosity scans) are taken 
to study and correct for this effect. During E00-116, luminosity scans 
were performed on both hydrogen and deuterium targets. It was found that the boiling effect 
gives a small correction of (0.35 $\pm$ 0.32)\% / 100 $\mu$A to the luminosity for both 
cryogenic targets as seen in Fig. 4. This parameterization was utilized on a run-by-run basis 
to correct for the boiling effect for both liquid targets. 
The majority of the data were collected at $\sim$100 $\mu$A. The density correction at this current 
is of the size of the uncertainty of the fit, therefore the normalization uncertainty was taken to be 
0.35\%. 
The point-to-point systematic uncertainty 
on the density correction, originating from the uncertainty in the current, is negligible. 
\begin{figure}
\centering 
\includegraphics[width=8.8cm]{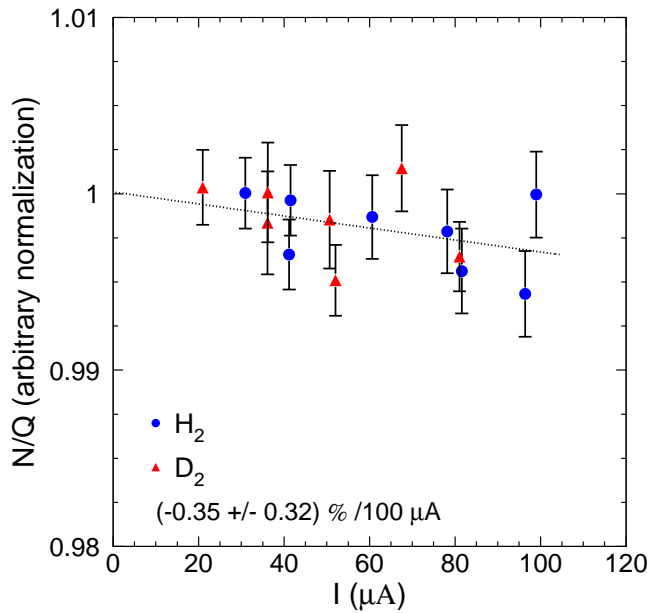}
\caption{(Color online) Relative hydrogen and deuterium target yield versus beam current. The correction for 
the boiling effect is obtained from a linear fit to the data shown as the dotted line.}
\end{figure}

Because of the circular geometry of the cryogenic target cell, a careful analysis is 
required to determine the effective target length that will enter in the 
calculation of the luminosity. If the beam was exactly 
aligned along the diameter of the target, the effective target length 
will simply be the outer diameter of the target cell minus the cell walls. If 
there is a displacement between the beam and the center of the target, then the 
effective target length will be $2\sqrt{r^{2} - dx^{2}}$, where $r$ is the 
inner radius of the cell and $dx$ is the beam offset from the center of the target.
 For E00-116, several sources of information were used to determine the effective 
target length: the target survey that provides measurements, at room temperature, 
of the outer diameter of the cell together with the thickness of the cell walls, 
a survey of the target position relative to the pivot, and dedicated data taken on a 
central Carbon foil that provide the beam offset relative to the 
pivot \cite{simona_thesis}. The effective target length used in the 
cross section extractions is listed in Table 2.
\begin{table}
  \squeezetable
  \label{cryo_lengths}
  \caption{E00-116 effective lengths of the cryogenic targets.}
  \begin{tabular}{cc}
    \hline
    \hline
    Target& Effective Target Length(cm)\\
    \hline
    Hydrogen     & 3.946 $\pm$ 0.029\\
    Deuterium    & 3.927 $\pm$ 0.029\\
    \hline
    \hline
  \end{tabular}
\end{table}

\subsubsection{Spectrometers}

In what follows, a summary of the main characteristics of Hall C spectrometers will be given with 
emphasis on the aspects relevant to E00-116. 
Detailed information about the Hall C HMS and SOS can be found in Ref. \cite{donprl,blok}.
The HMS is a magnetic spectrometer consisting of a 25 deg vertical bend dipole magnet for 
momentum dispersion and three quadrupole magnets for focusing. All magnets are 
superconducting. For this experiment, the HMS was operated in the point-to-point optical 
tune. The range used in the momentum ($E'$) acceptance, $\delta$ = $\frac{\Delta p}{p}$, 
was of $\pm$8\% while the range in the angular ($\theta$) acceptance, $\Delta(\theta)$ was $\pm$35 mrad. 
The SOS consists of a quadrupole magnet and two dipole magnets. 
For E00-116, the point-to-point optical tune was used. 
The range used in $\delta$ 
was of (-15,+20)\% while the range in the angular acceptance was $\pm$60 mrad.

The detector packages for the two spectrometers are very similar and consist of two 
drift chambers for track reconstruction, 
scintillators arrays for triggering, a threshold gas Cerenkov and an electromagnetic 
calorimeter, which were both used in this experiment for particle identification (PID) 
and pion rejection.  
\begin{figure}
\centering 
\includegraphics[width=8.6cm]{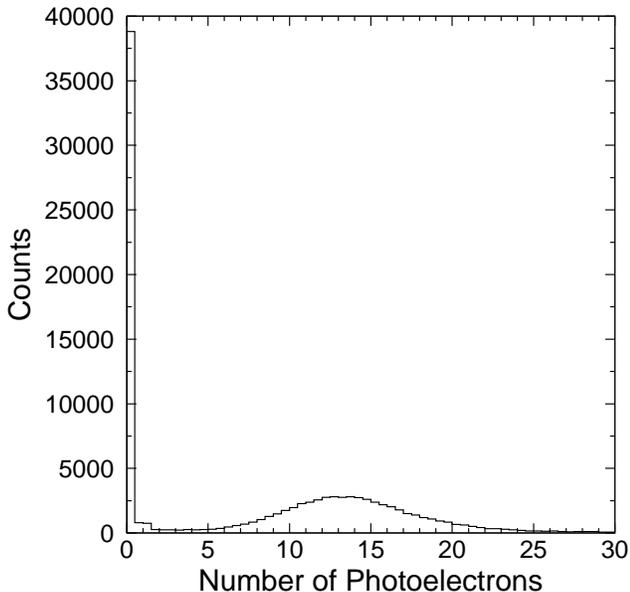}
\caption{An example of the distribution of number of photoelectrons collected 
in the HMS Cerenkov detector. The pion peak appears at zero while electrons 
produce on average about 13 photoelectrons. As it will be discussed later in the text, a cut of 
number of photoelectrons = 2 was used to separate electrons from pions.}
\end{figure}
\begin{figure}
\centering 
\includegraphics[width=8.6cm]{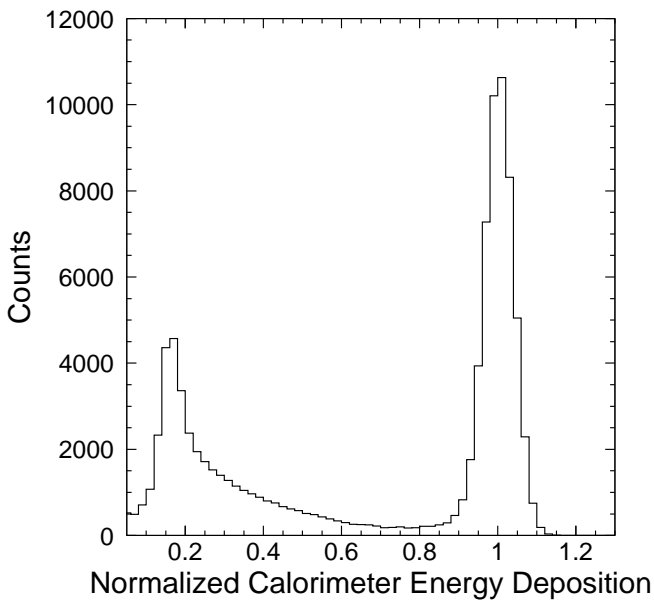}
\caption{An example of the distribution of fractional energy deposited in the HMS calorimeter. Electrons 
deposit their entire energy in the calorimeter peaking at 1 in the distribution while pions loose just 
a fraction of their energy. In the data analysis, a cut of fractional energy deposited = 0.7 was used 
to separate electrons from pions.}
\end{figure}
The HMS Cerenkov counter was used to distinguish between $e^{-}$ and $\pi^{-}$ with 
momenta between 0.8 and 2.3 GeV. For this purpose, the Cerenkov tank was filled with Perfluorobutane 
($C_{4}F_{10}$, $n$ = 1.00143 at 1 atm and 300K) at about 0.9 atm making the  
detector fully sensitive to $e^{-}$ but insensitive to $\pi^{-}$ in the momentum range specified above. 
The SOS Cerenkov 
counter was used to detect $e^{+}$ and reject $\pi^{+}$ with momenta ranging from 0.8 to 1.7 GeV.
The Cerenkov tank was filled with Freon-12 ($n$ = 1.00108 at 1 atm) at about 
1 atm giving a Cerenkov threshold of 3 GeV for $\pi^{+}$ and 11 MeV for $e^{+}$.
For E00-116, a typical spectrum of the HMS Cerenkov number of photoelectrons is shown in Fig. 5. In this 
distribution, the $\pi^{-}$ events peak at zero while the $e^{-}$ events give an average of about 
13 photoelectrons.  

The HMS and SOS calorimeters are identical except for their size. Each calorimeter consists 
of 10 cm $\times$ 10 cm $\times$ 70 cm blocks of TF-1-1000 type lead glass ($\rho$ = 3.86 
g/cm$^{3}$, $n$ = 1.67 and radiation length r.l. = 2.5 cm). 
The HMS calorimeter is 13 blocks high while the one in the SOS 
just 11. The calorimeters are rotated by 5 deg from the spectrometers optical axis in order to 
reduce eventual losses through the cracks between the blocks. For this experiment, the 
calorimeters were used to detect $e^{-}$ (HMS) and $e^{+}$ (SOS). The hadrons that could 
have reached the calorimeters were mostly $\pi^{-}$ or $\pi^{+}$. The $e^{-}$ ($e^{+}$) were 
distinguished from $\pi^{-}$ ($\pi^{+}$) according to their fractional energy, total energy 
deposited in the calorimeter, normalized by the momentum. The $e^{-}$ ($e^{+}$) deposit their 
entire energy in the detector peaking in the fractional energy spectrum at 1; the 
$\pi^{-}$ ($\pi^{+}$) deposit around 0.3 GeV and they will peak in the fractional energy 
distribution at 0.3 GeV/$E^{'}$. A typical distribution of the 
fractional energy deposited in the HMS calorimeter is shown in Fig. 6. 

\subsection{Data Analysis}
The inclusive electroproduction cross section can be expressed as:
\begin{equation}
\frac{d^{2} \sigma}{d\Omega dE^{'}}= (N_{measured} - BG)\frac{1}{N_{e} N_{t}} \frac{1}{d\Omega dE^{'}}\frac{1}{A}\frac{1}{\varepsilon} .
\end{equation}

Here, $N_{e}$ is the number of incident electrons and $N_{t}$ is the 
number of target particles per unit area, which can be calculated 
in terms of the mass density $\rho$, the atomic number A and the thickness $x$ from 
$N_{t} = \frac{\rho N_{A} x}{A}$ ($N_{A}$ is Avogadro's number). $N_{measured}$ 
is the number of scattered electrons $observed$ in the solid angle $d\Omega$ and 
in the energy range $dE^{'}$. $BG$ is the background, $A$ is the detector acceptance 
and $\varepsilon$ is the detector efficiency. The most significant corrections that were applied 
to $N_{measured}$ will be discussed below.

\subsubsection{PID Cut Efficiency}
The rejection of negatively charged pions was accomplished by placing requirements on both 
the number of photoelectrons collected by the Cerenkov detector, number of photoelectrons 
larger than 2, and the fractional energy deposited by the particle in the calorimeter, 
fractional energy larger than 0.7. In what follows, the efficiency of these cuts in not 
rejecting valid electrons will be discussed.

{\bf Cerenkov Cut Efficiency.}
In order to determine how many electrons are lost when applying the Cerenkov cut 
number of photoelectrons larger than 2, it is 
important to work with a clean sample of electrons (no pion contamination). Once a clean 
sample of electrons is selected, than the Cerenkov cut efficiency is determined from the ratio of number 
of events that pass the cut to the total number of events in the clean sample. If the sample 
is pion contaminated then the Cerenkov cut efficiency will be artificially 
lower. Unfortunately, it was impossible for this experiment to select a clean sample of 
electrons just with a calorimeter cut. For the particular kinematics of E00-116, 
the pion to electron ratio was rather large (up to 150:1). The pions can 
undergo charge-exchange reactions and deposit up to their entire energy in the calorimeter. This 
could result in a {\sl high-energy tail} for pions which could extend beyond 1 in the fractional energy 
spectrum making the selection of a clean sample of electrons practically impossible even with high 
calorimeter cuts. The unbiased electron cut 
efficiency for the Cerenkov was determined by extrapolating to zero pion to electron ratio. 
It was found to be (99.60 $\pm$ 0.24)\%, which in very good agreement with the findings of 
other experiments that ran at similar conditions \cite{edwin_exp}. This value was used as a 
correction for the data. The normalization 
systematic uncertainty was taken to be the uncertainty of the fit extrapolation at zero pion to 
electron ratio, 0.24\%.

{\bf Calorimeter Cut Efficiency.}
Just as for the Cerenkov, the estimation of how many valid electrons are lost when 
using a cut on the fractional energy 
deposited in the calorimeter was complicated by the fact that, for this experiment, 
the large pion to electron ratio made impossible the selection of a clean sample of electrons 
using just a cut on the number of photoelectrons acquired in the Cerenkov detector. The same 
approach was taken in this case as for the estimation of the Cerenkov cut efficiency: the 
calorimeter cut efficiency was extrapolated to zero pion to electron ratio in order to find the 
true electron efficiency. The extrapolation was done for each momentum setting separately in order 
to deconvolute the efficiency dependence on the pion to electron ratio from the dependence on the 
resolution of the calorimeter. An example of the efficiency extrapolation at zero pion to electron 
ratio for one momentum setting is shown in Fig. 7.
\begin{figure}
\centering 
\includegraphics[width=8.8cm]{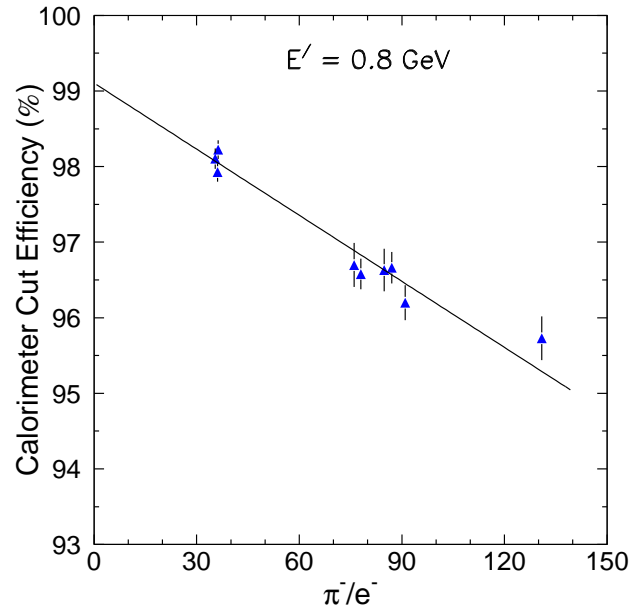}
\caption{(Color online) The electromagnetic calorimeter cut efficiency versus $\pi$/$e$ ratio. The fit 
represented by the solid line extrapolates the efficiency at zero $\pi$/$e$ ratio 
in order to obtain the true electron efficiency.}
\end{figure}
\begin{figure}
\centering 
\includegraphics[width=8.8cm]{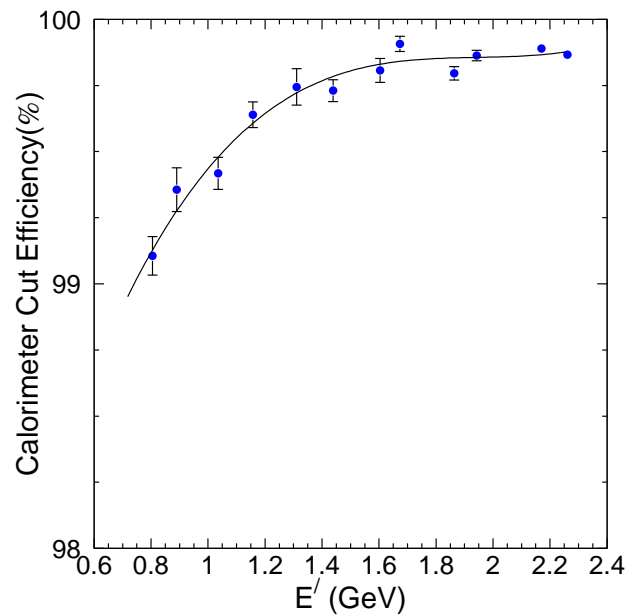}
\caption{(Color online) The electromagnetic calorimeter cut efficiency versus momentum. The solid line is a fit 
to the data and parameterizes the cut efficiency dependence on the momentum of the particle. 
This parameterization 
was used to correct for the loss of valid electrons due to the calorimeter cut inefficiency.}
\end{figure}
The efficiency, obtained in this manner, and parameterized as a function of momentum is shown in Fig. 8. 
The parameterization was used 
as a correction in the data analysis. Our parameterization was compared to the result obtained 
from an experiment that ran in similar experimental conditions but at 
different kinematics 
where the pion to electron ratio is small \cite{edwin_exp}. The two results were found to agree within 0.3\%. 
The normalization and point-to-point uncertainties on this correction were estimated to 
be 0.3\% and 0.25\%, respectively.

\subsubsection{Backgrounds}
There are three physical processes that are possible sources of background for this experiment: 
electrons scattered from the target aluminum walls, negatively charged pions that are not 
rejected by the PID cuts, and electrons originating from other processes like charge 
symmetric processes which produce equal number of positrons. Each of these possible sources 
of background will be discussed in what follows.

{\bf Target Cell Background.}
During data taking on the cryogenic targets, some of the incoming electrons scatter on 
the aluminum walls of the target cell and end up being detected at the same kinematics as the 
electrons that scatter from the cryogen. This background has to be determined and subtracted from 
the measured yields in order to obtain the yields for scattering from the cryogen only. To 
determine this background dedicated data were taken on a dummy target at exactly the same 
kinematics as on hydrogen and deuterium. In order to minimize the data acquisition time, 
the total thickness of the dummy target was about eight times the total cell wall thickness 
seen by the beam. The background coming from the scattering from the target cell walls $BTW(E^{'},\theta)$, 
was determined as:
\begin{equation}
BTW (E^{'}, \theta) = \frac{T_{w} Q_{w} R^{ext}_{d}}{T_{d} Q_{d} R^{ext}_{w}} N_{d} (E^{'}, \theta),
\end{equation}
where $\theta$ is the spectrometer angle, $Q_{w(d)}$ is the total charge incident on the cell 
walls (dummy), $T_{w(d)}$ is 
the total thickness of the cell walls (dummy), $N_{d}$ ($E^{'}$, $\theta$) is the number of events 
collected from the dummy run after applying efficiencies and dead time corrections and 
$R^{ext}_{w(d)}$ is the external radiative correction (external bremsstrahlung emission) for 
the cell walls (dummy).
The target cell background subtraction was performed for each hydrogen and deuterium 
run on a ($E^{'}$,$\theta$) bin-by-bin basis. The size of this background was at most 18\% 
and its uncertainty 
was dominated by the statistical uncertainty on $N_{d}(E^{'},\theta)$ and by the uncertainty 
in measuring the thickness of the cell walls and dummy. The thickness of the cell walls 
was known up to 1\% while, by comparison, the uncertainty in the dummy thickness measurement 
was negligible \cite{meekins}. This led to a systematic uncertainty in the 
cross section of at most 0.2\%. 
The statistical uncertainty on $N_{d}(E^{'},\theta)$ was propagated to the uncertainty 
of the cross section.

{\bf Pion Background.}
Even after applying the PID cuts, some pion background may still be present. 
Although pions do not produce Cerenkov light 
directly, they can generate, through ionization, $\delta$-rays in the 
materials preceding the Cerenkov detector (electron knockout). 
These knock-on electrons could have high enough energy to emit Cerenkov light 
and pass the PID Cerenkov cut. In the electromagnetic calorimeter, the pions give signal 
according to their energy loss but through a charge-exchange 
reaction they can produce neutral pions which decay into $\gamma$$\gamma$ or 
$e^{-}e^{+}\gamma$.
In this way, the entire energy of the pion can be deposited in the calorimeter. This process 
typically gives the high-energy ``tail'' for pions that extends to deposited fractional energy  
of 1. 
\begin{figure}
\centering 
\includegraphics[width=8.8cm]{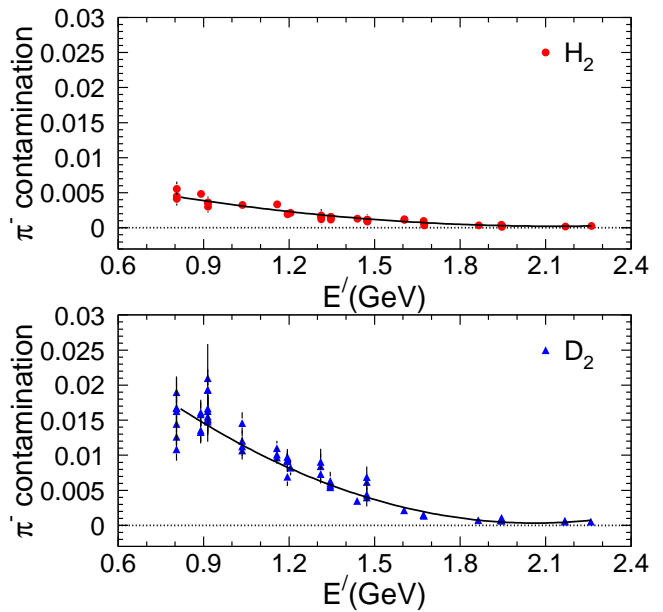}
\caption{(Color online) The pion contamination (see text) as a function of momentum. As expected, the pion 
contamination is larger for the deuterium target (bottom panel) than for the hydrogen target 
(top panel). The solid lines represent parameterizations of the pion contamination as a function 
of momentum. These parameterizations were used as corrections in the cross section extraction.}
\end{figure}
For this experiment, the pion background was estimated using a method developed for the 
Hall C E99-118 analysis \cite{vladas_thesis} in which the pion rejection factor is used to 
normalize the pion fractional energy distribution in the calorimeter. The number of events 
in this normalized distribution that pass the PID calorimeter cut of 0.7 represents 
the pion contamination. The result of the 
pion contamination estimation for this experiment is shown in Fig. 9 for both hydrogen and 
deuterium targets. As expected for a heavier target, the pion contamination for deuterium (maximum 
of about 1.7\%) is about 3 times larger than for hydrogen (maximum of 0.5\%) at the lowest momentum.
For this background subtraction, parameterizations as a function 
of momentum were used as corrections (no angle dependence was 
observed). The point-to-point uncertainty on this correction was determined to be 0.2\% for 
both targets.

{\bf Charge-Symmetric Background.}
For electron-proton scattering there is a significant probability to produce a 
neutral pion in the target which then decays into $\gamma$$\gamma$ or 
$e^{-}e^{+}\gamma$. The photons can further convert into electron-positron pairs in the target 
material or in the materials preceding the detectors. Photons can also be produced through 
the Bethe-Heitler process. However, the leptons resulting from Bethe-Heitler processes are 
very forward peaked so their contribution is significant only at forward angles. 
The outcome is that the secondary electrons will end up being detected together with the scattered 
electrons. For the kinematics of this experiment (backward angles) the 
dominant source of secondary electrons is the neutral pion production in the 
target and the subsequent decays. The electron production through Bethe-Heitler 
process is negligible. The fact that the background electrons are produced in 
pairs with positrons (charge-symmetric background) can be exploited experimentally 
and the background electrons can be disentangled from the scattered electrons by detecting 
positrons. 

For E00-116, due to the limited running period (less than a week), it was decided to take advantage of 
the availability of the SOS. The SOS has a larger momentum acceptance than the HMS 
and two SOS momentum settings could easily cover an HMS scan with three momentum settings. 
However, by using a different spectrometer 
for positron measurements the photon to electron-positron pairs conversion factor 
is different as the photons encounter different radiation lengths of material. In addition, the SOS 
acceptance function and the detector inefficiencies are different than the HMS ones. 
Taking into account these considerations, we decided that 
an accurate estimation of the charge-symmetric background would require the extraction of the 
positron cross sections rather than the yields as it was done for previous Hall C experiments. 
This way, the charge-symmetric background would be corrected by subtracting the measured 
positron cross section from the measured electron cross section bin-by-bin on a ($E^{'}$,$\theta$) 
grid. 

The first step in the positron cross section analysis was to perform the detectors calibrations. 
Once the calibrations were performed, the positron yield selected with 
PID cuts was binned in the ($\delta$,$\theta$) acceptance around the central values. 
The yield was corrected for the electronic and computer dead times and for the tracking inefficiency. 
In order to obtain the positron yields from the cryotargets alone, the endcap contributions had 
to be subtracted. Also the pion contamination was determined and parameterized as a function of 
momentum for each cryotarget and applied as correction to the yield. Next, the 
spectrometer acceptance corrections were calculated and applied to the yield. Thus, the positron 
cross section was obtained on a ($\delta$,$\theta$) grid. 

Our goal was to determine the cross section at the central angle as a function of momentum but 
still to keep the statistics accumulated. This could be done, in principle, by statistically 
averaging the measured cross section over the angular acceptance. However, the variation 
of the positron cross section across $\theta$ acceptance was non-negligible. Therefore, before 
averaging, a model was needed to remove the cross section $\theta$ dependence 
(the so called bin-centering correction). 
The positron cross section model used for this purpose was developed by 
P. Bosted \cite{pos_peter}. The model uses a fit to the charged pion production data 
accumulated at SLAC \cite{pion_data_slac}. The neutral pion production is estimated as the 
average of the positive and negative pion production. The positron cross section is calculated 
using the decay branching ratios for a neutral pion and the radiation length of the material 
where a photon that results from the decay can produce electron-positron pairs. Taking into 
account that the positron cross section model described above was used just for bin-centering, 
the main requirement was that the model should describe {\sl the shape} of the angular 
dependence of the positron cross section. To make sure that this 
requirement was met, first it was checked if, after applying the bin-centering correction, there 
is any angular dependence left across the acceptance. A typical example is shown in Fig. 10 where 
it can be seen that, within the statistical uncertainty, the bin-centering correction removes the 
angular dependence of the cross section. 
\begin{figure}
\centering 
\includegraphics[width=9.1cm]{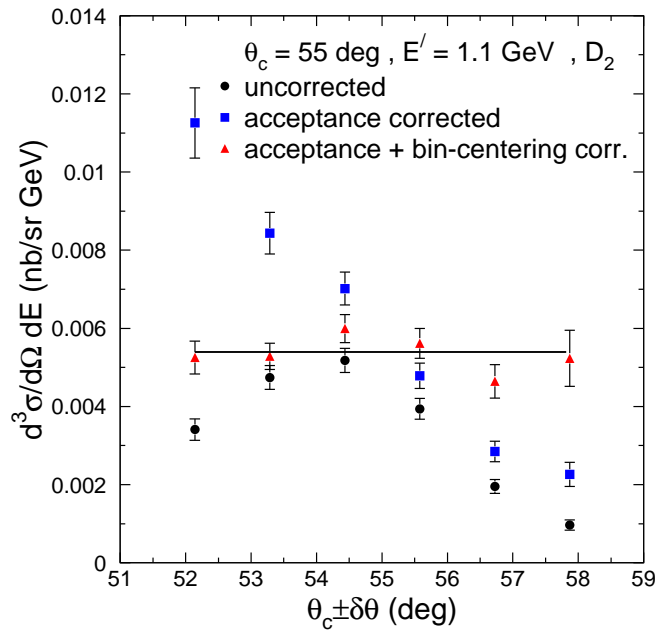}
\caption{(Color online) An example of the measured positron cross section across the SOS angular acceptance. The cross 
section is shown at various stages in the analysis. In black circles the cross section is depicted 
before acceptance and bin-centering corrections. The blue squares show the cross section after 
acceptance corrections were applied. The angular dependence of the positron cross section is obvious. 
The red triangles represent the cross section after both the acceptance and bin-centering corrections 
were applied. It can be seen that the bin-centering corrections removed the angular dependence of the 
cross section.}
\end{figure}

It was also checked that the data overlap in the angular acceptance from one central angle setting 
to the next if the model would be used to center the data at certain angle values in the acceptance. 
Good agreement was found for neighboring scans in the overlapping region.
\begin{figure}
\centering 
\includegraphics[width=8.9cm]{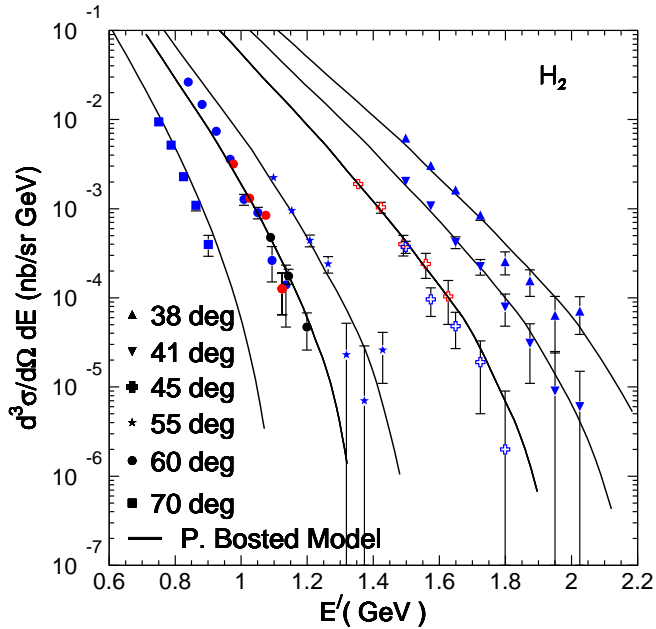}
\caption{(Color online) The measured positron cross section on hydrogen as a function of momentum compared 
to the model of P. Bosted \cite{pos_peter}.}
\end{figure}
\begin{figure}
\centering 
\includegraphics[width=8.9cm]{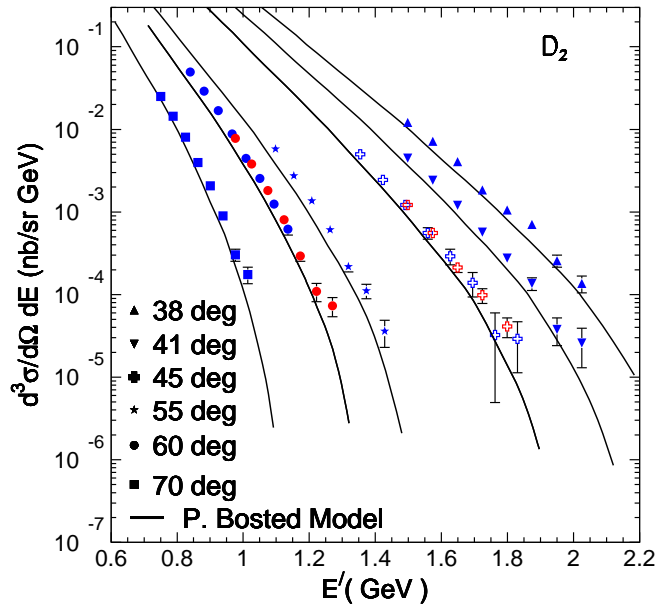}
\caption{(Color online) The measured positron cross section on deuterium as a function of momentum compared 
to the model of P. Bosted \cite{pos_peter}.}
\end{figure}

Finally, the positron cross section was extracted at fixed central angles as a function of 
momentum. Figures 11 and 12 show the positron cross section for both hydrogen and deuterium targets 
compared to the model of P. Bosted. It can be seen that the model describes qualitatively well 
the momentum dependence of the cross section. 

As stated previously, the electron data, both the scattered and the background electrons, were 
taken using HMS while the charge-symmetric background was measured using SOS. Therefore, at the 
end of the experiment, a setting was taken at the same kinematics in HMS and SOS 
(both spectrometers were set on negative polarity). 
The result of the analysis of this scan in the two spectrometers is shown in Fig. 13. 
It was found that the analysis in the two spectrometers agreed within 1.3\%. This translated in a 
normalization uncertainty in the scattered electron cross section below 0.2\%, considering that 
the relative contribution of the charge symmetric background to the measured cross section was 
at most 15\%.
\begin{figure}
\centering 
\includegraphics[width=8.8cm]{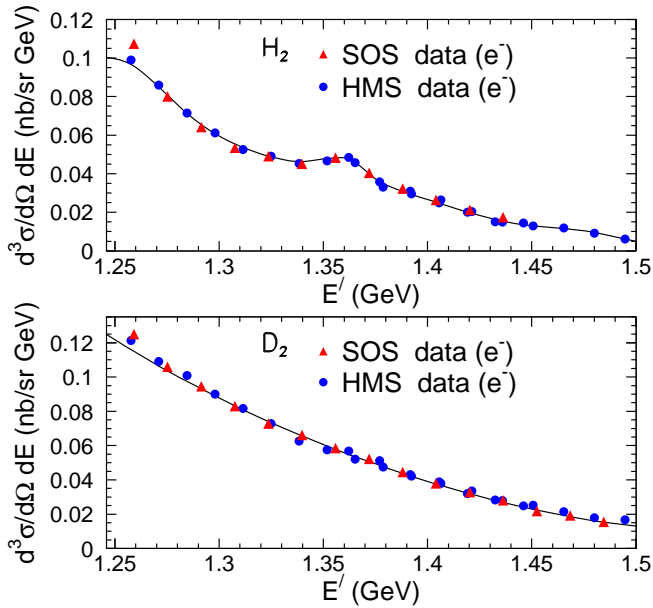}
\caption{(Color online) The comparison of SOS and HMS analyses for $H(e,e')$ (top panel) and $D(e,e')$ (bottom panel).}
\end{figure}

In the end, the charge symmetric background was subtracted bin-by-bin in ($E^{'}$,$\theta$). For the 
subtraction, the positron cross section had to be centered at the scattered electron data kinematics. The 
model of P. Bosted was used for this purpose. 
Quantitatively, it was found that the use of the model for 
bin-centering corrections in momentum introduced an uncertainty of 6\% in the positron cross section at 
38 deg, 41 deg, 45 deg, 55 deg and of 20\% at 60 deg and 70 deg. This translated in an uncertainty 
on the electron cross section up to 2\% at the lowest momentum at 60 deg and 70 deg, but below 
0.2\% for the rest of the data.

\subsubsection{Acceptance Corrections}

For a fixed angle and momentum setting, the spectrometers have a finite angle and 
momentum acceptance. 
This experiment used the same procedure of extracting the spectrometer's 
acceptance functions as previous Hall C experiments. This procedure is described in detail 
elsewhere \cite{eric_elastic}. 

For this analysis, the acceptance correction was applied on a bin-by-bin basis in ($\delta$,$\theta$). 
The point-to-point uncertainty on the acceptance correction in HMS was estimated to be 0.8\%. This is 
dominated by the position uncertainties on the target, collimator, magnets, and detector 
package. The normalization uncertainty on the acceptance correction 
was determined by combining in quadrature an uncertainty of 0.7\% coming from the reduction in the 
solid angle and an uncertainty of 0.4\% due to modeling of the HMS optics \cite{eric_elastic}.

\subsection{Cross Sections Extraction}
For this experiment, the electroproduction differential cross section $H(e,e')$ and 
$D(e,e')$ was extracted 
according to Eq. 6, binned in 16 and 20 bins in momentum and angle, respectively. Our goal was to 
obtain the one photon-exchange (Born) cross section at a fixed angle 
as a function of momentum. For this, two additional corrections were necessary: the 
bin-centering correction, which makes possible the extraction of the differential 
cross section at a fixed angle without sacrificing statistics and the radiative 
corrections which are necessary to obtain from the measured cross section 
the one photon-exchange contribution. Considering that both the bin-centering 
corrections and the radiative corrections are calculated using a model for the 
cross section, the sensitivity of our results to the model input was studied in detail.
All this will be discussed in what follows.

{\bf Bin-Centering Corrections}.
As previously mentioned, the measured cross section was initially extracted  binned in small  
momentum and angle bins corresponding to the acceptance intervals in $\delta$ and $\theta$, 
respectively. The goal, however, was to extract the cross section at a fixed angle, the central 
angle $\theta_{c}$, keeping all the statistics accumulated. If the cross section would not vary 
across the $\theta$ acceptance, then the cross section at $\theta_{c}$ could be simply obtained 
by statistically averaging the cross section over the angular acceptance. However, for the 
kinematics of this experiment, the variation of the cross section over the angular acceptance 
was not negligible. Thus, before statistically averaging, the so called bin-centering correction had 
to be applied in order to center the cross section measured in the $\theta$ acceptance interval at 
$\theta_{c}$. This correction was applied as:  

\begin{equation}
\sigma^{data}(E,E^{'},\theta_{c}) = \sigma^{data}(E,E^{'},\theta_{i}) \frac{\sigma^{model}(E,E^{'},\theta_{c})}{\sigma^{model}(E,E^{'},\theta_{i})},
\end{equation}

where $\sigma^{model}(E,E^{'},\theta_{i})$ and $\sigma^{model}(E,E^{'},\theta_{c})$ are the 
model cross sections calculated at $\theta_{i}$ and $\theta_{c}$, while 
$\sigma^{data}(E,E^{'},\theta_{i})$ and $\sigma^{data}(E,E^{'},\theta_{c})$ 
are the cross sections extracted from the data at $\theta_{i}$ and $\theta_{c}$.
The bin-centering correction was applied  
to the measured radiated hydrogen and deuterium cross sections using the radiated model cross section. 
The models used to calculate the correction will be discussed next. 

{\bf Radiative Corrections}.
In the perturbative picture, the lowest order process in $\alpha$ 
(the electromagnetic running coupling constant) that contributes to the cross section for 
inclusive electron-nucleon scattering is represented schematically in Fig. 14 (a). 
Besides this leading one photon exchange diagram (Born), there are higher order processes 
in $\alpha$ that contribute to the scattering. These diagrams are shown schematically 
in Fig. 14 (b, c, d, e) and include vacuum polarization (the exchanged photon 
creates particle-antiparticle pairs), vertex processes (emission and reabsorption of virtual 
photons), and Bremsstrahlung (emission of real photons in the field of the nucleon during 
interaction). In order to determine the differential cross section that accounts just for the 
one photon exchange process, all the other contributions from higher order processes in $\alpha$ 
have to be calculated and corrected for in the measured cross section.   
\begin{figure}
\centering \includegraphics[width=8 cm]{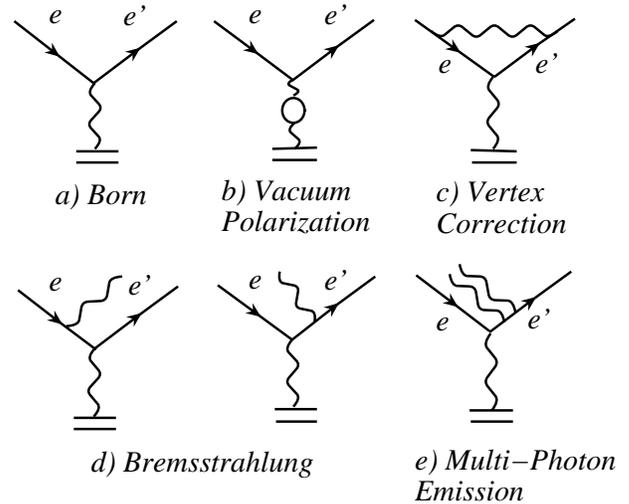}
\caption{Lowest order Feynman diagrams for inclusive lepton-nucleon scattering \cite{vladas_thesis}.}
\end{figure}
The radiative processes can be divided into two main categories: internal and external. 
The internal effects take place at the scattering vertex and include Bremsstrahlung, 
vacuum polarization, vertex processes and multiple photon exchange. External Bremsstrahlung occurs 
within the target material before or after the primary scattering takes place and is dependent 
on the target thickness. 
As a consequence, the energy of the incoming and/or the scattered electron will change.
\begin{figure}
\centering \includegraphics[width=8 cm]{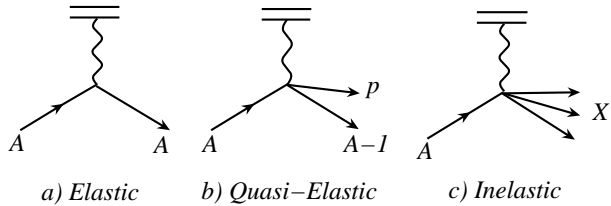}
\caption{Schematic representation of the processes that can contribute to the E00-116 measurements.}
\end{figure}

The measured cross section can be written as the sum of various processes (Fig. 15):
\begin{equation}
\sigma_{measured}^{hydrogen} = \sigma_{inelastic}^{radiated} + \sigma_{elastic}^{radiated} .
\end{equation}
\begin{equation}
\sigma_{measured}^{deuterium} = \sigma_{inelastic}^{radiated} + \sigma_{elastic}^{radiated} + \sigma_{quasielastic}^{radiated} .
\end{equation}
To obtain the Born inelastic cross section, the radiative tails from elastic/nuclear elastic 
and quasielastic 
cross sections were subtracted while the inelastic radiative effects were corrected 
multiplicatively.
In practice, the usual calculation of radiative corrections includes only the emission of 
one hard photon. However, there is a probability for the electron to emit two hard photons. 
Therefore additional corrections (the {\sl $\alpha^{2}$ term}) should be taken into account when 
estimating the radiative correction. 

There are two programs utilized for the radiative 
corrections calculations: one based on the Mo and Tsai formalism \cite{mo_tsai} and the other one based 
on the Bardin formalism \cite{bardin}, which includes in the calculations the two hard-photon radiation 
and has a different treatment of the soft photon contribution. The program 
based on Mo and Tsai formalism calculates both internal and external radiative corrections, 
unlike the Bardin one which calculates just internal radiative corrections. For this experiment, 
the radiative corrections (both internal and external) were estimated using the Mo and Tsai 
formalism. The $\alpha^{2}$ term correction was estimated using the Bardin formalism and at 
this experiment kinematics it proved to be below 0.5$\%$ (1$\%$) for hydrogen (deuterium). 
The size of the correction is within the theoretical uncertainty of the calculation and it 
was assigned as a point-to-point uncertainty. 

Additional uncertainties in the experimental cross section originate from the elastic 
(elastic and quasielastic) contribution subtractions. 
At the kinematics of this experiment, the elastic contribution to the total experimental cross section 
is negligible. The uncertainty coming from the quasielastic contribution subtraction was estimated 
by propagating the point-to-point model uncertainty into the experimental cross section. This 
kinematic dependent uncertainty was parametrized for each angle setting separately. 
Finally, the normalization uncertainty in the cross section coming from 
the theoretical uncertainty in the radiative corrections calculation was estimated to 
be 1$\%$ \cite{liang_thesis}.   

{\bf Iteration Procedure}
As stated previously, the bin-centering corrections were calculated using a model for the 
cross section. The same model was typically used to calculate the radiative corrections. In order to 
minimize the model dependence of the extracted cross section, an iterative procedure was followed. 
First, a starting model was used to calculate both the bin-centering and radiative corrections. Then, 
the extracted cross section was fit and the new fit was used to calculate the corrections and 
reextract the cross section. This process continued until the extracted cross section did not vary 
significantly (not more than 0.3\%) from one iteration to the next. Additionally, the iteration 
procedure was followed using 
two different starting models. After the last iteration, the two sets of cross sections were expected 
to be consistent. 

For $H(e,e')$ the two starting models (fits to previous data) used were the model of M.E. Christy and 
P. Bosted \cite{eric_fit} 
and $H_{2}$ model \cite{h2_model}. The fitting procedure used is described extensively in Ref. 
\cite{eric_fit}. Only two iterations were necessary and it was found that the difference in the cross 
section between the last iteration and the one before last was about 0.3\%. Also after each iteration 
the difference in the cross section when starting with the two models specified above was calculated. 
This difference after the last iteration was assigned as a kinematic dependent uncertainty accounting 
for the model dependence of the final result. 

For $D(e,e')$ measurements, the iteration 
was performed using the same fitting procedure as for $H(e,e')$ but with one modification: the fit 
form for the resonances used non-relativistic Breit-Wigners. The data seemed to be described better 
around pion threshold by such a fit. Three iterations were performed and the difference in the 
cross sections between the last iteration and the second last was around 0.3\%. Just like for 
the $H(e,e')$ data set, two different models were used in the iteration procedure: the Bodek model 
\cite{bodek_fit} and ALLM97 \cite{allm97} multiplied by the parameterization of the ratio of the 
deuterium and the proton electroproduction cross sections \cite{d_p_antje}. The difference between 
the two sets of cross sections after the last iteration was parameterized to give the uncertainty 
originating in the possible model dependence of the final result. 
\begin{figure}
\centering 
\includegraphics[width=8.9cm]{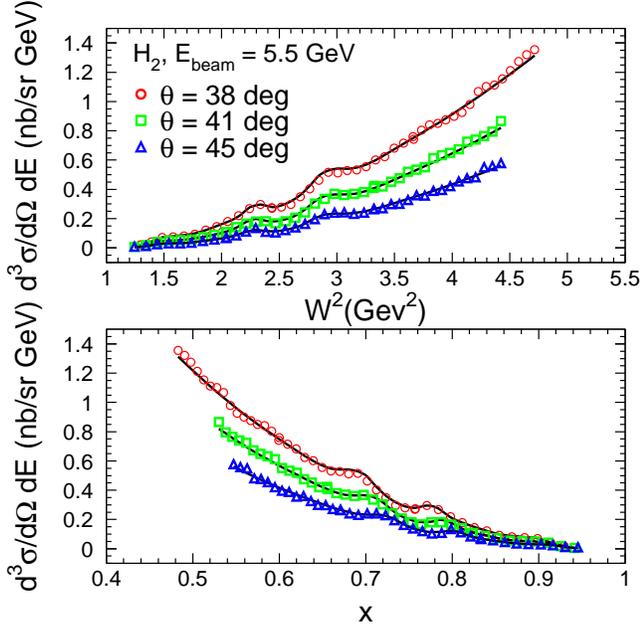}
\caption{(Color online) The $H(e,e')$ Born differential cross section extracted from E00-116 data 
at a beam energy of 5.5 GeV and spectrometer central angle of 38 deg (empty circles), 
41 deg (empty squares) and 45 deg (empty triangles) as a function of $W^{2}$ 
(top panel) and $x$ (bottom panel). Both the statistical and point-to-point systematic 
uncertainties are plotted. The curves shown represent the fit after the last iteration 
\cite{eric_fit,simona_thesis}.}
\end{figure}
\begin{figure}
\centering 
\includegraphics[width=8.9cm]{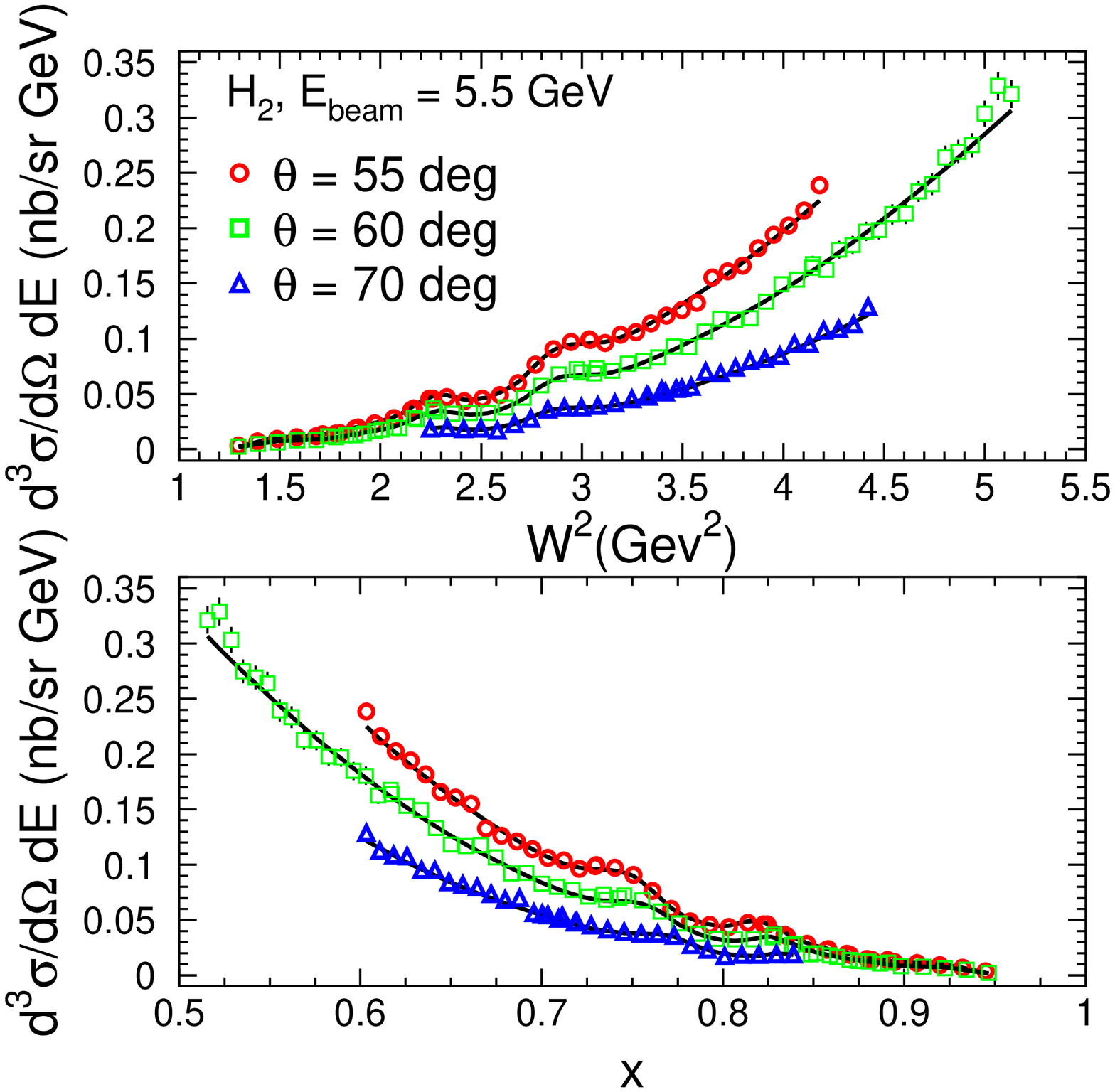}
\caption{(Color online) The $H(e,e')$ Born differential cross section extracted from E00-116 data 
at a beam energy of 5.5 GeV and spectrometer central angle of 55 deg (empty circles), 
60 deg (empty squares) and 70 deg (empty triangles) as a function of $W^{2}$ 
(top panel) and $x$ (bottom panel). Both the statistical and point-to-point systematic 
uncertainties are plotted. The curves shown represent the fit after the last iteration 
\cite{eric_fit,simona_thesis}.}
\end{figure}
\begin{figure}
\centering 
\includegraphics[width=8.9cm]{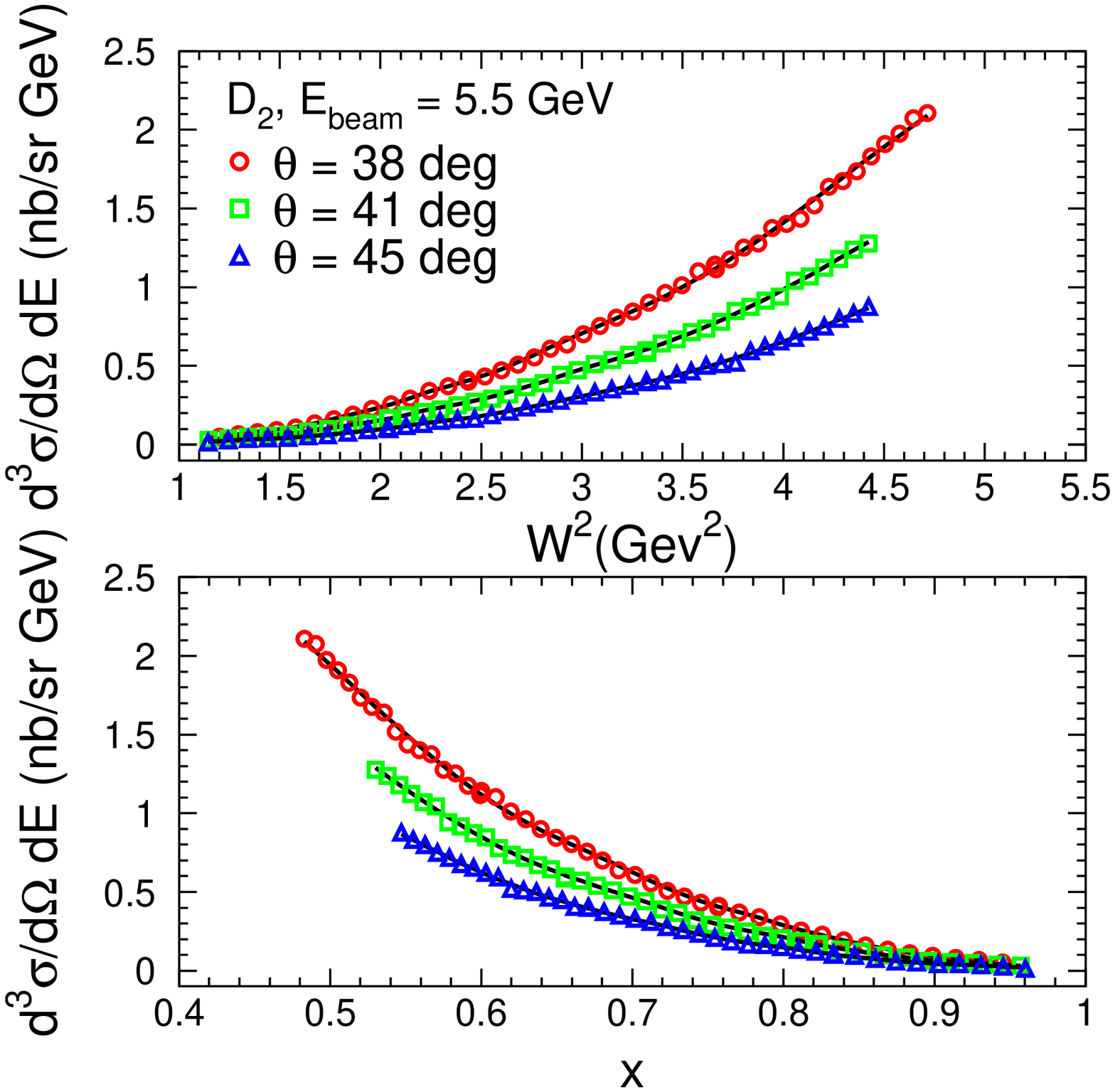}
\caption{(Color online) The $D(e,e')$ Born differential cross section extracted from E00-116 data 
at a beam energy of 5.5 GeV and spectrometer central angle of 38 deg (empty circles), 
41 deg (empty squares) and 45 deg (empty triangles) as a function of $W^{2}$ 
(top panel) and $x$ (bottom panel). Both the statistical and point-to-point systematic 
uncertainties are plotted. The curves shown represent the fit after the 
last iteration \cite{simona_thesis}.}
\end{figure}
\begin{figure}
\centering 
\includegraphics[width=8.9cm]{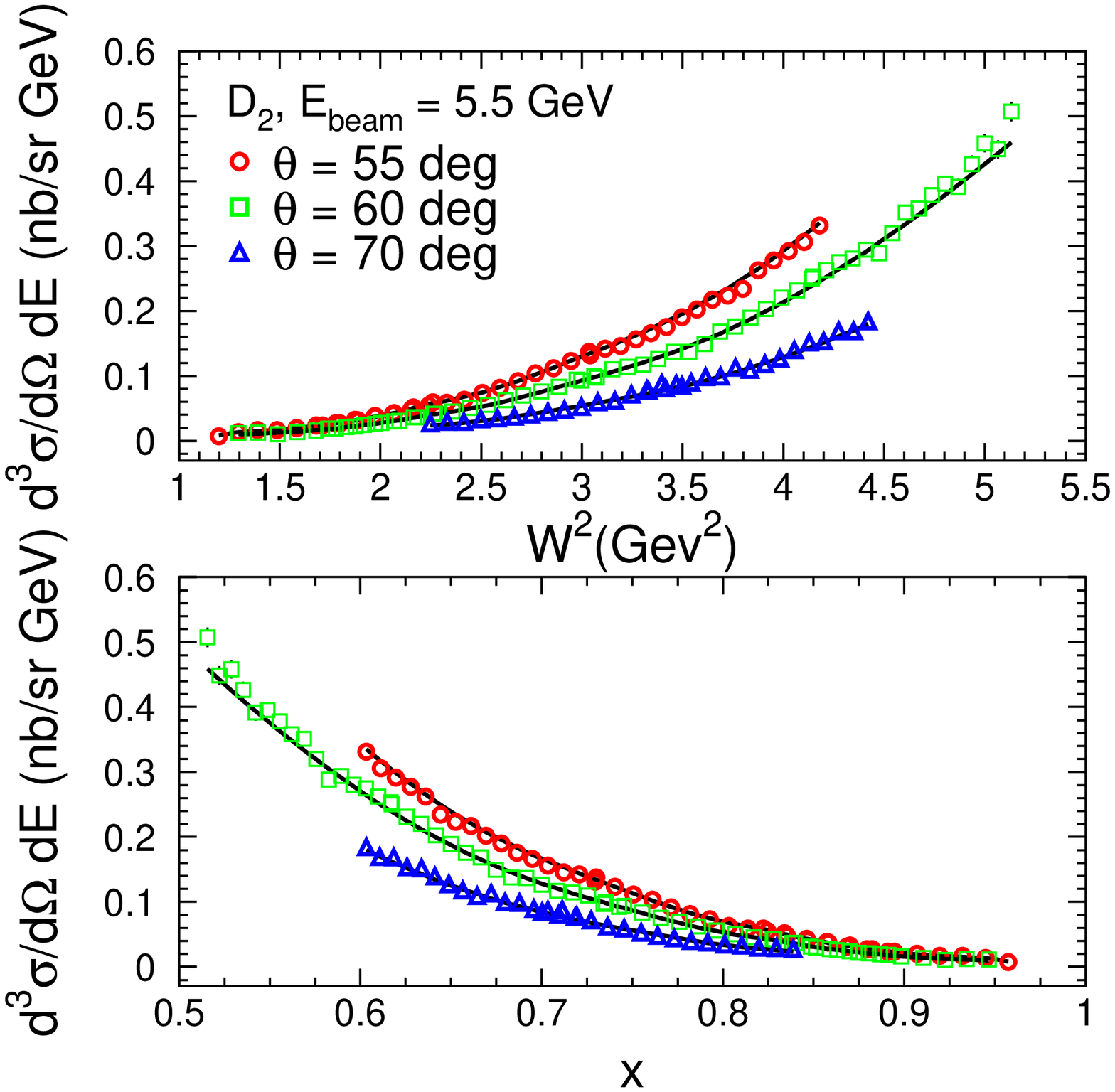}
\caption{(Color online) The $D(e,e')$ Born differential cross section extracted from E00-116 data 
at a beam energy of 5.5 GeV and spectrometer central angle of 55 deg (empty circles), 
60 deg (empty squares) and 70 deg (empty triangles) as a function of $W^{2}$ 
(top panel) and $x$ (bottom panel). Both the statistical and point-to-point systematic 
uncertainties are plotted. The curves shown represent the fit after the 
last iteration \cite{simona_thesis}.}
\end{figure}

{\bf Systematic Uncertainties}
The total point-to-point systematic uncertainty in the cross section extraction was taken as the sum in quadrature 
of the individual uncertainties. An overview of these uncertainties 
is given in Table 3. The total normalization systematic uncertainty amounted to about 1.75\% while the 
statistical uncertainty is below 3\%. 
\begin{table}
  \squeezetable
  \label{systematic_uncertainties}
  \caption{Point-to-point systematic uncertainties in the experimental parameters and the corresponding systematic 
uncertainties in the differential cross section.}
  \begin{tabular}{ccc}
    \hline
    \hline
    Quantity & Uncertainty & $\delta_{\sigma}$($\%$)\\
    \hline
Beam Energy & 5x$10^{-4}$ & 0.30$\%$\\
Scattered $e^{'}$ Energy & 5x$10^{-4}$ & 0.25$\%$ \\
Scattered $e^{'}$ Angle & 0.2 mrad & 0.26$\%$ \\
Beam Charge & 0.05$\%$ & 0.05$\%$ \\
Dead time & 0.25$\%$ & 0.25$\%$ \\
Trigger Efficiency & 0.2$\%$ & 0.2$\%$ \\
Tracking Efficiency & 0.2$\%$ & 0.2$\%$ \\
PID cut efficiency & 0.25$\%$ & 0.25$\%$ \\
Pion Cont. Subtraction & 0.2$\%$ & 0.2$\%$ \\
Charge-Symmetric Background & 6$\%$ - 20$\%$ & $<$2$\%$ \\
Acceptance Correction & 0.8$\%$ & 0.8$\%$ \\
Radiative Corrections & 0.5$\%$-3.6$\%$ & 0.5$\%$-3.6$\%$ \\
Model dependence & 0.2$\%$-5$\%$ & 0.2$\%$-5$\%$ \\
    \hline
    \hline
  \end{tabular}
\end{table}
The total and Born deradiated differential cross sections extracted from this experiment for 
both $H(e,e')$ and $D(e,e')$, together with the 
point-to-point associated uncertainties are given in Tables 5-9. The extracted 
Born differential cross sections are also shown as a function of $x$ and $W^{2}$ in Figs. 16-19. 
These data provide large $x$ and intermediate $Q^{2}$ high-precision measurements in the resonance 
region where the precision of existing data from SLAC is typically 5 to 30\% for the 
statistical uncertainty alone. This precision is not enough to distinguish between 
theoretical parameterizations of the structure function which at a $Q^{2}$ of 8 GeV$^{2}$ and 
$x$ = 0.8, for example, differ by at most 30\% as shown in Fig. 2. 

\subsection{$F_{2}$ Extraction}

The structure function $F_{2}$ was calculated utilizing the formula:
\begin{equation}
F_{2} = \frac{d^{2} \sigma}{d \Omega dE^{'}} \frac{1 + R}{1 + \varepsilon R} \frac{K \nu}{4 \pi \alpha} \frac{1}{\Gamma} \frac{1}{1 + \frac{\nu^{2}}{Q^{2}}},
\end{equation}
where $K = (W^{2} - M^{2})/(2M)$, $\nu = E -E^{'}$ and $\alpha$ is the electromagnetic coupling constant. 
The quantity $\varepsilon$ is the degree of polarization of the virtual photon:
\begin{equation} 
\varepsilon = (1 + 2 \frac{\nu^{2} + Q^{2}}{Q^{2}} tan^{2} \frac{\theta}{2})^{-1},
\end{equation}
and $\Gamma$ is the flux of the virtual photon:
\begin{equation}
\Gamma = \frac{\alpha K}{2 \pi^{2} Q^{2}} \frac{E^{'}}{E} \frac{1}{1 - \varepsilon}.
\end{equation}
The extraction of $F_{2}$ requires the knowledge of both the differential cross section 
and the quantity $R$ which is the ratio of the longitudinal to the transverse component of the 
cross section. For this experiment, it was not planned to measure $R$ but just the differential 
cross section since $R$ is not expected to be large in the $Q^{2}$ range of this data set. 
Thus, for the structure function extraction, it was needed to resort to an existing $R$ 
parameterization. In order to estimate the sensitivity of the $F_{2}$ structure function to different $R$ 
parameterization, the following approach was taken. $F_{2}$ was extracted at the lowest and highest 
$Q^{2}$ for this experiment using three different $R$ parameterizations: R from R1990 \cite{r1990}, R1998 
\cite{r1998} and E94-110 \cite{liang_thesis}, respectively. It was found that $F_{2}$ varies on average 
by 2\% when different $R$ parameterizations are used. R1990 and R1998 are parameterizations of $R$ 
extracted mainly from DIS measurements, while $R$ from E94-110 is extracted 
from resonance region measurements, only. The $R$ parameterization from E94-110 is kinematically 
limited to $W^{2}$ $<$ 3.85 GeV$^2$ and typically lower $Q^{2}$ than this experiment. 
However, it was shown to agree, where applicable kinematically, to R1990 and R1998. 
The $R$ parameterization R1998 was obtained using a larger data set than R1990 (R1998 used, in addition to 
the data set of R1990, other measurements extending the parameterization to lower and higher $x$), and 
also had a better confidence level of the fit (73\%) than R1990 (61\%). For all these reasons, $R$ from 
R1998 was used for the structure function calculation and an additional uncertainty of 2\% was assigned to 
$F_{2}$ in order to account for the sensitivity of the extraction to $R$. 
  
\section{Results}
In this section we present our results with a preamble of previous quark-hadron duality studies. 
With this very brief summary of previous studies we intend to create the apropriate context 
for the detailed discusssion of our results and the conclusions that will be drawn.

\subsection{Previous Quark-Hadron Duality Studies with Electron Scattering}

Over three decades ago, Bloom and Gilman acknowledged the resonance-scaling connection in inclusive 
electron-nucleon scattering. After more than 20 years, it was an early JLab 
experiment that revived the interest in the phenomenon of quark-hadron duality 
\cite{ioana_thesis,F2JL1,F2JL3}. This experiment confirmed the observations of Bloom and Gilman and, 
in addition, acknowledged the onset of duality also locally. The findings of these studies prompted 
interest in a more detailed analysis of duality, one within the QCD formalism. 

Comparisons of the resonance region data to some QCD predictions were performed using measurements 
from JLab experiment E94-110 \cite{liang_thesis,F2JL2}. The $F_{1}^{p}$ and $F_{2}^{p}$ extracted from 
E94-110 were compared to QCD fits from the 
MRST \cite{yong_mrst} and CTEQ6 \cite{yong_cteq} collaborations evaluated at the same $Q^{2}$ as the data 
and with the inclusion of target mass corrections. It was observed that the QCD fits seem to 
describe on average the resonance 
strength at each $Q^{2}$ value investigated - $Q^{2}$ of 1.5, 2.5 and 3.5 GeV$^{2}$- as if the resonances 
would follow, on average, the same perturbative $Q^{2}$ evolution as the QCD fits. 
On a more quantitative level, quark-hadron duality was investigated by computing ratios 
of the integrals of the structure function over $x$ in the resonance region at fixed $Q^{2}$ values from 
both the data and the scaling curves. It was found that duality seems to hold better than 
5\% above $Q^{2}$ = 1 GeV$^{2}$ when compared to MRST with target mass corrections. 
However, at the highest $Q^{2}$ of 3.5 GeV$^{2}$ and the highest $x$ the ratio to MRST 
was noticed to rise above unity up to 18\% \cite{wally_duality}. 

This finding came in strong contradiction with the expectation that duality should 
work even better at higher $Q^{2}$. If the higher-twist contributions seem to be small or cancel 
to some degree at low $Q^{2}$ then, considering that these terms are weighted by powers of 1/$Q^{2}$ in the 
operator product expansion, it is expected that this must be even more the case at a higher $Q^{2}$. 
Moreover, the observation of increasing discrepancy between data and some QCD fits with increasing $Q^{2}$ 
(and increasing $x$) 
is not unique to the resonance region: DIS data from SLAC exhibit the same behavior \cite{thia_priv}. In consequence, 
this rise has been ascribed not to a violation of duality but rather to an underestimation of the 
large $x$ strength in some QCD fits. 

These studies made obvious the utility of high precision resonance region data at an 
even higher $Q^{2}$ (and thus larger $x$). This extension 
of resonance region measurements was crucial for the verification of 
QCD fit behavior in this kinematic regime. Considering that most of the currently available large $x$ 
data lie in the resonance region, the confirmation of quark-hadron duality as an effective tool would 
offer much needed experimental constraints for theoretical predictions in the region of $x$ $\rightarrow$ 1. 

In what follows, therefore, we present quark-hadron duality studies performed using the $F_{2}$ structure 
function extracted from this experiment, as well as from earlier 
Hall C \cite{ioana_thesis,liang_thesis} and SLAC 
\cite{whitlow_f2,riordan_thesis,poucher_thesis,bodek_thesis,bodek_paper,mestayer_paper,poucher_paper,atwood_paper} 
measurements in the resonance region. 

\subsection{Quark-Hadron Duality: the $Q^{2}$ Dependence}

An exhaustive description of nucleon structure in terms of parton distribution functions requires 
knowledge of the strength of the PDFs for the entire $x$ regime. 
Most global QCD fits are essentially unconstrained at large $x$ \cite{Schienbein:2007fs}. 
The quark-hadron duality phenomenon could be the key for providing experimental constraints in the 
large $x$ region by the use of properly averaged resonance data. This avenue relies, however, on our 
ability to unravel the $Q^{2}$ dependence of the data in a region where the perturbative mechanisms are 
not the only ones to be taken into account.  

In this context, a comparison of the  $Q^{2}$ dependence of various theoretical predictions to the 
one exhibited by averaged resonance region data was studied in a similar fashion to that of 
\cite{ioana_thesis,liang_thesis}. 
Ratios of the integrals of the $F_{2}$ structure function were considered:
\begin{equation}
I = \frac{\int_{x_{min}}^{x_{max}} F_{2}^{data}(x,Q^{2})dx}{\int_{x_{min}}^{x_{max}} F_{2}^{param.}(x,Q^{2})dx}.
\end{equation}
The integrand in the numerator -$F_{2}^{data}$- is the $F_{2}$ structure function extracted from the 
experimental cross sections. The integrand in the denominator -$F_{2}^{param.}$- is the $F_{2}$ structure 
function as given by the parameterizations introduced in Sect. 2: 
CTEQ6M+TM, MRST2004+TM, ALEKHIN, ALLM97. 
It is important to note that, for this analysis, $F_{2}^{param.}$ was generated at the same 
values of $x$ and $Q^{2}$ as the data and was integrated over the same range in $x$ as the data 
using the same integration procedure. This, by dint of the $W^{2}$ cuts used to obtain these global fits, 
by definition extends them into regions where they are not constrained in $x$, and only their $Q^{2}$ 
dependence is determined.

For {\sl global} duality studies, the limits of the integrals, $x_{min}$ and $x_{max}$, were the experimental 
$x$ values corresponding to $W^{2}_{min}$ = 1.3 GeV$^{2}$ and $W^{2}_{max}$ = 4.5 GeV$^{2}$, respectively. 
To compare the $Q^{2}$ dependence of theoretical predictions to individual resonance structures, for 
{\sl local} duality studies, the 
resonance regions were delimited using the same $W^{2}$ cuts as in a previous analysis \cite{F2JL1}:
\begin{enumerate}
\item{first region ($1^{st}$) $\to$ $W^{2}$ $\in$ [1.3 , 1.9] GeV$^2$ }
\item{second region ($2^{nd}$) $\to$ $W^{2}$ $\in$ [1.9 , 2.5] GeV$^2$ }
\item{third region ($3^{rd}$) $\to$ $W^{2}$ $\in$ [2.5 , 3.1] GeV$^2$ }
\item{fourth region ($4^{th}$) $\to$ $W^{2}$ $\in$ [3.1 , 3.9] GeV$^2$ }
\item{DIS region ($DIS$) $\to$ $W^{2}$ $\in$ [3.9 , 4.5] GeV$^2$ }
\end{enumerate}
These $W^{2}$ limits translate in the integrals of Eq. 14, to $x_{min}$ and $x_{max}$ values according to:
\begin{equation}
x = \frac{Q^{2}}{W^{2} + Q^{2} - M},
\end{equation} 
where $M$ is the proton mass. As an example, the x range covered by different resonance regions 
for two Q$^{2}$ values are given in Table 4. At a given $Q^{2}$, the lowest $W^{2}$ region 
(the first region) corresponds to the highest $x$ regime while for a fixed $W^{2}$ region, the larger 
the $Q^{2}$, the larger the x regime. 

Figures 20-23 depict the results of the global and local duality studies performed for $H(e,e')$. 
The quantity $I$ is shown for each resonance region individually as 
well as integrated over the full region specified above. The uncertainties shown are 
obtained by adding in quadrature the statistical and systematic uncertainties on the numerator alone. 
No parameterization uncertainties are plotted. The latter are typically substantial at large $x$, on 
the order of 100\% \cite{cteq}. 
\begin{table}
  \squeezetable
  \label{x_regimes}
  \caption{An example of the x ranges covered by different resonance regions for two $Q^{2}$ values.}
  \begin{tabular}{|c|c|c|} \hline
&\multicolumn{2}{|c|}{$x$ range} \\ \cline{2-3}
\hline
$W^{2}$ region & $Q^{2}$ = 2 (GeV$^{2}$) & $Q^{2}$ = 6 (GeV$^{2}$) \\
\hline
$1^{st}$ & 0.66 - 0.83 & 0.85 - 0.93   \\
\hline
$2^{nd}$ & 0.55 - 0.66 & 0.79 - 0.85   \\
\hline
$3^{rd}$ & 0.47 - 0.55 & 0.73 - 0.79  \\
\hline
$4^{th}$ & 0.40 - 0.47 & 0.67 - 0.73   \\
\hline
$DIS$   & 0.35 - 0.40 & 0.62 - 0.67   \\
\hline
  \end{tabular}
\end{table}
Overall, the data of this experiment (blue circles), 
previous JLab data (the data represented by black stars and red triangles are from \cite{ioana_thesis} 
and \cite{liang_thesis}, respectively), 
and SLAC data (green squares \cite{whitlow_f2,riordan_thesis,poucher_thesis,bodek_thesis,bodek_paper,mestayer_paper,poucher_paper,atwood_paper}), are found to be in good agreement. 
A slight disagreement could be observed between this experiment and the SLAC experiment E-8920 which 
is singled out from the other SLAC data sets (empty green square). The disagreement becomes smaller 
as we approach the DIS region. This is possibly related to the fact, that for E-8920, the radiative 
corrections were rescaled to bring the DIS data in agreement with the other SLAC experiments which 
could however have resulted in a possible incorrect estimation of the radiative corrections 
in the resonance region.

Figure 20 presents our results when we compare the integral of the $F_{2}^{p}$ extracted from the data to the 
integral of $F_{2}^{p}$ obtained from the CTEQ6M PDFs with the inclusion of target mass effects as 
explained in Sect. 2. 
The quantity $I$ is close 
to unity at a $Q^2$ of about 1.5 GeV$^2$, and then rises above unity with increasing $Q^2$. However, 
$I$ reaches a plateau at a $Q^2$ of about 4 GeV$^2$ and, above this value, the $Q^{2}$ dependence saturates.
This behavior is displayed when the integration is done globally as seen in the bottom right panel, but 
also for ``all'' the individual resonance regions except for the first resonance region (upper left panel). 
The saturation of the $Q^{2}$ dependence indicates that the discrepancy between data and parameterization is 
not a $Q^{2}$ dependent effect. It is most likely, therefore, due to the fact that CTEQ6M+TM does not model 
accurately the strength of the PDFs at large $x$. Put differently, $I$ being greater than 1 and, 
reaching a constant 
value above $Q^{2}$ $\sim$ 4 GeV$^{2}$ most likely does not represent the failure of QCD in describing the $Q^{2}$ 
evolution of the averaged resonance region data but rather a paucity in the strength 
of the PDFs at large $x$. 
The resonance region data do display on average a QCD type $Q^{2}$-dependence.

The ratio $I$ seems to become constant at a slightly different value 
for each resonance region. In fact, as we move from the fourth resonance region ($I$ $\sim$ 1.1) to the 
third ($I$ $\sim$ 1.28) and then to the second ($I$ $\sim$ 1.35) the discrepancy increases. 
This is possibly related to the growing uncertainty associated with PDF strengths at large $x$. 
For a fixed $Q^{2}$, the second resonance region probes a larger $x$ regime than 
the third and, the third larger than the fourth, etc. Since the $x$ dependence of the PDFs is 
less and less constrained at larger and larger $x$, we expect this to be reflected in a more obvious 
way when we study the second resonance region than the third, for example.

It was reported before that the $N$-$\Delta$ transition 
region provides a different behavior when compared to the rest of the resonance region \cite{simonetta}. 
This could be related to the fact that this region is the only one with a single resonant state and there 
are arguments that more than one state is necessary to approximate closure and duality 
\cite{isgur_close}. It 
should also be pointed out that the first resonance region probes the highest $x$ regime where the PDFs 
are expected to be least constrained.

Figure 21 shows the ratio of the integrals of $F_{2}^{p}$ from the data and MRST2004 
with target mass corrections. The observed $Q^2$ dependence of $I$ yields similar 
conclusions to those drawn from the comparison to CTEQ6M: we encounter the same rise of $I$ with $Q^{2}$ 
which eventually saturates for all resonance regions except for the first one and also globally. 
And just as for CTEQ6M, $I$ saturates at a different values for each resonance region. This is not 
surprising considering that the extraction procedure of PDFs for MRST2004 is rather similar to the one 
employed by CTEQ6M. There are, however, few features that set apart the comparison with MRST2004. 
This parameterization undershoots the data by an even larger amount and $I$ 
saturates at a larger value of $Q^2$ than for CTEQ6M+TM. This most likely results from the 
differences in the $x$ dependence modelling of the PDFs between the two parameterizations. 

Figure 22 shows the comparison of our averaged $F_{2}^{p}$ resonance data to the parameterization of ALEKHIN. 
It should be reminded that it is 
not just the leading twist that is considered in this parameterization, as is the case with 
CTEQ6M and MRST2004, but also the 
higher-twists. By explicitly accounting for higher-twist terms, Alekhin can extend the 
validity of his fit to an $x$ as large as 0.75 and a $W^{2}$ as low as 3.24 GeV$^{2}$. 
Though this $W^{2}$ cut 
practically excludes resonance region data, is still more permissive than the cuts employed by CTEQ6M or 
MRST2004, ensuring that the ALEKHIN fit is far better constrained at large $x$. 
Indeed, the agreement between the averaged resonance data and ALEKHIN is obvious in Fig. 22. For 
the fourth resonance and the DIS regions (upper and middle right panel, respectively) the quantity 
$I$ is very close to unity across the entire $Q^{2}$ range investigated. Good agreement is obtained when 
analyzing the second and third resonance 
regions: $I$ deviates from unity by only 5\% or less and, for most part, seems independent of $Q^{2}$. 
This finding is quite remarkable: according to ALEKHIN higher-twist coefficients in the 
resonance region, on average, differ from the ones extracted from the DIS region by at most 5\%. 
It should be pointed out that not all of this already small discrepancy can be attributed to the 
contribution from higher-twist terms: ALEKHIN does not include resonance data in his fit 
therefore the $x$-dependence of 
the PDFs is unconstrained in this region, though to a far lesser extent than for CTEQ6M or MRST2004. 
This finding is quite different from what was observed when comparing to CTEQ6M, for example. There 
$I$ deviates from unity by about 10\% in the DIS region but by almost 35\% in the second resonance region.
When compared to ALEKHIN, the first resonance region (upper left panel) behaves differently but in this 
kinematic regime the fit validity is questionable. The data are well described on average also globally. 
Thus the higher-twist terms contributions in the resonance region is shown to be quantitatively comparable 
on average with the ones extracted from the deep inelastic scattering data pointing to 
the onset of quark-hadron duality.  
        
The comparison of our integrated $F_{2}^{p}$ resonance data to ALLM97 is presented in Fig. 23. 
The $Q^{2}$ dependence of $I$ shows very good agreement between data and this parameterization in the 
fourth resonance (upper right panel) and DIS (middle right panel) regions and also globally. 
If quark-hadron duality holds, this is to be expected considering that ALLM97 successfully fits 
data down to a $W^{2}$ as low as 3 GeV$^{2}$. The agreement 
slightly worsens as we move to the third and second resonance regions. Though $I$ is about 7\% above 
unity, it seems to be independent of $Q^{2}$ for the third resonance region. A 
familiar pattern emerges when analyzing the $Q^{2}$ dependence of $I$ in the second resonance region: $I$ 
rises with increasing $Q^{2}$ but this rise eventually saturates around $Q^{2}$ of 4 GeV$^{2}$. It is to 
be expected that as the larger $x$ and lower $W^{2}$ region is probed, 
the comparison between averaged resonance data and this parameterization to unravel some of its 
shortcomings like unconstrained $x$ and $Q^{2}$ dependence or inability to fully account for target 
mass effects.  
The averaged resonance data in the first region (upper left panel) compare surprisingly well with 
ALLM97 but no definite conclusions could be drawn within QCD 
framework considering that ALLM97 accounts for the $x$ and $Q^{2}$ dependence empirically and this is a
region far from the domain of validity of this parameterization. Overall, the 
comparison of averaged resonance data to ALLM97 confirms that, quantitatively, higher-twist terms 
contributions in the resonance region seem to be comparable, on average, with the ones in the deep 
inelastic scattering data.

A similar pattern as for $H(e,e')$ is observed when studying the $Q^{2}$ dependence of $I$ 
for $D(e,e')$ (Figs. 24-27). It should be reminded that, as discussed in Sect. 2, there is an 
additional factor to consider when analyzing the results from our global and local quark-hadron 
duality studies for $D(e,e')$: 
all of the three QCD-based parameterizations utilized in 
our analysis provide PDFs from which the proton structure function is constructed. 
Also ALLM97 is a fit to only proton data. So, in order to obtain parameterizations 
for the deuteron structure function we used the $F_{2}^{p}$ parameterizations 
and the d/p parameterization from \cite{d_p_antje} introduced in Sect. 2. 

Figure 24 shows our results when we compare the integral of $F_{2}^{d}$ extracted from the data to the integral 
of $F_{2}^{d}$ obtained from CTEQ6M PDFs as explained above. The $Q^{2}$ dependence of $I$ displays similar 
characteristics to the ones acknowledged in our study of the proton data presented in Fig. 20: $I$ rises 
above unity with increasing $Q^{2}$ but a plateau is reached at a $Q^{2}$ of about 4 GeV$^{2}$. Above this 
value $I$ is practically independent of $Q^{2}$. 
As observed before, the first 
resonance region (upper left panel) stands out and, in addition to the aspects discussed for $H(e,e')$, 
the complication of having to resort to extrapolations of the d/p parameterization should be taken into 
account. 

A similar behavior is acknowledged when the data are compared with MRST2004 (Fig. 25). However, the $Q^{2}$ 
dependence of $I$ saturates at a larger $Q^{2}$ value than for CTEQ6M just as it happened for $H(e,e')$. 
This trend is even more accentuated as we probe larger $x$ regimes (second resonance region in the middle 
left panel, for example) where the reliability of the d/p parameterization is questionable.  

Just as for $H(e,e')$, good agreement is observed when the $D(e,e')$ data are compared to ALEKHIN, 
as seen in Fig. 26. In fact, except for the first resonance region (upper left panel) where both the 
PDFs and the d/p parameterization are expected to be largely unconstrained, the $D(e,e')$ data are 
described by this parameterization down to the lowest $Q^{2}$ analyzed. 
Similar conclusions can be drawn from the comparison of the data to ALLM97 which is presented in Fig. 27.

To summarize, our studies showed that above a $Q^{2}$ of about 4 GeV$^{2}$ for CTEQ6M and 
slightly higher for MRST2004 the ratio of the integrals of resonance data and parameterizations becomes 
independent of $Q^{2}$. This is a very important finding which suggests that, above a surprisingly low 
$Q^{2}$ value, most of the disagreement between the averaged resonance data and the above mentioned 
parameterizations is unrelated to the violation of the $Q^{2}$ evolution by contributions from the 
higher-twist terms in the resonance region. In fact, the comparison of our data to ALEKHIN and ALLM97 
confirmed that higher-twist contributions to deep inelastic scattering and averaged resonance region data 
are comparable. All these findings point to the unconstrained strength of the CTEQ6M and MRST2004 PDFs at 
large $x$ as major source for the disagreement between data and the above mentioned parameterizations in this 
kinematic regime.   


\subsection{Quark-Hadron Duality: the $x$-Dependence}

Our quark-hadron duality studies discussed above indicate that there are small rather than large 
violations of the $Q^{2}$ evolution in the resonance region {\sl on average}. 
Thus when referring to disagreements between data and theory, the ability of PDF-based calculations 
to describe the $x$-dependence of the data in particular at large $x$ is brought into discussion. 

We used the averaged proton structure function data for the five $W^{2}$ regions to draw a comparison to 
the theoretical calculations at fixed $Q_{0}^{2}$ as a function of $x$. 
The data averaging was done as follows:
\begin{equation}
F_{2}^{p,ave} = \frac{\int_{x_{min}}^{x_{max}} F_{2,data}^{p} dx}{x_{max} - x_{min}},
\end{equation}
where $x_{min}$ and $x_{max}$ are the integration limits corresponding to the $W^{2}$ limits defined in the 
previous section. 

The averaged structure function data, $F_{2}^{p,ave}$ were then centered at a fixed $Q_{0}^{2}$. 
Data within small $Q^{2}$ intervals were chosen for centering: for example, all $F_{2}^{p,ave}$ 
data in the $Q^{2}$ interval of 2 GeV$^{2}$ to 4 GeV$^{2}$ were evaluated at $Q_{0}^{2}$ = 3 GeV$^{2}$, etc. 
This was done as follows:
\begin{equation}
F_{2}^{p,ave}(x,Q^{2}_{0}) = F_{2}^{p,ave}(x,Q^{2}) \frac{F_{2,param.}^{p}(x,Q^{2}_{0})}{F_{2,param.}^{p}(x,Q^{2})}.
\end{equation}
The parameterization of M.E. Christy \cite{eric_param} was used for bin centering because it describes 
the $Q^{2}$ dependence of the data to better than 3\%. In addition, to study the sensitivity of 
the results to the choice of parameterization, the CTEQ6M fit was also utilized and the difference 
in the results when the two parameterizations are used was assigned as a systematic uncertainty. 
Just as for the quark-hadron duality studies discussed in the previous section our data were 
compared with all of the four parameterizations introduced in Sect. 2 and the results are 
presented in Figs. 28-31. 

Given that the deviations between our locally averaged resonance region data and the expectations based 
on PDF parameterizations such as CTEQ6M and MRST2004 seem related to the uncertainty of these PDFs at 
large $x$, we start a comparison of the $x$-dependence of our averaged resonance data with 
the phenomenological ALEKHIN and ALLM97 structure function parameterizations.
Figure 28 shows the ratio of the $F_{2}^{p}$ structure function extracted from the data as explained above 
and the parameterization of ALEKHIN at four values of $Q^{2}$ as a function of $x$. At $Q^{2}$ of 3 and 5 
GeV$^{2}$ (upper right and lower left panel, respectively) the parameterization describes the $x$ 
dependence of the data well, except for the largest $x$ regime where measurements from the first resonance 
region are used. There is a small shift between the DIS+fourth resonance region data and the rest but no 
obvious disagreement which depends on $x$ is observed. At $Q^{2}$ of 7 GeV$^{2}$ (lower right panel) our data 
probe the largest $x$ regime where ALEKHIN is least constrained and we acknowledge a growing discrepancy 
between data and parameterization with increasing $x$. At $Q^{2}$ of 1 GeV$^{2}$ ALEKHIN fails to describe 
the $x$-dependence of our data as $x$ increases: the data probe here a regime where both the $x$ and $Q^{2}$ 
limits of applicability are reached for this parameterization. Similar conclusions can be drawn from the 
comparison of our data with ALLM97 presented in Fig. 29.
   
Figure 30 shows the comparison of the data to CTEQ6M+TM. The CTEQ6M uncertainties are plotted also as a 
band. The parameterization fails to describe the $x$-dependence of the data. The discrepancy is much 
larger than observed in our comparisons with ALEKHIN and ALLM97 and it also grows strongly with 
increasing $x$. The same conclusion can be drawn from the comparison of the data to 
MRST2004+TM presented in Fig. 31. Again, this is not surprising considering that for both calculations 
the strength of the PDFs is largely unconstrained in the kinematic regime that we study. 
On the other hand, the ALEKHIN parameterization accounts explicitly for higher-twist, 
which allows it to include data with lower $W^{2}$ (and larger $x$) than CTEQ6M and MRST2004. 
This offers better constraints to the $x$ functional form which reflects in a more realistic description 
of the data. 

Figure 32 shows the comparison of the $F_{2}^{p}$ structure function from Bourrely et {\it al.} \cite{bourrely} 
to CTEQ6M+TM and ALLM97 at a fixed $Q^{2}$ value of 5 GeV$^{2}$. The target mass effects where included 
in the parameterization of Bourrely {\it et al.} in an identical fashion as for CTEQ6M, according 
to Georgi and Politzer prescription \cite{georgi_politzer_tmc}. At low $x$, $x$ $<$ 0.4, the 
three parameterizations agree reasonably well. At large $x$ however, significant discrepancies arise: 
up to 30\% when compared to CTEQ6M+TM and larger for ALLM97, as shown in the insert of Fig. 32. 
Without the inclusion of target mass 
effects, the ratio of CTEQ6M+TM and ALLM97 to the parameterization from Bourrely {\it et al.} would 
be even larger at large $x$ ($x$ $>$ 0.5). The major cause of this discrepancy is, 
most likely, the scarcity of high $W^{2}$ and high $x$ data which could constrain the $x$ dependence 
of the parton distribution functions. The parameterization from Bourrely {\it et al.} undershoots 
the $F_{2}^{p}$ structure function from ALLM97 which was found to be in fairly good agreement with our 
averaged resonance region data.

We conclude that what appeared to be a violation of quark-hadron duality when 
we compared our averaged resonance data with CTEQ6M and MRST2004 is actually, for most part, a 
reflection of the inability of these parameterizations to realistically model the large $x$ strength 
of their PDFs.     


\section{Conclusions}
We have performed high precision measurements of the $H(e,e')$ and $D(e,e')$ cross sections in the resonance 
region at large $x$ and intermediate $Q^{2}$. In this work, both global and local quark-hadron duality was 
quantified for the proton and deuteron using our new large $x$ data as well as previous 
resonance region measurements 
from JLab and SLAC. Previous studies \cite{F2JL2,liang_thesis} indicated quark-hadron duality in the 
$F_{2}^{p}$ structure function to 
hold better than 5\% above $Q^{2}$ = 1 GeV$^{2}$ when compared to a typical QCD fit like MRST,  
with a growing discrepancy observed as regions of higher $Q^{2}$, about 3.5 GeV$^{2}$, and higher $x$  
were explored. This finding came in strong contradiction with the expectation that duality should 
work best with increasing $Q^{2}$. The question arose whether this growing discrepancy was really a violation 
of duality by contributions from higher-twist terms or mostly a consequence of the well-known issue of 
highly unconstrained PDFs at large $x$. 

We found that, when compared to CTEQ6M and MRST2004, the ratio of integrals of resonance data and these 
parameterizations becomes independent of $Q^{2}$ starting with a value of about 4 GeV$^{2}$ for CTEQ6M, and 
slightly higher for MRST2004. This is an indication that, as expected from quark-hadron duality, 
there are only small violations of the $Q^{2}$ evolution by data in 
the resonance region, on average. This ratio 
saturates above unity at increasing values for regions with decreasing $W^{2}$ (and increasing $x$), but 
remains constant in $Q^{2}$, likely due to the uncertainty in the PDFs extraction at large $x$.

The comparison to ALEKHIN revealed that the higher-twist contributions to the 
averaged resonance region data are comparable to the ones in the low $W^{2}$ DIS region at the 
level of 5\% or less. This points as well to the unconstrained PDFs at large $x$ as a major source of the 
observed discrepancy between data and CTEQ6M and MRST2004. This argument is further supported by our studies 
of the $x$ dependence of the data and parameterizations. 

This analysis concludes that, with a careful study of the $Q^{2}$-dependent contributions, 
properly averaged resonance region data could be used to provide much needed constraints for PDFs 
at large $x$, shedding light on the parton dynamics in this regime. In view of quark-hadron duality, 
a CTEQ subgroup has begun to attempt the improvement of PDFs at large $x$ expanding the possible 
data sets by lowering the $W^{2}$ cut \cite{alberto_hallc_talk}.


\section{Acknowledgements}
This work is supported in part by research grants from the 
National Science Foundation (NSF) and the US Department of Energy, including 
NSF awards 0400332 and 0653508. We thank the Jefferson Lab Hall C 
scientific and 
engineering staff for their outstanding support. The Southeastern Universities 
Research Association operates the Thomas Jefferson National Accelerator 
Facility under the U.S. Department of Energy contract DE-AC05-84ER40150.

\begin{figure*}
\centering 
\includegraphics[width=10cm]{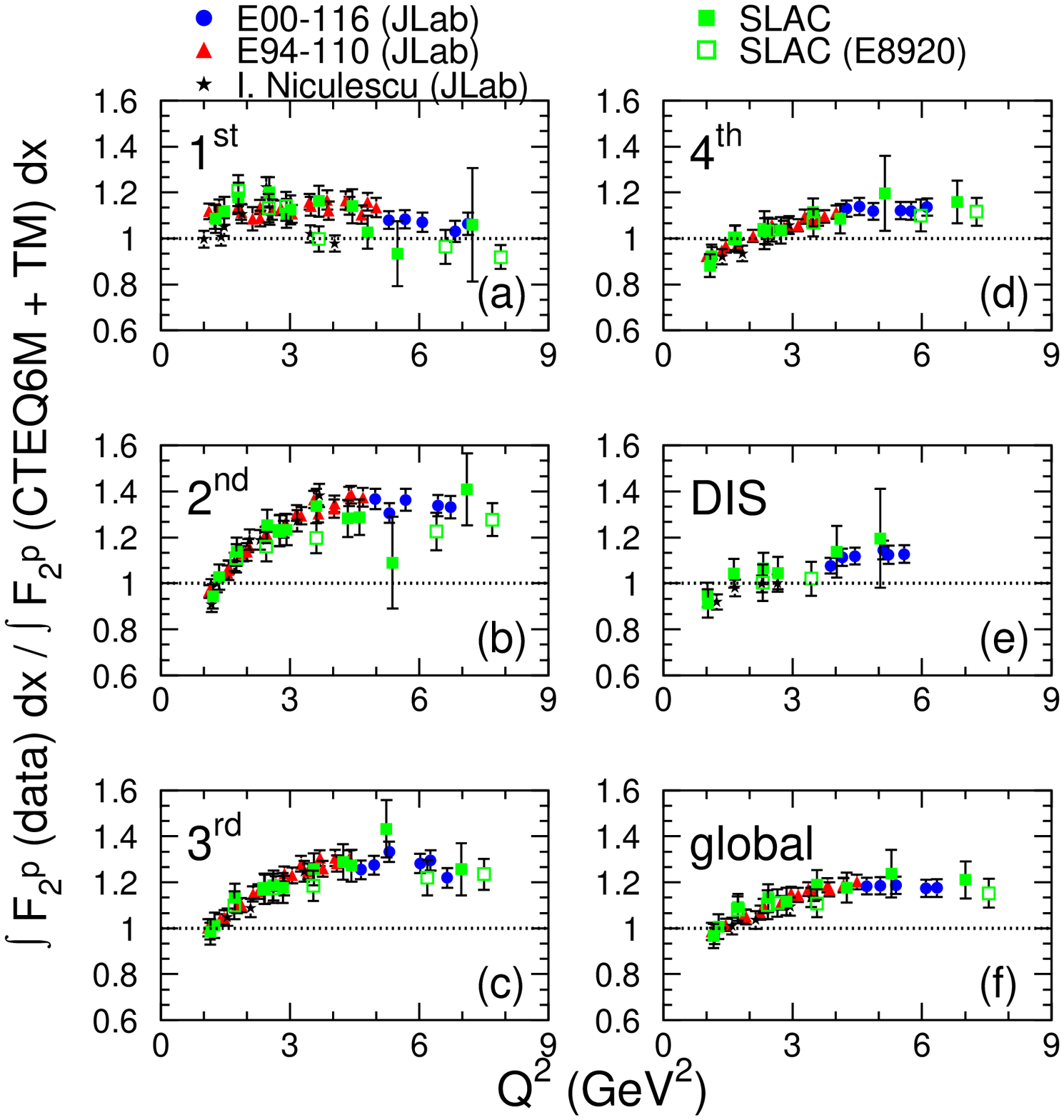}
\caption{(Color online) Local and global duality studies for $H(e,e')$ where CTEQ6M+TM \cite{cteq} was used for comparison. Together with the data of 
this experiment (blue circles) are also plotted the results using measurements from two previous Hall C experiments, I. Niculescu 
(black stars) \cite{ioana_thesis} and E94-110 (red triangles) \cite{liang_thesis}, and from SLAC \cite{whitlow_f2,riordan_thesis,poucher_thesis,bodek_thesis,bodek_paper,mestayer_paper,poucher_paper,atwood_paper}.}
\end{figure*}
\begin{figure*}
\centering 
\includegraphics[width=10cm]{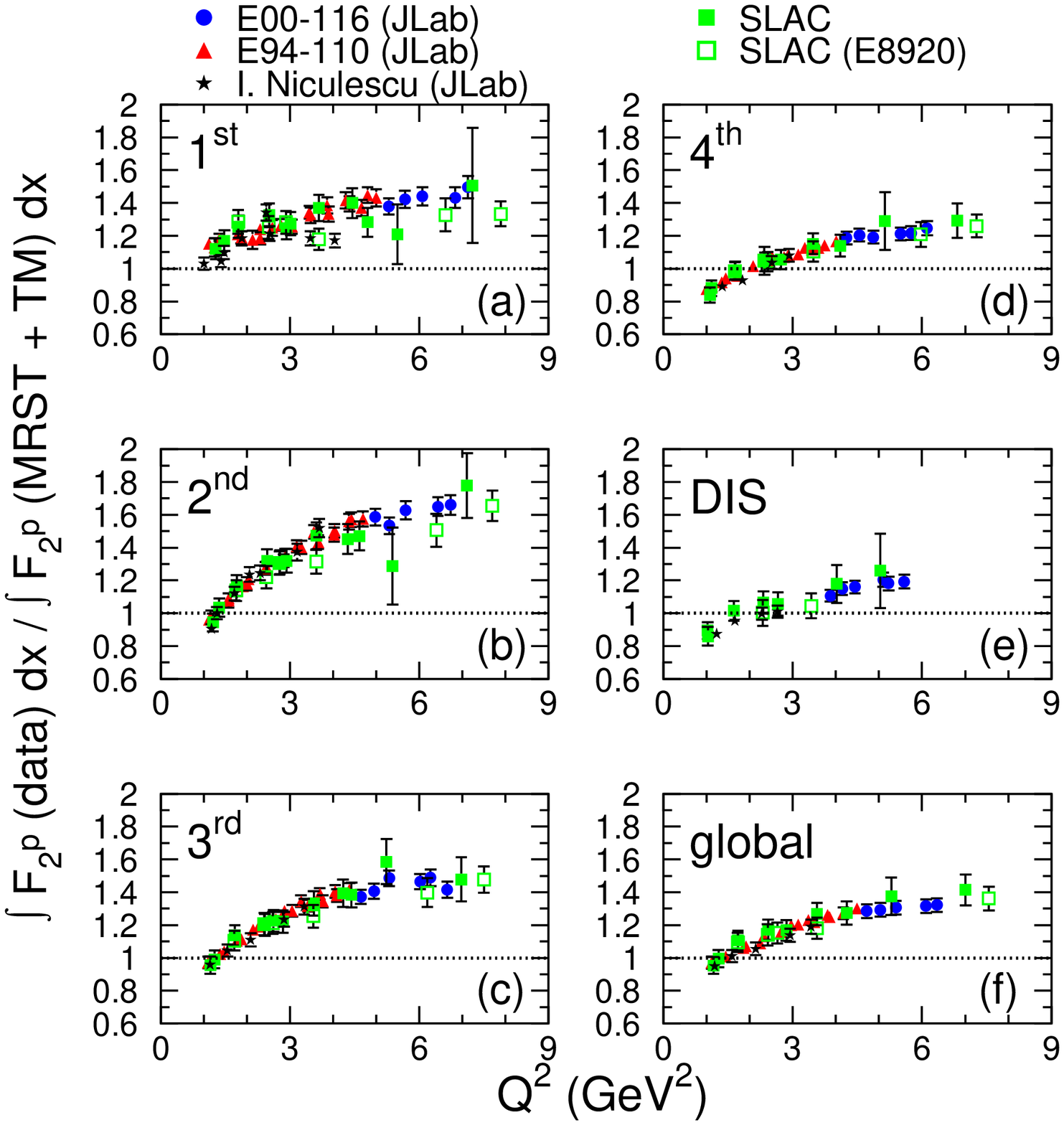}
\caption{(Color online) Local and global duality studies for $H(e,e')$ where MRST2004+TM \cite{mrst} was used for comparison. Together with the data of 
this experiment (blue circles) are also plotted the results using measurements from two previous Hall C experiments, I. Niculescu 
(black stars) \cite{ioana_thesis} and E94-110 (red triangles) \cite{liang_thesis}, and from SLAC \cite{whitlow_f2,riordan_thesis,poucher_thesis,bodek_thesis,bodek_paper,mestayer_paper,poucher_paper,atwood_paper}.}
\end{figure*}
\begin{figure*}
\centering 
\includegraphics[width=10cm]{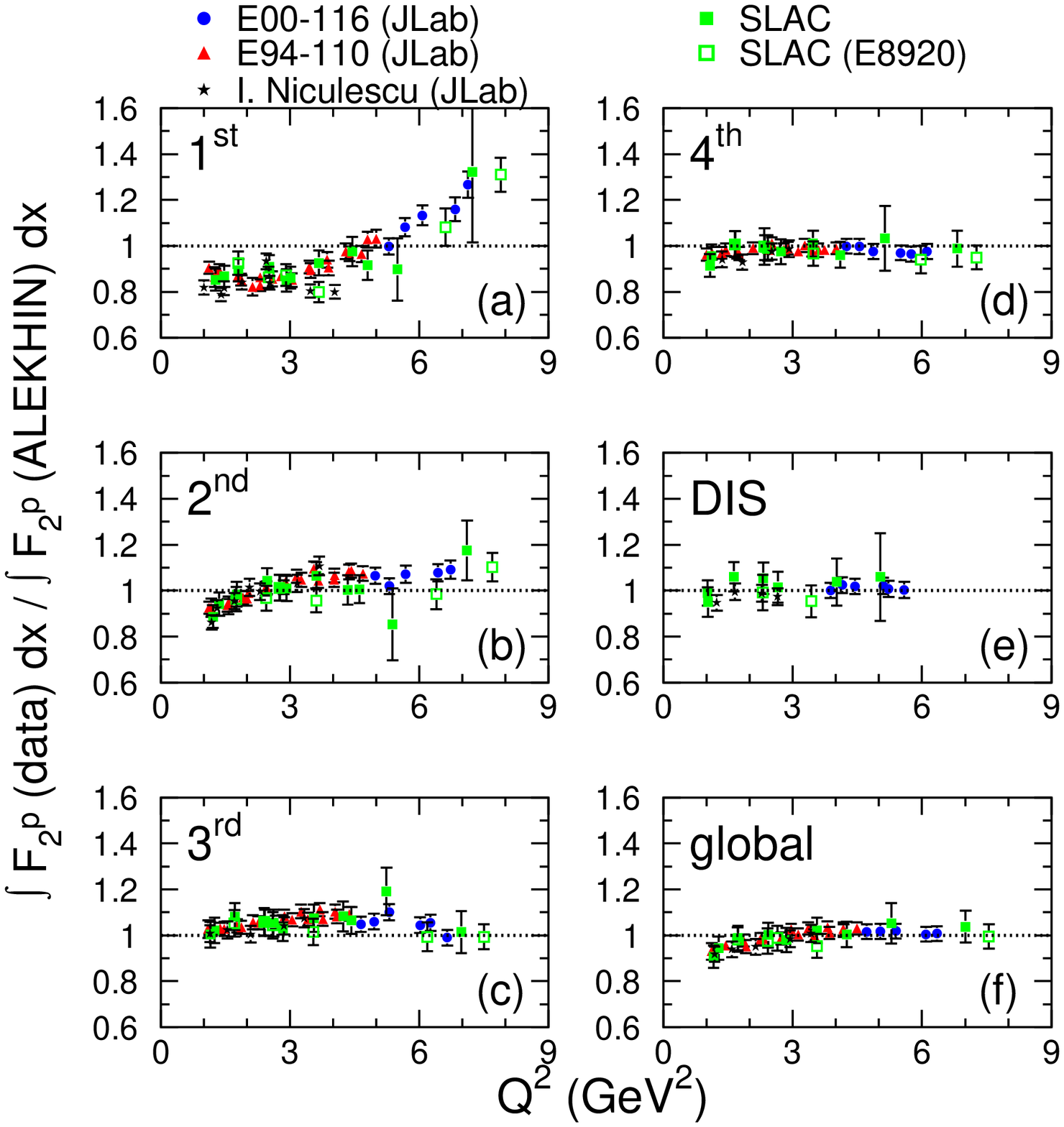}
\caption{(Color online) Local and global duality studies for $H(e,e')$ where ALEKHIN \cite{alekhin_05,alekhin_03} was used for comparison. Together with the data of 
this experiment (blue circles) are also plotted the results using measurements from two previous Hall C experiments, I. Niculescu 
(black stars) \cite{ioana_thesis} and E94-110 (red triangles) \cite{liang_thesis}, and from SLAC \cite{whitlow_f2,riordan_thesis,poucher_thesis,bodek_thesis,bodek_paper,mestayer_paper,poucher_paper,atwood_paper}.}
\end{figure*}
\begin{figure*}
\centering 
\includegraphics[width=10cm]{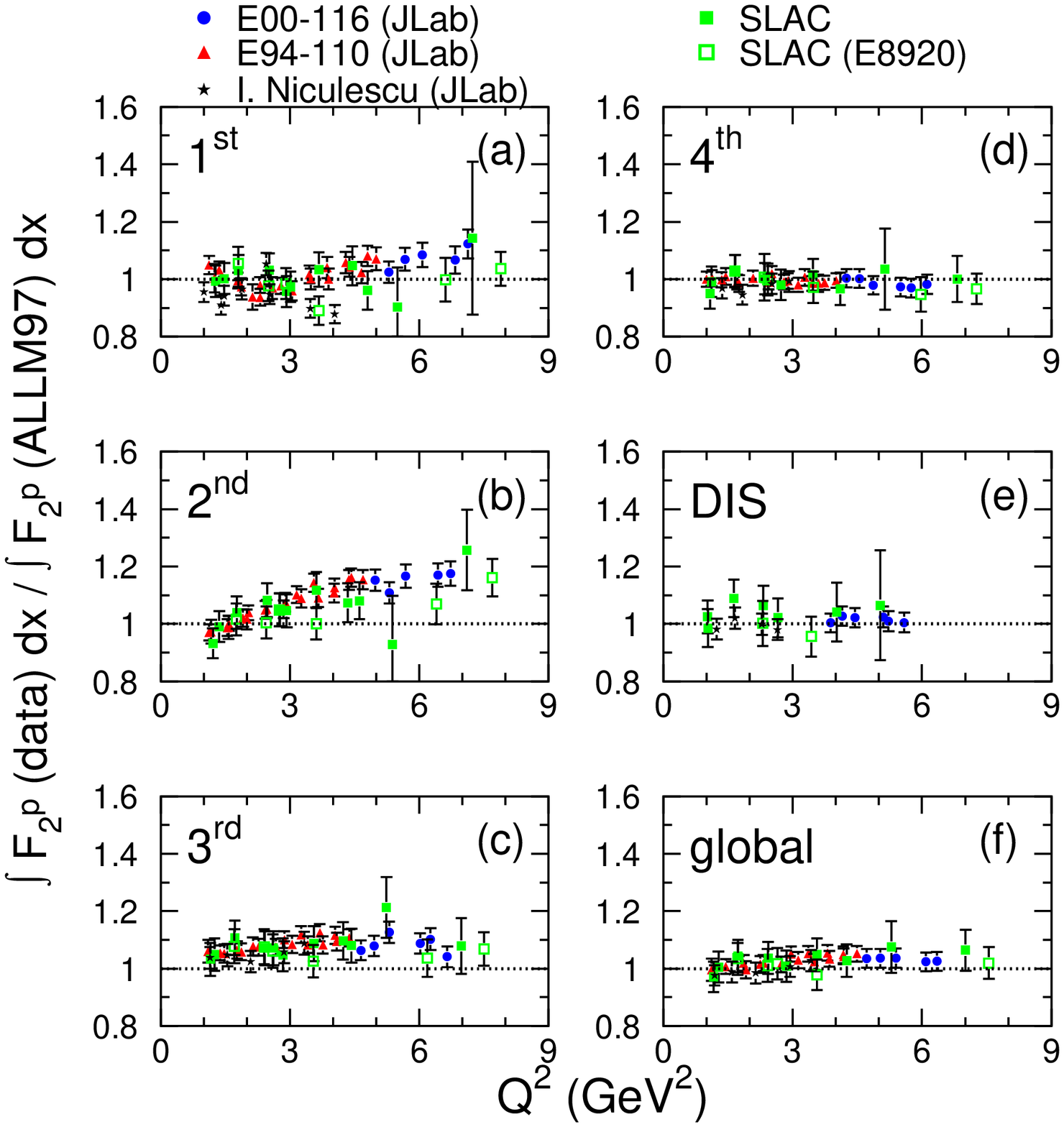}
\caption{(Color online) Local and global duality studies for $H(e,e')$ where ALLM97 \cite{allm97} was used for comparison. Together with the data of 
this experiment (blue circles) are also plotted the results using measurements from two previous Hall C experiments, I. Niculescu 
(black stars) \cite{ioana_thesis} and E94-110 (red triangles) \cite{liang_thesis}, and from SLAC \cite{whitlow_f2,riordan_thesis,poucher_thesis,bodek_thesis,bodek_paper,mestayer_paper,poucher_paper,atwood_paper}.}
\end{figure*}
\begin{figure*}
\centering 
\includegraphics[width=9.6cm]{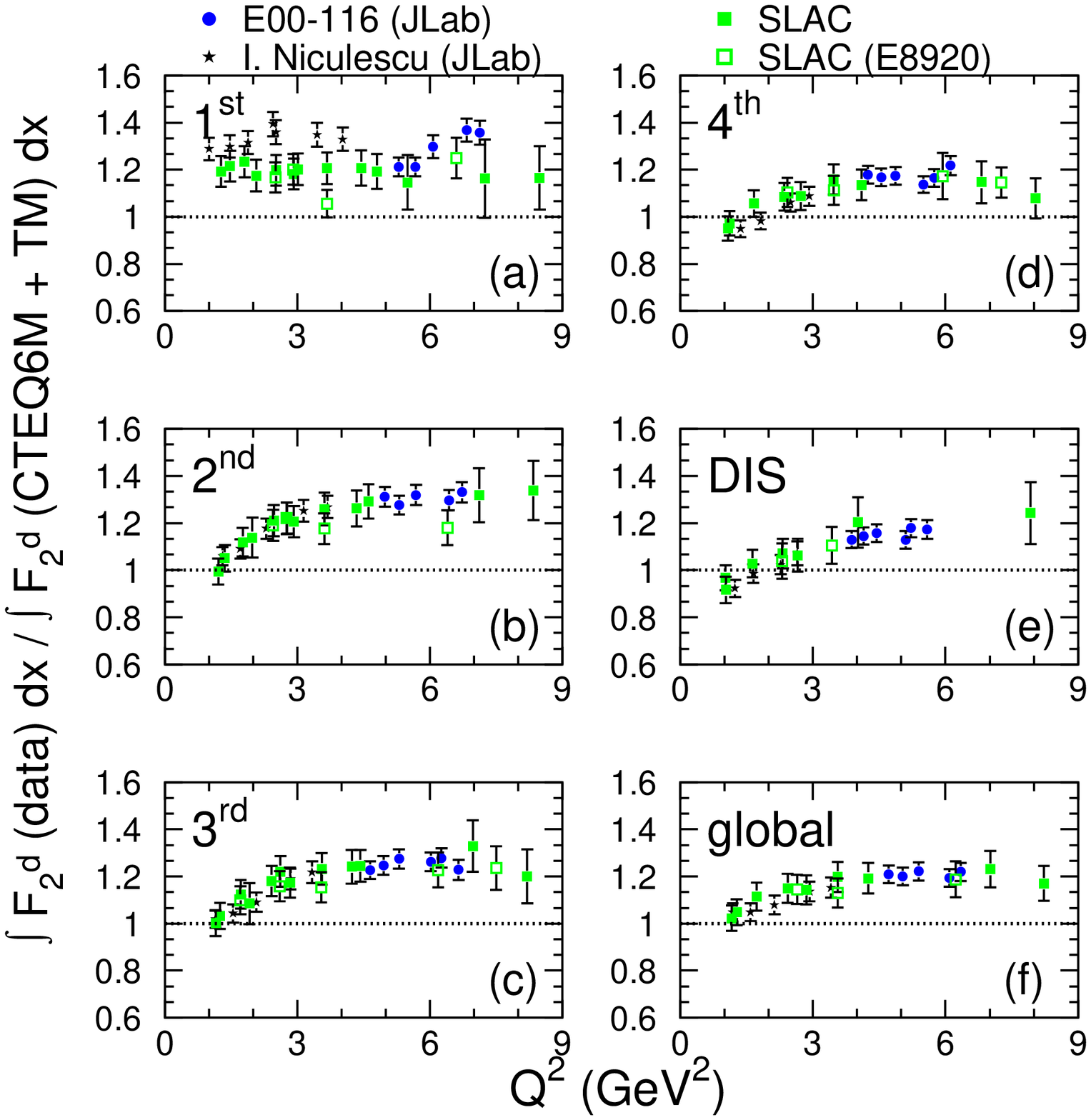}
\caption{(Color online) Local and global duality studies for $D(e,e')$ where CTEQ6M+TM \cite{cteq} multiplied by d/p ratio from \cite{d_p_antje} was used for comparison. Together with the data of 
this experiment (blue circles) are also plotted the results using measurements from two previous Hall C experiments, I. Niculescu 
(black stars) \cite{ioana_thesis} and E94-110 (red triangles) \cite{liang_thesis}, and from SLAC \cite{whitlow_f2,riordan_thesis,poucher_thesis,bodek_thesis,bodek_paper,mestayer_paper,poucher_paper,atwood_paper}.}
\end{figure*}
\begin{figure*}
\centering 
\includegraphics[width=9.6cm]{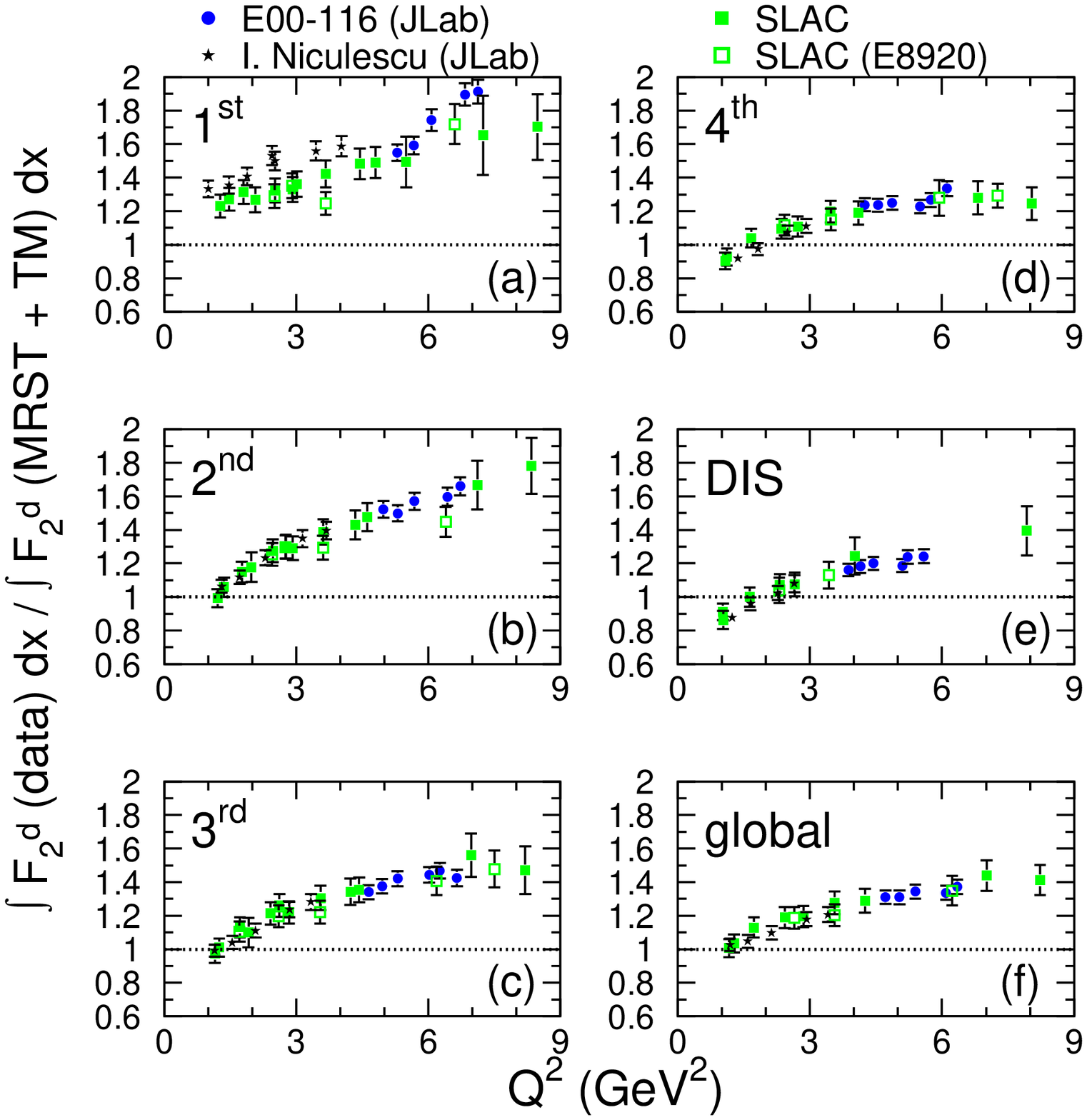}
\caption{(Color online) Local and global duality studies for $D(e,e')$ where MRST2004+TM \cite{mrst} multiplied by d/p ratio from \cite{d_p_antje} was used for comparison. Together with the data of 
this experiment (blue circles) are also plotted the results using measurements from two previous Hall C experiments, I. Niculescu 
(black stars) \cite{ioana_thesis} and E94-110 (red triangles) \cite{liang_thesis}, and from SLAC \cite{whitlow_f2,riordan_thesis,poucher_thesis,bodek_thesis,bodek_paper,mestayer_paper,poucher_paper,atwood_paper}.}
\end{figure*}
\begin{figure*}
\centering 
\includegraphics[width=9.6cm]{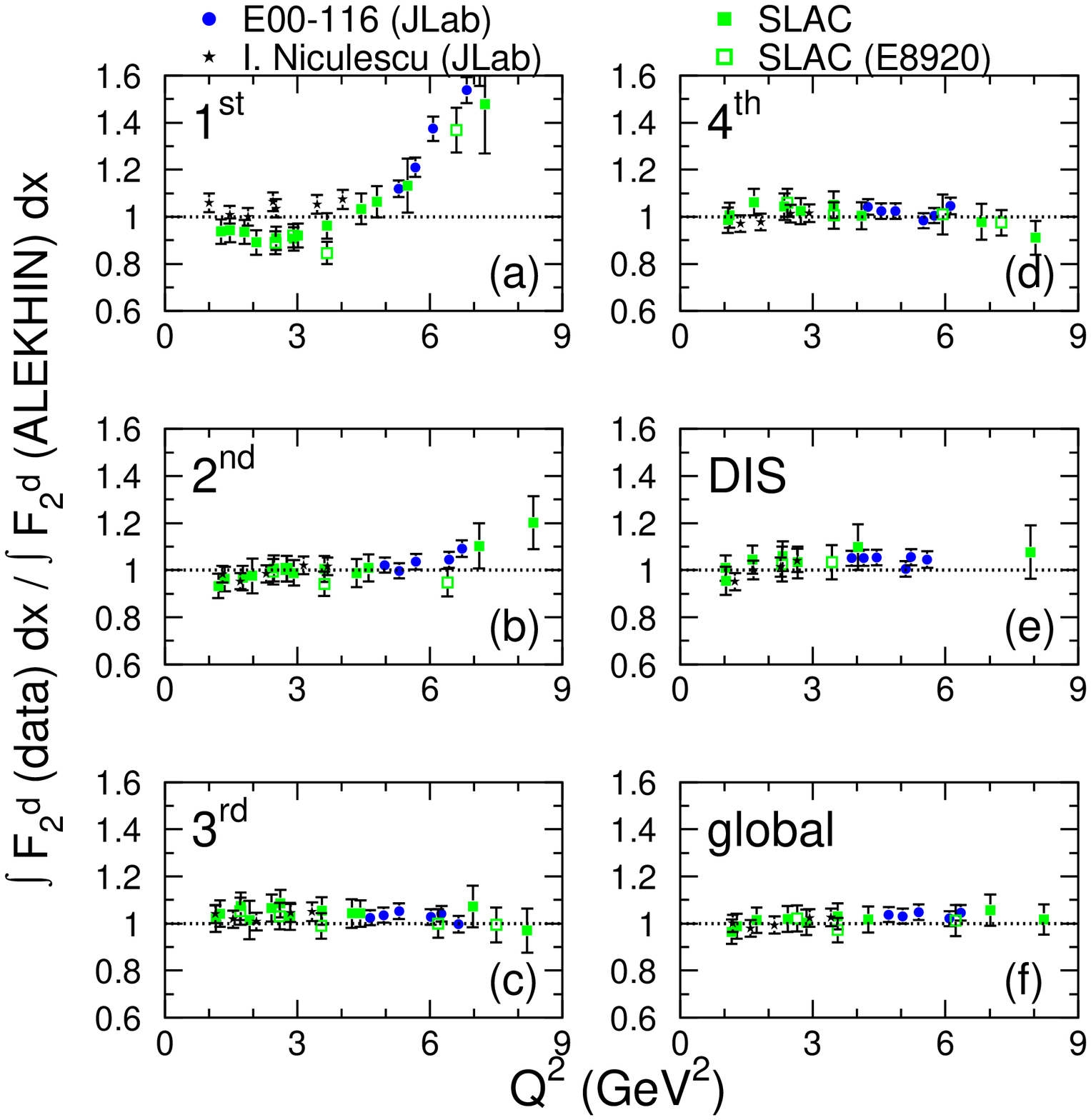}
\caption{(Color online) Local and global duality studies for $D(e,e')$ where ALEKHIN \cite{alekhin_05,alekhin_03} 
multiplied by d/p ratio from \cite{d_p_antje} was used for comparison. Together with the data of 
this experiment (blue circles) are also plotted the results using measurements from two previous 
Hall C experiments, I. Niculescu 
(black stars) \cite{ioana_thesis} and E94-110 (red triangles) \cite{liang_thesis}, and from 
SLAC \cite{whitlow_f2,riordan_thesis,poucher_thesis,bodek_thesis,bodek_paper,mestayer_paper,poucher_paper,atwood_paper}.}
\end{figure*}
\begin{figure*}
\centering 
\includegraphics[width=9.6cm]{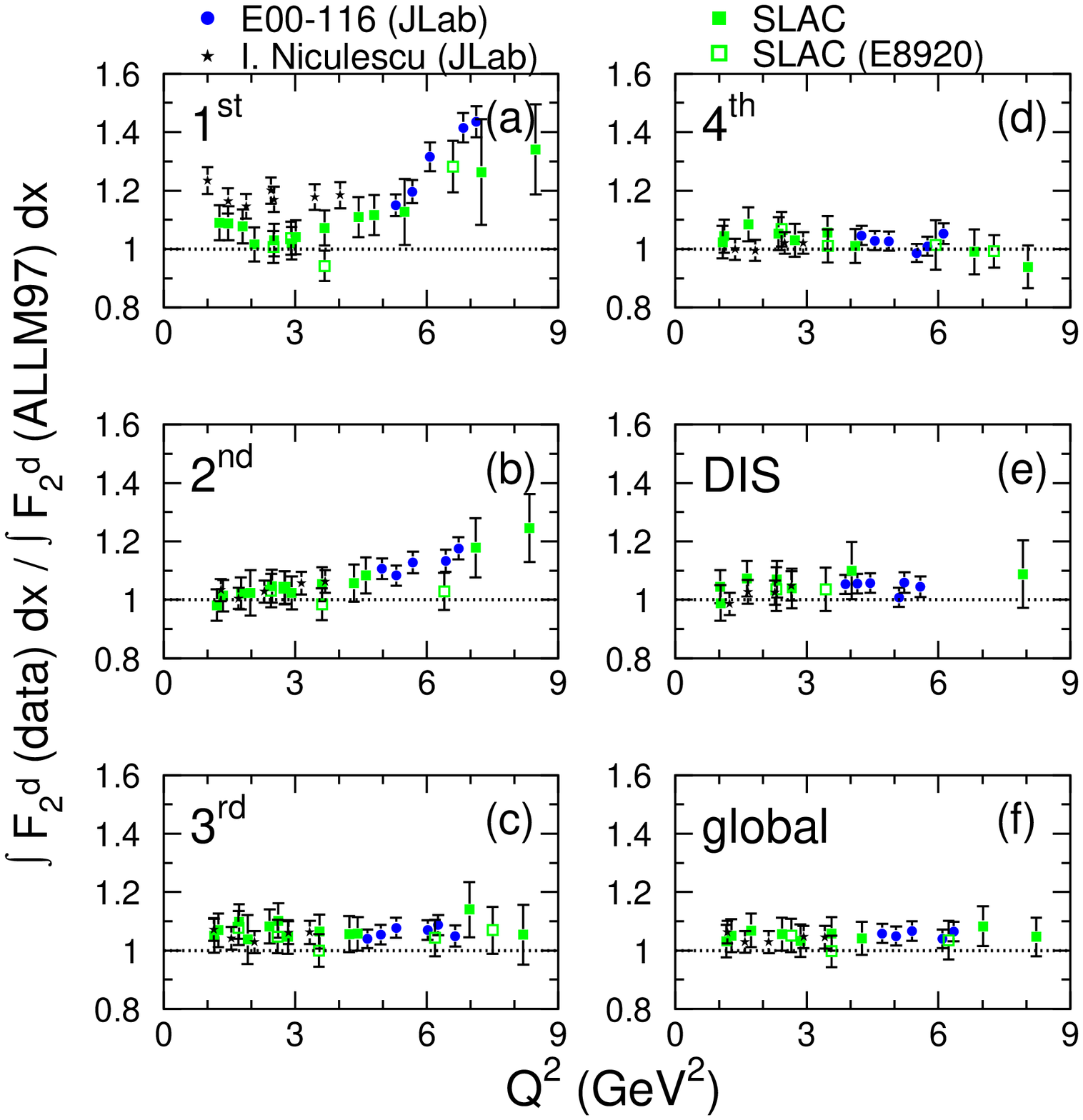}
\caption{(Color online) Local and global duality studies for $D(e,e')$ where ALLM97 \cite{allm97} multiplied by d/p ratio from \cite{d_p_antje} was used for comparison. Together with the data of 
this experiment (blue circles) are also plotted the results using measurements from two previous Hall C experiments, I. Niculescu 
(black stars) \cite{ioana_thesis} and E94-110 (red triangles) \cite{liang_thesis}, and from SLAC \cite{whitlow_f2,riordan_thesis,poucher_thesis,bodek_thesis,bodek_paper,mestayer_paper,poucher_paper,atwood_paper}.}
\end{figure*}
\begin{figure*}
\centering 
\includegraphics[width=10cm]{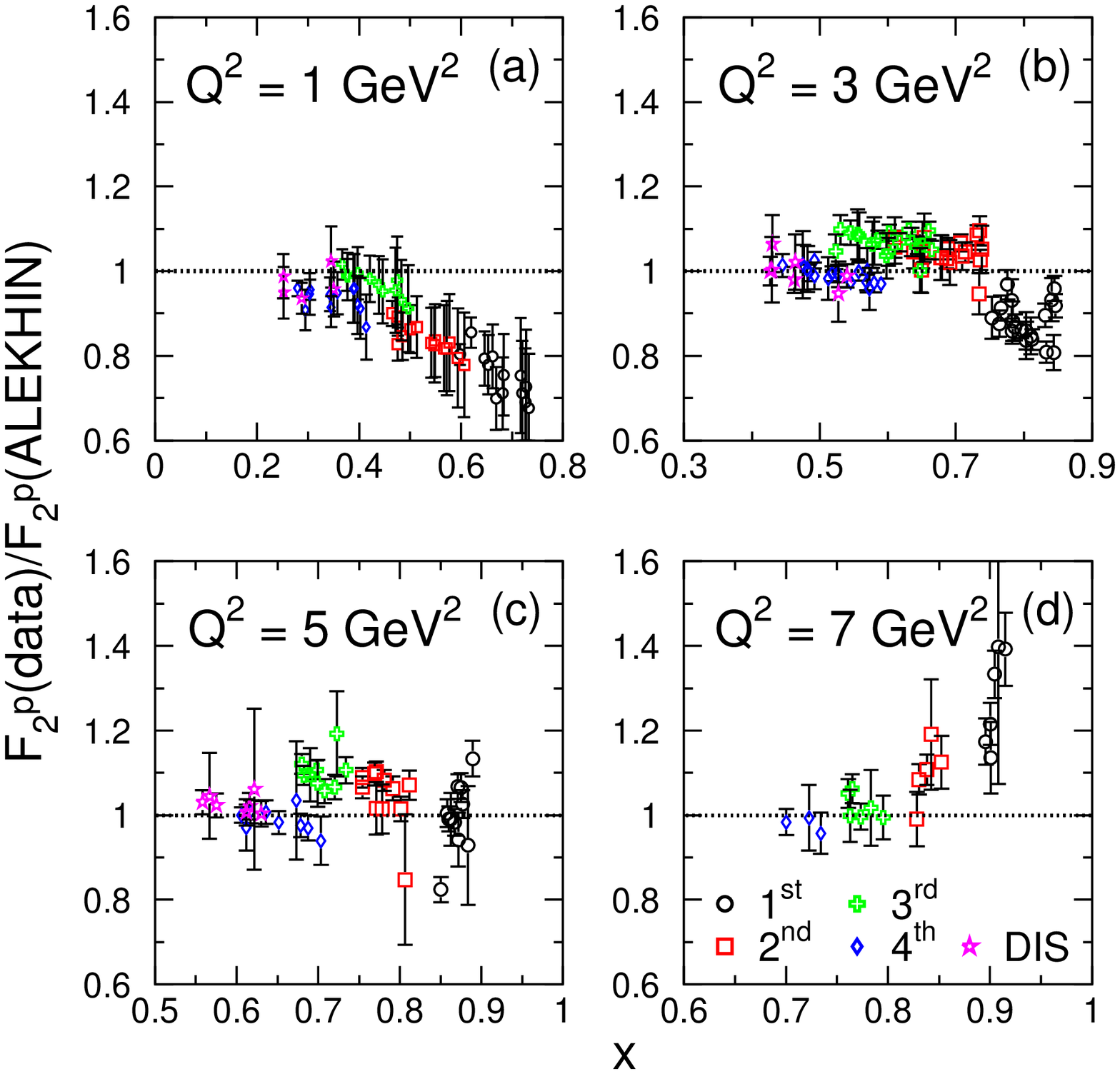}
\caption{(Color online) The ratio of $F_2$ structure function from data to $F_2$ from ALEKHIN \cite{alekhin_05,alekhin_03} 
versus $x$ at fixed $Q^{2}$.}
\end{figure*}
\begin{figure*}
\centering 
\includegraphics[width=10cm]{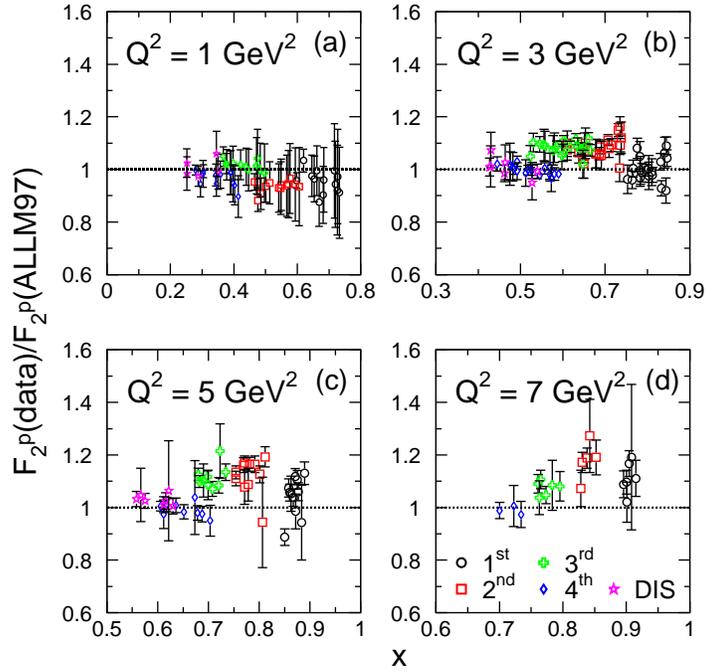}
\caption{(Color online) The ratio of $F_{2}$ structure function from data to $F_{2}$ from ALLM97 \cite{allm97} versus $x$ at 
fixed $Q^{2}$.}
\end{figure*}
\begin{figure*}
\centering 
\includegraphics[width=10cm]{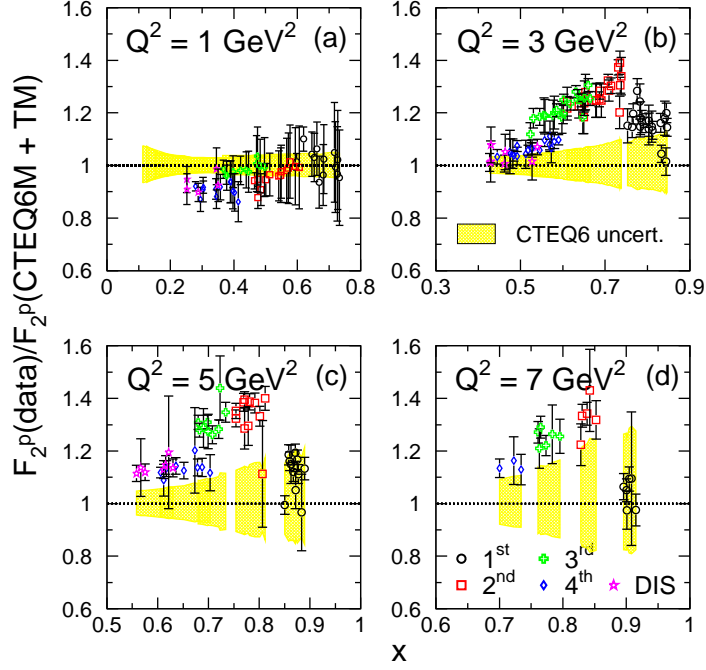}
\caption{(Color online) The ratio of $F_{2}$ structure function from data to $F_{2}$ from CTEQ6M+TM \cite{cteq} versus $x$ at 
fixed $Q^{2}$.}
\end{figure*}
\begin{figure*}
\centering 
\includegraphics[width=10cm]{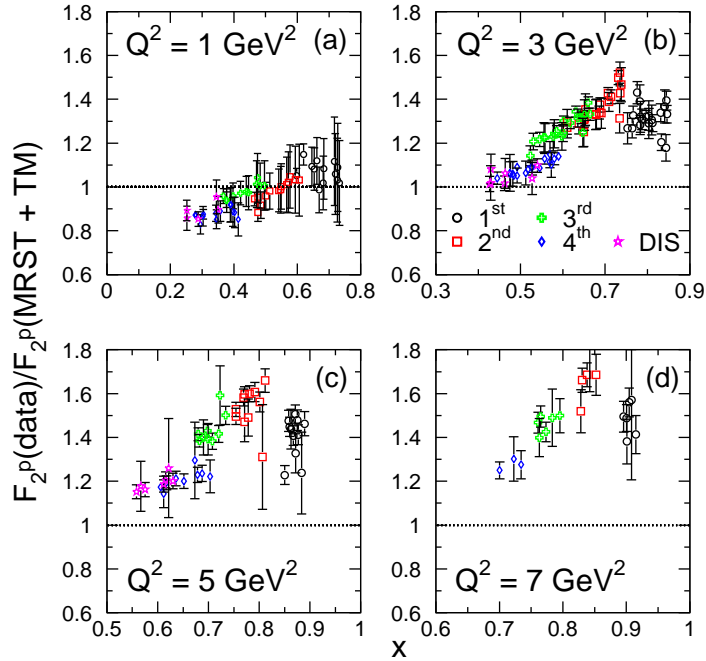}
\caption{(Color online) The ratio of $F_{2}$ structure function from data to $F_{2}$ from MRST2004+TM \cite{mrst} versus $x$ at 
fixed $Q^{2}$.}
\end{figure*}
\begin{figure*}
\centering 
\includegraphics[width=12cm]{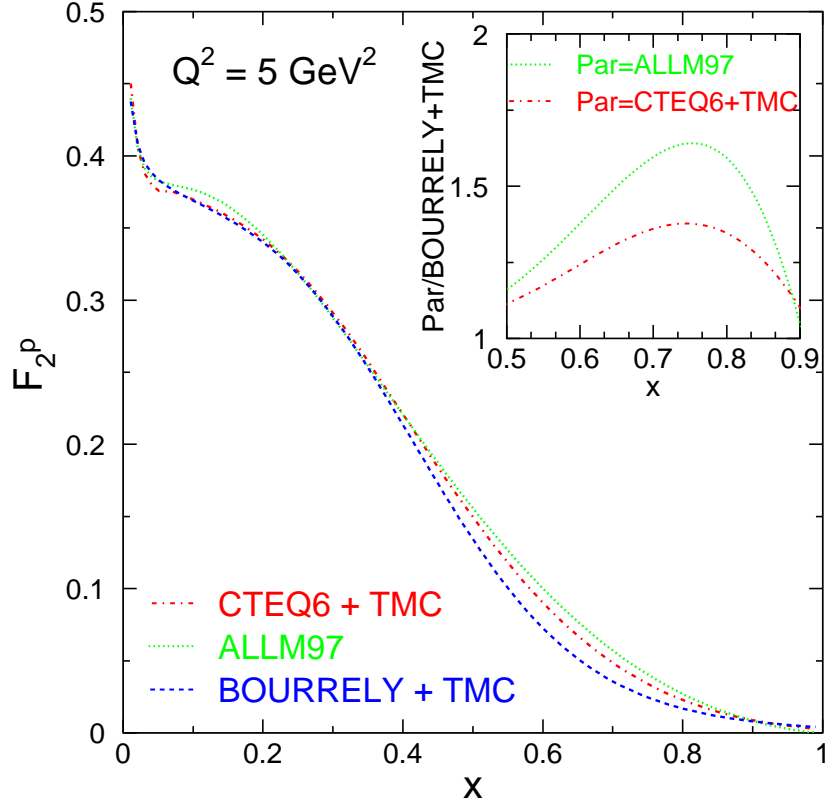}
\caption{(Color online) A comparison of the $F_{2}$ parameterization from Bourrely {\it et al.} \cite{bourrely} 
to CTEQ6M \cite{cteq} and ALLM97 \cite{allm97} at a $Q^{2}$ value of 5 GeV$^{2}$. In the insert, the ratio 
of the two parameterizations, ALLM97 and CTEQ6+TMC, to the parameterization from Bourrely {\it et al.} 
is shown.}
\end{figure*}

\begin{table*}
  \label{cross_sections}	
  \caption{Differential cross sections extracted from the measurements of E00-116. The normalization uncertainty is 1.75\%.}
  \begin{tabular}{ccccccccccccccc}
    \hline
    \hline
    $E^{'}$ & $Q^{2}$ & $W^{2}$ & $\frac{d \sigma^{Born}}{dE^{'}d \Omega}$ (H) & stat & syst & $\frac{d\sigma^{rad}}{dE^{'}d \Omega}$ (H) & stat & syst &
    $\frac{d \sigma^{Born}}{dE^{'}d \Omega}$ (D) & stat & syst & $\frac{d \sigma^{rad}}{dE^{'}d \Omega}$ (D) & stat & syst \\
    \hline
1.5456  & 3.5853  & 4.7162   & 1.3552  & 0.0215  & 0.0208  & 1.3921  & 0.0215  & 0.0156  & 2.1071  & 0.0276  & 0.0319  & 2.1383  & 0.0278  & 0.0242 \\
1.5623  & 3.6241  & 4.6461   & 1.3206  & 0.0209  & 0.0203  & 1.3507  & 0.0210  & 0.0152  & 2.0745  & 0.0268  & 0.0314  & 2.0941  & 0.0271  & 0.0237 \\
1.5790  & 3.6629  & 4.5760   & 1.2747  & 0.0200  & 0.0196  & 1.2985  & 0.0202  & 0.0146  & 1.9751  & 0.0255  & 0.0299  & 1.9844  & 0.0259  & 0.0224 \\
1.5957  & 3.7016  & 4.5059   & 1.2117  & 0.0194  & 0.0186  & 1.2309  & 0.0196  & 0.0138  & 1.9098  & 0.0247  & 0.0289  & 1.9096  & 0.0252  & 0.0216 \\
1.6124  & 3.7404  & 4.4358   & 1.1546  & 0.0187  & 0.0177  & 1.1680  & 0.0190  & 0.0131  & 1.8305  & 0.0238  & 0.0277  & 1.8221  & 0.0244  & 0.0206 \\
1.6291  & 3.7791  & 4.3656   & 1.1127  & 0.0182  & 0.0171  & 1.1212  & 0.0185  & 0.0126  & 1.7357  & 0.0229  & 0.0263  & 1.7200  & 0.0236  & 0.0194 \\
1.6458  & 3.8179  & 4.2955   & 1.1025  & 0.0179  & 0.0169  & 1.1045  & 0.0184  & 0.0124  & 1.6753  & 0.0223  & 0.0254  & 1.6534  & 0.0230  & 0.0187 \\
1.6625  & 3.8567  & 4.2254   & 1.0683  & 0.0174  & 0.0164  & 1.0668  & 0.0179  & 0.0120  & 1.6388  & 0.0219  & 0.0248  & 1.6102  & 0.0227  & 0.0182 \\
1.6793  & 3.8954  & 4.1553   & 0.9795  & 0.0165  & 0.0150  & 0.9756  & 0.0170  & 0.0109  & 1.5186  & 0.0207  & 0.0230  & 1.4874  & 0.0216  & 0.0168 \\
1.6960  & 3.9342  & 4.0852   & 0.9257  & 0.0161  & 0.0142  & 0.9189  & 0.0166  & 0.0103  & 1.4352  & 0.0200  & 0.0218  & 1.4014  & 0.0209  & 0.0158 \\
1.7127  & 3.9729  & 4.0151   & 0.8999  & 0.0157  & 0.0138  & 0.8897  & 0.0162  & 0.0100  & 1.4014  & 0.0195  & 0.0212  & 1.3632  & 0.0205  & 0.0154 \\
1.7294  & 4.0117  & 3.9450   & 0.8757  & 0.0153  & 0.0135  & 0.8623  & 0.0160  & 0.0097  & 1.3753  & 0.0193  & 0.0208  & 1.3332  & 0.0203  & 0.0151 \\
1.7461  & 4.0505  & 3.8748   & 0.8500  & 0.0149  & 0.0131  & 0.8337  & 0.0155  & 0.0094  & 1.2775  & 0.0183  & 0.0194  & 1.2358  & 0.0193  & 0.0140 \\
1.7628  & 4.0892  & 3.8047   & 0.8408  & 0.0147  & 0.0129  & 0.8209  & 0.0154  & 0.0092  & 1.2525  & 0.0180  & 0.0190  & 1.2076  & 0.0190  & 0.0137 \\
1.7795  & 4.1280  & 3.7346   & 0.8035  & 0.0142  & 0.0123  & 0.7816  & 0.0149  & 0.0088  & 1.1757  & 0.0172  & 0.0178  & 1.1318  & 0.0182  & 0.0128 \\
1.7962  & 4.1667  & 3.6645   & 0.7409  & 0.0135  & 0.0114  & 0.7188  & 0.0142  & 0.0081  & 1.1152  & 0.0166  & 0.0169  & 1.0705  & 0.0176  & 0.0121 \\
1.7973  & 4.1692  & 3.6601   & 0.7589  & 0.0102  & 0.0117  & 0.7352  & 0.0108  & 0.0082  & 1.1433  & 0.0107  & 0.0173  & 1.0962  & 0.0113  & 0.0124 \\
1.8167  & 4.2143  & 3.5785   & 0.7148  & 0.0097  & 0.0110  & 0.6903  & 0.0102  & 0.0077  & 1.1020  & 0.0102  & 0.0167  & 1.0524  & 0.0109  & 0.0119 \\
1.8361  & 4.2593  & 3.4970   & 0.6820  & 0.0093  & 0.0105  & 0.6565  & 0.0098  & 0.0074  & 1.0121  & 0.0095  & 0.0153  & 0.9639  & 0.0101  & 0.0109 \\
1.8556  & 4.3044  & 3.4155   & 0.6232  & 0.0087  & 0.0096  & 0.5993  & 0.0093  & 0.0067  & 0.9622  & 0.0090  & 0.0146  & 0.9125  & 0.0097  & 0.0103 \\
1.8750  & 4.3495  & 3.3339   & 0.6015  & 0.0085  & 0.0092  & 0.5773  & 0.0091  & 0.0065  & 0.8998  & 0.0087  & 0.0136  & 0.8496  & 0.0093  & 0.0096 \\
1.8944  & 4.3946  & 3.2524   & 0.5551  & 0.0080  & 0.0085  & 0.5320  & 0.0086  & 0.0060  & 0.8456  & 0.0082  & 0.0128  & 0.7941  & 0.0089  & 0.0090 \\
1.9139  & 4.4396  & 3.1709   & 0.5318  & 0.0077  & 0.0082  & 0.5070  & 0.0083  & 0.0057  & 0.8058  & 0.0079  & 0.0122  & 0.7520  & 0.0086  & 0.0085 \\
1.9333  & 4.4847  & 3.0893   & 0.5271  & 0.0077  & 0.0081  & 0.4961  & 0.0083  & 0.0056  & 0.7552  & 0.0075  & 0.0114  & 0.7002  & 0.0082  & 0.0079 \\
1.9527  & 4.5298  & 3.0078   & 0.5116  & 0.0074  & 0.0079  & 0.4707  & 0.0082  & 0.0053  & 0.7004  & 0.0070  & 0.0106  & 0.6454  & 0.0078  & 0.0073 \\
1.9721  & 4.5748  & 2.9263   & 0.5150  & 0.0072  & 0.0079  & 0.4637  & 0.0082  & 0.0052  & 0.6366  & 0.0066  & 0.0097  & 0.5837  & 0.0073  & 0.0066 \\
1.9916  & 4.6199  & 2.8447   & 0.4624  & 0.0066  & 0.0071  & 0.4090  & 0.0076  & 0.0046  & 0.6079  & 0.0063  & 0.0092  & 0.5545  & 0.0071  & 0.0063 \\
2.0110  & 4.6650  & 2.7632   & 0.4032  & 0.0061  & 0.0062  & 0.3600  & 0.0070  & 0.0040  & 0.5547  & 0.0060  & 0.0084  & 0.5046  & 0.0067  & 0.0057 \\
2.0304  & 4.7101  & 2.6817   & 0.3493  & 0.0055  & 0.0054  & 0.3173  & 0.0062  & 0.0036  & 0.5063  & 0.0056  & 0.0077  & 0.4597  & 0.0063  & 0.0052 \\
2.0499  & 4.7551  & 2.6001   & 0.2953  & 0.0050  & 0.0045  & 0.2722  & 0.0056  & 0.0031  & 0.4710  & 0.0053  & 0.0071  & 0.4269  & 0.0060  & 0.0048 \\
2.0693  & 4.8002  & 2.5186   & 0.2786  & 0.0050  & 0.0043  & 0.2585  & 0.0055  & 0.0029  & 0.4320  & 0.0050  & 0.0066  & 0.3906  & 0.0056  & 0.0044 \\
2.0887  & 4.8453  & 2.4370   & 0.2701  & 0.0049  & 0.0042  & 0.2491  & 0.0055  & 0.0028  & 0.4007  & 0.0047  & 0.0061  & 0.3610  & 0.0054  & 0.0041 \\
2.0898  & 4.8477  & 2.4327   & 0.2737  & 0.0035  & 0.0042  & 0.2519  & 0.0039  & 0.0028  & 0.4123  & 0.0036  & 0.0063  & 0.3707  & 0.0041  & 0.0042 \\
2.1124  & 4.9001  & 2.3379   & 0.2959  & 0.0035  & 0.0046  & 0.2601  & 0.0041  & 0.0029  & 0.3702  & 0.0033  & 0.0056  & 0.3306  & 0.0038  & 0.0037 \\
2.1349  & 4.9525  & 2.2431   & 0.2671  & 0.0032  & 0.0041  & 0.2266  & 0.0038  & 0.0026  & 0.3394  & 0.0030  & 0.0052  & 0.2998  & 0.0035  & 0.0034 \\
2.1575  & 5.0049  & 2.1483   & 0.2188  & 0.0027  & 0.0034  & 0.1875  & 0.0033  & 0.0021  & 0.2923  & 0.0027  & 0.0045  & 0.2584  & 0.0032  & 0.0029 \\
2.1801  & 5.0573  & 2.0535   & 0.1789  & 0.0024  & 0.0028  & 0.1548  & 0.0029  & 0.0018  & 0.2545  & 0.0024  & 0.0039  & 0.2246  & 0.0029  & 0.0026 \\
2.2027  & 5.1097  & 1.9587   & 0.1442  & 0.0021  & 0.0023  & 0.1252  & 0.0026  & 0.0014  & 0.2271  & 0.0023  & 0.0035  & 0.2002  & 0.0027  & 0.0023 \\
2.2253  & 5.1621  & 1.8639   & 0.1218  & 0.0019  & 0.0019  & 0.1063  & 0.0023  & 0.0013  & 0.1912  & 0.0020  & 0.0030  & 0.1696  & 0.0024  & 0.0020 \\
2.2479  & 5.2145  & 1.7691   & 0.0925  & 0.0017  & 0.0015  & 0.0823  & 0.0020  & 0.0010  & 0.1595  & 0.0018  & 0.0026  & 0.1431  & 0.0022  & 0.0017 \\
2.2705  & 5.2669  & 1.6743   & 0.0816  & 0.0016  & 0.0014  & 0.0727  & 0.0020  & 0.0010  & 0.1342  & 0.0016  & 0.0022  & 0.1227  & 0.0020  & 0.0015 \\
2.2931  & 5.3193  & 1.5795   & 0.0750  & 0.0016  & 0.0015  & 0.0657  & 0.0020  & 0.0011  & 0.1093  & 0.0015  & 0.0019  & 0.1032  & 0.0018  & 0.0013 \\
2.3157  & 5.3718  & 1.4847   & 0.0696  & 0.0017  & 0.0017  & 0.0582  & 0.0023  & 0.0013  & 0.0929  & 0.0013  & 0.0018  & 0.0912  & 0.0017  & 0.0012 \\
2.3383  & 5.4242  & 1.3899   & 0.0405  & 0.0012  & 0.0014  & 0.0354  & 0.0018  & 0.0012  & 0.0796  & 0.0013  & 0.0019  & 0.0816  & 0.0017  & 0.0012 \\
2.3609  & 5.4766  & 1.2951   & 0.0203  & 0.0009  & 0.0011  & 0.0218  & 0.0014  & 0.0012  & 0.0670  & 0.0011  & 0.0019  & 0.0723  & 0.0015  & 0.0013 \\
1.4831  & 3.9954  & 4.4235   & 0.8663  & 0.0185  & 0.0133  & 0.8764  & 0.0187  & 0.0098  & 1.2778  & 0.0208  & 0.0194  & 1.2761  & 0.0213  & 0.0144 \\
1.4991  & 4.0386  & 4.3502   & 0.7946  & 0.0176  & 0.0122  & 0.8019  & 0.0179  & 0.0090  & 1.2380  & 0.0202  & 0.0188  & 1.2305  & 0.0208  & 0.0139 \\
1.5151  & 4.0818  & 4.2769   & 0.7645  & 0.0167  & 0.0117  & 0.7684  & 0.0170  & 0.0086  & 1.1795  & 0.0193  & 0.0179  & 1.1689  & 0.0199  & 0.0132 \\
1.5312  & 4.1250  & 4.2036   & 0.7400  & 0.0164  & 0.0114  & 0.7408  & 0.0168  & 0.0083  & 1.1221  & 0.0186  & 0.0170  & 1.1068  & 0.0192  & 0.0125 \\
1.5472  & 4.1682  & 4.1303   & 0.7246  & 0.0161  & 0.0111  & 0.7221  & 0.0166  & 0.0081  & 1.0686  & 0.0179  & 0.0162  & 1.0499  & 0.0187  & 0.0119 \\
1.5632  & 4.2114  & 4.0571   & 0.6731  & 0.0152  & 0.0103  & 0.6686  & 0.0157  & 0.0075  & 1.0422  & 0.0174  & 0.0158  & 1.0200  & 0.0182  & 0.0115 \\
1.5793  & 4.2546  & 3.9838   & 0.6490  & 0.0149  & 0.0100  & 0.6418  & 0.0155  & 0.0072  & 0.9415  & 0.0164  & 0.0143  & 0.9193  & 0.0172  & 0.0104 \\
1.5953  & 4.2978  & 3.9105   & 0.6325  & 0.0147  & 0.0097  & 0.6228  & 0.0153  & 0.0070  & 0.9152  & 0.0161  & 0.0139  & 0.8904  & 0.0169  & 0.0101 \\
1.6113  & 4.3410  & 3.8372   & 0.6119  & 0.0143  & 0.0094  & 0.5998  & 0.0150  & 0.0067  & 0.8736  & 0.0155  & 0.0132  & 0.8473  & 0.0163  & 0.0096 \\
1.6273  & 4.3842  & 3.7639   & 0.5512  & 0.0135  & 0.0085  & 0.5390  & 0.0142  & 0.0060  & 0.8480  & 0.0152  & 0.0129  & 0.8196  & 0.0161  & 0.0093 \\
1.6434  & 4.4274  & 3.6907   & 0.5340  & 0.0132  & 0.0082  & 0.5199  & 0.0139  & 0.0058  & 0.7789  & 0.0145  & 0.0118  & 0.7510  & 0.0154  & 0.0085 \\
1.6594  & 4.4706  & 3.6174   & 0.5193  & 0.0131  & 0.0080  & 0.5035  & 0.0138  & 0.0056  & 0.7367  & 0.0141  & 0.0112  & 0.7082  & 0.0149  & 0.0080 \\
1.6754  & 4.5138  & 3.5441   & 0.4758  & 0.0124  & 0.0073  & 0.4603  & 0.0132  & 0.0052  & 0.7148  & 0.0138  & 0.0108  & 0.6846  & 0.0147  & 0.0077 \\
1.6915  & 4.5570  & 3.4708   & 0.4524  & 0.0120  & 0.0070  & 0.4367  & 0.0128  & 0.0049  & 0.6694  & 0.0133  & 0.0102  & 0.6393  & 0.0142  & 0.0072 \\
	    \hline
    	\hline
  	\end{tabular}
	\end{table*}

\begin{table*}
  \label{cross_sections}	
  \caption{Differential cross sections extracted from the measurements of E00-116. The normalization uncertainty is 1.75\%.}
  \begin{tabular}{ccccccccccccccc}
    \hline
    \hline
    $E^{'}$ & $Q^{2}$ & $W^{2}$ & $\frac{d \sigma^{Born}}{dE^{'}d \Omega}$ (H) & stat & syst & $\frac{d\sigma^{rad}}{dE^{'}d \Omega}$ (H) & stat & syst &
    $\frac{d \sigma^{Born}}{dE^{'}d \Omega}$ (D) & stat & syst & $\frac{d \sigma^{rad}}{dE^{'}d \Omega}$ (D) & stat & syst \\
    \hline
1.7075  & 4.6001  & 3.3975   & 0.4201  & 0.0116  & 0.0065  & 0.4054  & 0.0123  & 0.0045  & 0.6431  & 0.0128  & 0.0098  & 0.6117  & 0.0137  & 0.0069 \\
1.7235  & 4.6433  & 3.3243   & 0.4230  & 0.0113  & 0.0065  & 0.4071  & 0.0120  & 0.0046  & 0.5836  & 0.0120  & 0.0089  & 0.5534  & 0.0129  & 0.0063 \\
1.7244  & 4.6456  & 3.3204   & 0.4093  & 0.0064  & 0.0063  & 0.3939  & 0.0068  & 0.0044  & 0.5993  & 0.0068  & 0.0091  & 0.5674  & 0.0073  & 0.0064 \\
1.7430  & 4.6958  & 3.2352   & 0.3807  & 0.0060  & 0.0058  & 0.3659  & 0.0064  & 0.0041  & 0.5712  & 0.0065  & 0.0087  & 0.5377  & 0.0071  & 0.0061 \\
1.7617  & 4.7460  & 3.1500   & 0.3693  & 0.0058  & 0.0057  & 0.3528  & 0.0062  & 0.0040  & 0.5378  & 0.0061  & 0.0082  & 0.5031  & 0.0066  & 0.0057 \\
1.7803  & 4.7963  & 3.0648   & 0.3606  & 0.0057  & 0.0055  & 0.3394  & 0.0062  & 0.0038  & 0.5111  & 0.0059  & 0.0078  & 0.4748  & 0.0065  & 0.0054 \\
1.7990  & 4.8465  & 2.9796   & 0.3687  & 0.0056  & 0.0057  & 0.3379  & 0.0062  & 0.0038  & 0.4707  & 0.0055  & 0.0071  & 0.4345  & 0.0061  & 0.0049 \\
1.8176  & 4.8967  & 2.8944   & 0.3350  & 0.0052  & 0.0051  & 0.3003  & 0.0059  & 0.0034  & 0.4412  & 0.0052  & 0.0067  & 0.4048  & 0.0058  & 0.0046 \\
1.8362  & 4.9469  & 2.8092   & 0.3029  & 0.0047  & 0.0047  & 0.2682  & 0.0055  & 0.0030  & 0.3902  & 0.0048  & 0.0059  & 0.3568  & 0.0054  & 0.0040 \\
1.8549  & 4.9972  & 2.7240   & 0.2416  & 0.0042  & 0.0037  & 0.2176  & 0.0047  & 0.0024  & 0.3643  & 0.0046  & 0.0055  & 0.3320  & 0.0052  & 0.0038 \\
1.8735  & 5.0474  & 2.6388   & 0.2065  & 0.0038  & 0.0032  & 0.1900  & 0.0042  & 0.0021  & 0.3191  & 0.0042  & 0.0048  & 0.2906  & 0.0048  & 0.0033 \\
1.8922  & 5.0976  & 2.5536   & 0.1809  & 0.0035  & 0.0028  & 0.1687  & 0.0039  & 0.0019  & 0.2912  & 0.0040  & 0.0044  & 0.2646  & 0.0045  & 0.0030 \\
1.9108  & 5.1478  & 2.4684   & 0.1664  & 0.0035  & 0.0026  & 0.1556  & 0.0039  & 0.0017  & 0.2737  & 0.0038  & 0.0042  & 0.2479  & 0.0044  & 0.0028 \\
1.9294  & 5.1980  & 2.3832   & 0.1810  & 0.0037  & 0.0028  & 0.1656  & 0.0042  & 0.0019  & 0.2485  & 0.0036  & 0.0038  & 0.2239  & 0.0041  & 0.0025 \\
1.9481  & 5.2483  & 2.2980   & 0.1802  & 0.0036  & 0.0028  & 0.1562  & 0.0043  & 0.0018  & 0.2217  & 0.0034  & 0.0034  & 0.1983  & 0.0039  & 0.0022 \\
1.9667  & 5.2985  & 2.2128   & 0.1649  & 0.0033  & 0.0025  & 0.1395  & 0.0040  & 0.0016  & 0.2068  & 0.0032  & 0.0032  & 0.1828  & 0.0037  & 0.0021 \\
1.9854  & 5.3487  & 2.1276   & 0.1410  & 0.0028  & 0.0022  & 0.1206  & 0.0035  & 0.0014  & 0.1870  & 0.0029  & 0.0029  & 0.1654  & 0.0034  & 0.0019 \\
2.0040  & 5.3989  & 2.0424   & 0.1089  & 0.0024  & 0.0017  & 0.0947  & 0.0029  & 0.0011  & 0.1648  & 0.0027  & 0.0025  & 0.1464  & 0.0032  & 0.0017 \\
2.0051  & 5.4019  & 2.0373   & 0.1060  & 0.0014  & 0.0016  & 0.0923  & 0.0017  & 0.0011  & 0.1679  & 0.0019  & 0.0026  & 0.1488  & 0.0022  & 0.0017 \\
2.0268  & 5.4603  & 1.9382   & 0.0885  & 0.0013  & 0.0014  & 0.0774  & 0.0015  & 0.0009  & 0.1345  & 0.0016  & 0.0021  & 0.1201  & 0.0020  & 0.0014 \\
2.0485  & 5.5187  & 1.8391   & 0.0718  & 0.0011  & 0.0011  & 0.0632  & 0.0014  & 0.0008  & 0.1214  & 0.0015  & 0.0019  & 0.1086  & 0.0018  & 0.0012 \\
2.0702  & 5.5771  & 1.7401   & 0.0560  & 0.0010  & 0.0009  & 0.0503  & 0.0012  & 0.0006  & 0.0981  & 0.0013  & 0.0016  & 0.0892  & 0.0016  & 0.0010 \\
2.0918  & 5.6355  & 1.6410   & 0.0520  & 0.0010  & 0.0009  & 0.0465  & 0.0012  & 0.0007  & 0.0830  & 0.0012  & 0.0014  & 0.0772  & 0.0015  & 0.0009 \\
2.1135  & 5.6939  & 1.5419   & 0.0503  & 0.0010  & 0.0011  & 0.0438  & 0.0012  & 0.0008  & 0.0625  & 0.0010  & 0.0011  & 0.0611  & 0.0013  & 0.0008 \\
2.1352  & 5.7523  & 1.4428   & 0.0407  & 0.0010  & 0.0012  & 0.0338  & 0.0014  & 0.0009  & 0.0547  & 0.0010  & 0.0012  & 0.0556  & 0.0013  & 0.0008 \\
2.1569  & 5.8107  & 1.3437   & 0.0182  & 0.0006  & 0.0008  & 0.0173  & 0.0009  & 0.0007  & 0.0487  & 0.0009  & 0.0013  & 0.0510  & 0.0012  & 0.0009 \\
2.1785  & 5.8691  & 1.2447   & 0.0062  & 0.0004  & 0.0004  & 0.0129  & 0.0007  & 0.0009  & 0.0389  & 0.0008  & 0.0014  & 0.0435  & 0.0011  & 0.0010 \\
1.3308  & 4.2817  & 4.4229   & 0.5780  & 0.0128  & 0.0089  & 0.5884  & 0.0130  & 0.0066  & 0.8757  & 0.0147  & 0.0133  & 0.8806  & 0.0150  & 0.0100 \\
1.3452  & 4.3280  & 4.3496   & 0.5583  & 0.0123  & 0.0086  & 0.5660  & 0.0125  & 0.0063  & 0.8315  & 0.0141  & 0.0126  & 0.8329  & 0.0145  & 0.0094 \\
1.3596  & 4.3743  & 4.2763   & 0.5494  & 0.0118  & 0.0084  & 0.5543  & 0.0121  & 0.0062  & 0.7993  & 0.0135  & 0.0121  & 0.7973  & 0.0139  & 0.0090 \\
1.3740  & 4.4206  & 4.2031   & 0.4809  & 0.0111  & 0.0074  & 0.4847  & 0.0113  & 0.0054  & 0.7488  & 0.0130  & 0.0114  & 0.7443  & 0.0134  & 0.0084 \\
1.3883  & 4.4668  & 4.1298   & 0.4764  & 0.0109  & 0.0073  & 0.4777  & 0.0112  & 0.0054  & 0.7165  & 0.0125  & 0.0109  & 0.7094  & 0.0129  & 0.0080 \\
1.4027  & 4.5131  & 4.0565   & 0.4544  & 0.0105  & 0.0070  & 0.4537  & 0.0108  & 0.0051  & 0.6773  & 0.0121  & 0.0103  & 0.6682  & 0.0126  & 0.0076 \\
1.4171  & 4.5594  & 3.9832   & 0.4213  & 0.0100  & 0.0065  & 0.4192  & 0.0104  & 0.0047  & 0.6539  & 0.0118  & 0.0099  & 0.6426  & 0.0123  & 0.0073 \\
1.4315  & 4.6057  & 3.9099   & 0.3955  & 0.0098  & 0.0061  & 0.3922  & 0.0101  & 0.0044  & 0.6223  & 0.0114  & 0.0094  & 0.6096  & 0.0119  & 0.0069 \\
1.4459  & 4.6520  & 3.8366   & 0.3889  & 0.0096  & 0.0060  & 0.3838  & 0.0100  & 0.0043  & 0.5932  & 0.0110  & 0.0090  & 0.5790  & 0.0115  & 0.0066 \\
1.4603  & 4.6983  & 3.7633   & 0.3559  & 0.0090  & 0.0055  & 0.3502  & 0.0094  & 0.0039  & 0.5211  & 0.0102  & 0.0079  & 0.5080  & 0.0107  & 0.0057 \\
1.4747  & 4.7446  & 3.6901   & 0.3590  & 0.0090  & 0.0055  & 0.3513  & 0.0095  & 0.0039  & 0.5106  & 0.0101  & 0.0077  & 0.4957  & 0.0106  & 0.0056 \\
1.4891  & 4.7909  & 3.6168   & 0.3255  & 0.0086  & 0.0050  & 0.3178  & 0.0090  & 0.0036  & 0.5001  & 0.0100  & 0.0076  & 0.4837  & 0.0105  & 0.0055 \\
1.5034  & 4.8372  & 3.5435   & 0.2999  & 0.0082  & 0.0046  & 0.2923  & 0.0086  & 0.0033  & 0.4644  & 0.0095  & 0.0070  & 0.4478  & 0.0101  & 0.0051 \\
1.5178  & 4.8834  & 3.4702   & 0.2983  & 0.0081  & 0.0046  & 0.2896  & 0.0085  & 0.0032  & 0.4466  & 0.0092  & 0.0068  & 0.4290  & 0.0098  & 0.0049 \\
1.5322  & 4.9297  & 3.3969   & 0.2710  & 0.0076  & 0.0042  & 0.2632  & 0.0081  & 0.0030  & 0.4049  & 0.0086  & 0.0061  & 0.3879  & 0.0092  & 0.0044 \\
1.5466  & 4.9760  & 3.3236   & 0.2653  & 0.0074  & 0.0041  & 0.2571  & 0.0079  & 0.0029  & 0.3978  & 0.0084  & 0.0060  & 0.3791  & 0.0090  & 0.0043 \\
1.5474  & 4.9787  & 3.3194   & 0.2636  & 0.0049  & 0.0041  & 0.2553  & 0.0052  & 0.0029  & 0.4027  & 0.0052  & 0.0061  & 0.3833  & 0.0056  & 0.0043 \\
1.5642  & 5.0325  & 3.2342   & 0.2402  & 0.0046  & 0.0037  & 0.2325  & 0.0050  & 0.0026  & 0.3733  & 0.0049  & 0.0057  & 0.3535  & 0.0053  & 0.0040 \\
1.5809  & 5.0863  & 3.1490   & 0.2311  & 0.0045  & 0.0036  & 0.2224  & 0.0048  & 0.0025  & 0.3494  & 0.0047  & 0.0053  & 0.3288  & 0.0051  & 0.0037 \\
1.5976  & 5.1402  & 3.0637   & 0.2419  & 0.0045  & 0.0037  & 0.2288  & 0.0049  & 0.0026  & 0.3297  & 0.0044  & 0.0050  & 0.3081  & 0.0048  & 0.0035 \\
1.6143  & 5.1940  & 2.9785   & 0.2407  & 0.0044  & 0.0037  & 0.2219  & 0.0049  & 0.0025  & 0.3108  & 0.0042  & 0.0047  & 0.2883  & 0.0047  & 0.0033 \\
1.6311  & 5.2478  & 2.8933   & 0.2297  & 0.0042  & 0.0035  & 0.2067  & 0.0047  & 0.0023  & 0.2786  & 0.0039  & 0.0042  & 0.2573  & 0.0044  & 0.0029 \\
1.6478  & 5.3016  & 2.8081   & 0.1955  & 0.0037  & 0.0030  & 0.1742  & 0.0043  & 0.0020  & 0.2588  & 0.0038  & 0.0039  & 0.2379  & 0.0042  & 0.0027 \\
1.6645  & 5.3555  & 2.7229   & 0.1634  & 0.0034  & 0.0025  & 0.1480  & 0.0038  & 0.0017  & 0.2356  & 0.0035  & 0.0036  & 0.2160  & 0.0039  & 0.0024 \\
1.6813  & 5.4093  & 2.6377   & 0.1392  & 0.0031  & 0.0021  & 0.1288  & 0.0035  & 0.0014  & 0.2098  & 0.0033  & 0.0032  & 0.1920  & 0.0037  & 0.0022 \\
1.6980  & 5.4631  & 2.5525   & 0.1200  & 0.0029  & 0.0018  & 0.1124  & 0.0032  & 0.0013  & 0.1855  & 0.0030  & 0.0028  & 0.1696  & 0.0034  & 0.0019 \\
1.7147  & 5.5169  & 2.4672   & 0.1038  & 0.0028  & 0.0016  & 0.0978  & 0.0030  & 0.0011  & 0.1656  & 0.0028  & 0.0025  & 0.1510  & 0.0032  & 0.0017 \\
1.7315  & 5.5707  & 2.3820   & 0.1088  & 0.0029  & 0.0017  & 0.1002  & 0.0032  & 0.0011  & 0.1615  & 0.0028  & 0.0025  & 0.1460  & 0.0032  & 0.0017 \\
1.7482  & 5.6246  & 2.2968   & 0.1343  & 0.0031  & 0.0021  & 0.1165  & 0.0038  & 0.0013  & 0.1506  & 0.0026  & 0.0023  & 0.1350  & 0.0031  & 0.0015 \\
1.7649  & 5.6784  & 2.2116   & 0.1111  & 0.0027  & 0.0017  & 0.0946  & 0.0032  & 0.0011  & 0.1320  & 0.0024  & 0.0020  & 0.1175  & 0.0028  & 0.0013 \\
1.7816  & 5.7322  & 2.1264   & 0.0846  & 0.0022  & 0.0013  & 0.0728  & 0.0026  & 0.0008  & 0.1194  & 0.0022  & 0.0018  & 0.1066  & 0.0026  & 0.0012 \\
1.7984  & 5.7860  & 2.0412   & 0.0708  & 0.0020  & 0.0011  & 0.0618  & 0.0024  & 0.0007  & 0.0999  & 0.0020  & 0.0015  & 0.0896  & 0.0024  & 0.0010 \\
	    \hline
    	\hline
  	\end{tabular}
	\end{table*}

\begin{table*}
  \label{cross_sections}	
  \caption{Differential cross sections extracted from the measurements of E00-116. The normalization uncertainty is 1.75\%.}
  \begin{tabular}{ccccccccccccccc}
    \hline
    \hline
    $E^{'}$ & $Q^{2}$ & $W^{2}$ & $\frac{d \sigma^{Born}}{dE^{'}d \Omega}$ (H) & stat & syst & $\frac{d\sigma^{rad}}{dE^{'}d \Omega}$ (H) & stat & syst &
    $\frac{d \sigma^{Born}}{dE^{'}d \Omega}$ (D) & stat & syst & $\frac{d \sigma^{rad}}{dE^{'}d \Omega}$ (D) & stat & syst \\
    \hline
1.7994  & 5.7894  & 2.0359   & 0.0658  & 0.0010  & 0.0010  & 0.0576  & 0.0012  & 0.0007  & 0.1043  & 0.0024  & 0.0016  & 0.0932  & 0.0028  & 0.0011 \\
1.8189  & 5.8520  & 1.9368   & 0.0568  & 0.0009  & 0.0009  & 0.0498  & 0.0010  & 0.0006  & 0.0929  & 0.0023  & 0.0014  & 0.0830  & 0.0027  & 0.0010 \\
1.8383  & 5.9145  & 1.8377   & 0.0456  & 0.0008  & 0.0007  & 0.0403  & 0.0009  & 0.0005  & 0.0752  & 0.0019  & 0.0012  & 0.0679  & 0.0023  & 0.0008 \\
1.8578  & 5.9771  & 1.7386   & 0.0346  & 0.0007  & 0.0006  & 0.0312  & 0.0008  & 0.0004  & 0.0594  & 0.0017  & 0.0010  & 0.0546  & 0.0021  & 0.0006 \\
1.8772  & 6.0397  & 1.6395   & 0.0307  & 0.0007  & 0.0006  & 0.0277  & 0.0008  & 0.0004  & 0.0524  & 0.0016  & 0.0009  & 0.0491  & 0.0019  & 0.0006 \\
1.8967  & 6.1023  & 1.5404   & 0.0294  & 0.0007  & 0.0006  & 0.0258  & 0.0008  & 0.0005  & 0.0424  & 0.0014  & 0.0008  & 0.0411  & 0.0018  & 0.0005 \\
1.9161  & 6.1649  & 1.4413   & 0.0238  & 0.0007  & 0.0007  & 0.0200  & 0.0010  & 0.0005  & 0.0404  & 0.0014  & 0.0008  & 0.0398  & 0.0018  & 0.0005 \\
1.9356  & 6.2275  & 1.3422   & 0.0118  & 0.0005  & 0.0005  & 0.0112  & 0.0008  & 0.0005  & 0.0351  & 0.0014  & 0.0009  & 0.0353  & 0.0018  & 0.0005 \\
1.9550  & 6.2901  & 1.2431   & 0.0083  & 0.0005  & 0.0006  & 0.0115  & 0.0007  & 0.0008  & 0.0258  & 0.0010  & 0.0008  & 0.0278  & 0.0015  & 0.0005 \\
1.0699  & 5.0150  & 4.1791   & 0.2387  & 0.0054  & 0.0037  & 0.2444  & 0.0055  & 0.0027  & 0.3316  & 0.0049  & 0.0050  & 0.3379  & 0.0050  & 0.0038  \\
1.0815  & 5.0692  & 4.1032   & 0.2160  & 0.0051  & 0.0033  & 0.2208  & 0.0052  & 0.0025  & 0.3061  & 0.0047  & 0.0046  & 0.3110  & 0.0048  & 0.0035  \\
1.0931  & 5.1234  & 4.0273   & 0.2025  & 0.0050  & 0.0031  & 0.2064  & 0.0051  & 0.0023  & 0.2916  & 0.0045  & 0.0044  & 0.2951  & 0.0046  & 0.0033  \\
1.1046  & 5.1776  & 3.9513   & 0.1942  & 0.0047  & 0.0030  & 0.1969  & 0.0048  & 0.0022  & 0.2775  & 0.0043  & 0.0042  & 0.2797  & 0.0044  & 0.0032  \\
1.1162  & 5.2319  & 3.8754   & 0.1818  & 0.0046  & 0.0028  & 0.1838  & 0.0047  & 0.0021  & 0.2624  & 0.0042  & 0.0040  & 0.2635  & 0.0043  & 0.0030  \\
1.1278  & 5.2861  & 3.7995   & 0.1659  & 0.0043  & 0.0025  & 0.1672  & 0.0044  & 0.0019  & 0.2342  & 0.0039  & 0.0036  & 0.2346  & 0.0040  & 0.0027  \\
1.1393  & 5.3403  & 3.7236   & 0.1609  & 0.0042  & 0.0025  & 0.1613  & 0.0043  & 0.0018  & 0.2236  & 0.0038  & 0.0034  & 0.2233  & 0.0040  & 0.0025  \\
1.1509  & 5.3945  & 3.6477   & 0.1551  & 0.0040  & 0.0024  & 0.1548  & 0.0041  & 0.0017  & 0.2171  & 0.0037  & 0.0033  & 0.2154  & 0.0038  & 0.0024  \\
1.1625  & 5.4487  & 3.5717   & 0.1325  & 0.0038  & 0.0020  & 0.1322  & 0.0039  & 0.0015  & 0.2020  & 0.0035  & 0.0031  & 0.1998  & 0.0037  & 0.0023  \\
1.1741  & 5.5029  & 3.4958   & 0.1260  & 0.0036  & 0.0019  & 0.1254  & 0.0038  & 0.0014  & 0.1900  & 0.0034  & 0.0029  & 0.1871  & 0.0036  & 0.0021  \\
1.1856  & 5.5572  & 3.4199   & 0.1209  & 0.0035  & 0.0019  & 0.1201  & 0.0037  & 0.0013  & 0.1756  & 0.0032  & 0.0027  & 0.1722  & 0.0034  & 0.0020  \\
1.1972  & 5.6114  & 3.3440   & 0.1139  & 0.0035  & 0.0017  & 0.1129  & 0.0036  & 0.0013  & 0.1661  & 0.0031  & 0.0025  & 0.1623  & 0.0033  & 0.0018  \\
1.2088  & 5.6656  & 3.2680   & 0.1060  & 0.0033  & 0.0016  & 0.1050  & 0.0034  & 0.0012  & 0.1559  & 0.0030  & 0.0024  & 0.1512  & 0.0032  & 0.0017  \\
1.2203  & 5.7198  & 3.1921   & 0.1036  & 0.0032  & 0.0016  & 0.1022  & 0.0034  & 0.0011  & 0.1459  & 0.0029  & 0.0022  & 0.1411  & 0.0031  & 0.0016  \\
1.2319  & 5.7740  & 3.1162   & 0.0962  & 0.0030  & 0.0015  & 0.0943  & 0.0032  & 0.0011  & 0.1421  & 0.0028  & 0.0022  & 0.1362  & 0.0030  & 0.0015  \\
1.2435  & 5.8282  & 3.0403   & 0.0987  & 0.0029  & 0.0015  & 0.0950  & 0.0032  & 0.0011  & 0.1313  & 0.0026  & 0.0020  & 0.1252  & 0.0028  & 0.0014  \\
1.2442  & 5.8318  & 3.0353   & 0.0994  & 0.0014  & 0.0015  & 0.0955  & 0.0015  & 0.0011  & 0.1373  & 0.0015  & 0.0021  & 0.1305  & 0.0016  & 0.0015  \\
1.2577  & 5.8949  & 2.9470   & 0.0972  & 0.0014  & 0.0015  & 0.0908  & 0.0015  & 0.0010  & 0.1232  & 0.0013  & 0.0019  & 0.1165  & 0.0015  & 0.0013  \\
1.2711  & 5.9579  & 2.8587   & 0.0907  & 0.0013  & 0.0014  & 0.0826  & 0.0015  & 0.0009  & 0.1116  & 0.0013  & 0.0017  & 0.1050  & 0.0014  & 0.0012  \\
1.2846  & 6.0210  & 2.7704   & 0.0765  & 0.0011  & 0.0012  & 0.0697  & 0.0013  & 0.0008  & 0.1037  & 0.0012  & 0.0016  & 0.0971  & 0.0013  & 0.0011  \\
1.2980  & 6.0840  & 2.6821   & 0.0598  & 0.0010  & 0.0009  & 0.0561  & 0.0011  & 0.0006  & 0.0918  & 0.0011  & 0.0014  & 0.0858  & 0.0012  & 0.0010  \\
1.3115  & 6.1471  & 2.5938   & 0.0488  & 0.0009  & 0.0008  & 0.0468  & 0.0010  & 0.0005  & 0.0816  & 0.0010  & 0.0012  & 0.0761  & 0.0012  & 0.0009  \\
1.3249  & 6.2101  & 2.5055   & 0.0456  & 0.0009  & 0.0007  & 0.0440  & 0.0010  & 0.0005  & 0.0737  & 0.0010  & 0.0011  & 0.0685  & 0.0011  & 0.0008 \\
1.3384  & 6.2732  & 2.4172   & 0.0435  & 0.0009  & 0.0007  & 0.0417  & 0.0010  & 0.0005  & 0.0632  & 0.0009  & 0.0010  & 0.0587  & 0.0010  & 0.0007 \\
1.3518  & 6.3362  & 2.3290   & 0.0472  & 0.0009  & 0.0007  & 0.0428  & 0.0011  & 0.0005  & 0.0587  & 0.0009  & 0.0009  & 0.0541  & 0.0010  & 0.0006 \\
1.3653  & 6.3992  & 2.2407   & 0.0458  & 0.0009  & 0.0007  & 0.0399  & 0.0011  & 0.0004  & 0.0547  & 0.0008  & 0.0008  & 0.0499  & 0.0009  & 0.0006 \\
1.3787  & 6.4623  & 2.1524   & 0.0355  & 0.0008  & 0.0005  & 0.0308  & 0.0009  & 0.0003  & 0.0481  & 0.0008  & 0.0007  & 0.0437  & 0.0009  & 0.0005 \\
1.3922  & 6.5253  & 2.0641   & 0.0281  & 0.0006  & 0.0004  & 0.0249  & 0.0008  & 0.0003  & 0.0418  & 0.0007  & 0.0006  & 0.0384  & 0.0008  & 0.0004 \\
1.4056  & 6.5884  & 1.9758   & 0.0231  & 0.0006  & 0.0004  & 0.0207  & 0.0007  & 0.0002  & 0.0381  & 0.0007  & 0.0006  & 0.0348  & 0.0007  & 0.0004 \\
1.4191  & 6.6514  & 1.8875   & 0.0195  & 0.0005  & 0.0003  & 0.0174  & 0.0006  & 0.0002  & 0.0314  & 0.0006  & 0.0005  & 0.0290  & 0.0007  & 0.0003 \\
1.4325  & 6.7145  & 1.7992   & 0.0147  & 0.0005  & 0.0002  & 0.0133  & 0.0006  & 0.0002  & 0.0272  & 0.0006  & 0.0004  & 0.0253  & 0.0007  & 0.0003 \\
1.4460  & 6.7775  & 1.7109   & 0.0137  & 0.0005  & 0.0002  & 0.0124  & 0.0006  & 0.0001  & 0.0233  & 0.0005  & 0.0004  & 0.0220  & 0.0006  & 0.0003 \\
1.3622  & 6.3850  & 2.2605   & 0.0458  & 0.0010  & 0.0007  & 0.0399  & 0.0012  & 0.0005  & 0.0592  & 0.0009  & 0.0009  & 0.0540  & 0.0010  & 0.0006 \\
1.3770  & 6.4541  & 2.1639   & 0.0368  & 0.0008  & 0.0006  & 0.0316  & 0.0010  & 0.0004  & 0.0517  & 0.0008  & 0.0008  & 0.0469  & 0.0009  & 0.0005 \\
1.3917  & 6.5231  & 2.0672   & 0.0282  & 0.0007  & 0.0004  & 0.0249  & 0.0008  & 0.0003  & 0.0429  & 0.0007  & 0.0007  & 0.0392  & 0.0009  & 0.0005 \\
1.4064  & 6.5921  & 1.9706   & 0.0235  & 0.0007  & 0.0004  & 0.0209  & 0.0008  & 0.0003  & 0.0380  & 0.0007  & 0.0006  & 0.0348  & 0.0008  & 0.0004 \\
1.4212  & 6.6612  & 1.8739   & 0.0187  & 0.0006  & 0.0003  & 0.0165  & 0.0007  & 0.0002  & 0.0326  & 0.0006  & 0.0005  & 0.0300  & 0.0007  & 0.0004 \\
1.4359  & 6.7302  & 1.7772   & 0.0139  & 0.0005  & 0.0002  & 0.0125  & 0.0006  & 0.0002  & 0.0271  & 0.0006  & 0.0004  & 0.0252  & 0.0007  & 0.0003 \\
1.4506  & 6.7992  & 1.6806   & 0.0119  & 0.0005  & 0.0002  & 0.0109  & 0.0006  & 0.0001  & 0.0234  & 0.0005  & 0.0004  & 0.0222  & 0.0006  & 0.0003 \\
1.4653  & 6.8682  & 1.5839   & 0.0110  & 0.0005  & 0.0002  & 0.0099  & 0.0006  & 0.0002  & 0.0199  & 0.0005  & 0.0003  & 0.0192  & 0.0006  & 0.0002 \\
1.4801  & 6.9373  & 1.4872   & 0.0094  & 0.0005  & 0.0002  & 0.0082  & 0.0006  & 0.0002  & 0.0169  & 0.0005  & 0.0003  & 0.0167  & 0.0006  & 0.0002 \\
1.4948  & 7.0063  & 1.3906   & 0.0070  & 0.0003  & 0.0002  & 0.0059  & 0.0005  & 0.0002  & 0.0168  & 0.0005  & 0.0004  & 0.0165  & 0.0006  & 0.0002 \\
1.5095  & 7.0753  & 1.2939   & 0.0034  & 0.0002  & 0.0002  & 0.0034  & 0.0003  & 0.0002  & 0.0137  & 0.0004  & 0.0004  & 0.0137  & 0.0006  & 0.0002 \\
0.8234  & 4.5252  & 5.1315   & 0.3212  & 0.0115  & 0.0049  & 0.3569  & 0.0111  & 0.0040  & 0.5071  & 0.0130  & 0.0077  & 0.5602  & 0.0125  & 0.0064 \\
0.8323  & 4.5741  & 5.0659   & 0.3289  & 0.0115  & 0.0051  & 0.3623  & 0.0111  & 0.0041  & 0.4487  & 0.0122  & 0.0068  & 0.4966  & 0.0118  & 0.0056 \\
0.8412  & 4.6230  & 5.0003   & 0.3035  & 0.0107  & 0.0047  & 0.3335  & 0.0104  & 0.0037  & 0.4580  & 0.0120  & 0.0070  & 0.5030  & 0.0116  & 0.0057 \\
0.8501  & 4.6719  & 4.9347   & 0.2749  & 0.0100  & 0.0042  & 0.3017  & 0.0097  & 0.0034  & 0.4263  & 0.0114  & 0.0065  & 0.4668  & 0.0111  & 0.0053 \\
0.8590  & 4.7209  & 4.8690   & 0.2692  & 0.0098  & 0.0041  & 0.2934  & 0.0096  & 0.0033  & 0.3912  & 0.0108  & 0.0060  & 0.4270  & 0.0105  & 0.0049 \\
0.8679  & 4.7698  & 4.8034   & 0.2640  & 0.0097  & 0.0041  & 0.2859  & 0.0095  & 0.0032  & 0.3956  & 0.0106  & 0.0060  & 0.4278  & 0.0104  & 0.0049 \\
0.8768  & 4.8187  & 4.7378   & 0.2396  & 0.0092  & 0.0037  & 0.2589  & 0.0091  & 0.0029  & 0.3781  & 0.0105  & 0.0058  & 0.4065  & 0.0103  & 0.0046 \\
	    \hline
    	\hline
  	\end{tabular}
	\end{table*}

\begin{table*}
  \label{cross_sections}	
  \caption{Differential cross sections extracted from the measurements of E00-116. The normalization uncertainty is 1.75\%.}
  \begin{tabular}{ccccccccccccccc}
    \hline
    \hline
    $E^{'}$ & $Q^{2}$ & $W^{2}$ & $\frac{d \sigma^{Born}}{dE^{'}d \Omega}$ (H) & stat & syst & $\frac{d\sigma^{rad}}{dE^{'}d \Omega}$ (H) & stat & syst &
    $\frac{d \sigma^{Born}}{dE^{'}d \Omega}$ (D) & stat & syst & $\frac{d \sigma^{rad}}{dE^{'}d \Omega}$ (D) & stat & syst \\
    \hline
0.8857  & 4.8676  & 4.6722   & 0.2333  & 0.0090  & 0.0036  & 0.2507  & 0.0089  & 0.0028  & 0.3579  & 0.0099  & 0.0054  & 0.3828  & 0.0098  & 0.0044 \\
0.8947  & 4.9165  & 4.6065   & 0.2130  & 0.0085  & 0.0033  & 0.2283  & 0.0085  & 0.0026  & 0.3515  & 0.0096  & 0.0053  & 0.3736  & 0.0095  & 0.0043 \\
0.9036  & 4.9655  & 4.5409   & 0.2124  & 0.0084  & 0.0033  & 0.2261  & 0.0084  & 0.0025  & 0.3200  & 0.0093  & 0.0049  & 0.3392  & 0.0093  & 0.0039 \\
0.9125  & 5.0144  & 4.4753   & 0.1982  & 0.0081  & 0.0030  & 0.2104  & 0.0081  & 0.0024  & 0.2887  & 0.0087  & 0.0044  & 0.3055  & 0.0086  & 0.0035 \\
0.9214  & 5.0633  & 4.4096   & 0.1970  & 0.0079  & 0.0030  & 0.2078  & 0.0079  & 0.0023  & 0.2940  & 0.0087  & 0.0045  & 0.3087  & 0.0087  & 0.0035 \\
0.9303  & 5.1122  & 4.3440   & 0.1848  & 0.0076  & 0.0028  & 0.1944  & 0.0076  & 0.0022  & 0.2807  & 0.0085  & 0.0043  & 0.2936  & 0.0085  & 0.0033 \\
0.9392  & 5.1611  & 4.2784   & 0.1804  & 0.0076  & 0.0028  & 0.1890  & 0.0076  & 0.0021  & 0.2750  & 0.0083  & 0.0042  & 0.2861  & 0.0084  & 0.0033 \\
0.9481  & 5.2101  & 4.2128   & 0.1625  & 0.0070  & 0.0025  & 0.1700  & 0.0071  & 0.0019  & 0.2625  & 0.0081  & 0.0040  & 0.2720  & 0.0081  & 0.0031 \\
0.9570  & 5.2590  & 4.1471   & 0.1675  & 0.0069  & 0.0026  & 0.1740  & 0.0070  & 0.0020  & 0.2532  & 0.0078  & 0.0039  & 0.2614  & 0.0079  & 0.0030 \\
0.9576  & 5.2623  & 4.1428   & 0.1640  & 0.0047  & 0.0025  & 0.1702  & 0.0048  & 0.0019  & 0.2501  & 0.0046  & 0.0038  & 0.2578  & 0.0046  & 0.0029 \\
0.9679  & 5.3191  & 4.0664   & 0.1534  & 0.0044  & 0.0024  & 0.1587  & 0.0045  & 0.0018  & 0.2313  & 0.0043  & 0.0035  & 0.2379  & 0.0044  & 0.0027 \\
0.9783  & 5.3760  & 3.9901   & 0.1493  & 0.0043  & 0.0023  & 0.1537  & 0.0044  & 0.0017  & 0.2202  & 0.0041  & 0.0033  & 0.2254  & 0.0042  & 0.0026 \\
0.9886  & 5.4329  & 3.9138   & 0.1333  & 0.0040  & 0.0020  & 0.1372  & 0.0041  & 0.0015  & 0.2029  & 0.0039  & 0.0031  & 0.2070  & 0.0040  & 0.0024 \\
0.9990  & 5.4898  & 3.8375   & 0.1183  & 0.0037  & 0.0018  & 0.1216  & 0.0038  & 0.0014  & 0.1894  & 0.0037  & 0.0029  & 0.1926  & 0.0038  & 0.0022 \\
1.0093  & 5.5467  & 3.7612   & 0.1170  & 0.0037  & 0.0018  & 0.1194  & 0.0038  & 0.0013  & 0.1758  & 0.0035  & 0.0027  & 0.1781  & 0.0037  & 0.0020 \\
1.0197  & 5.6036  & 3.6849   & 0.1178  & 0.0036  & 0.0018  & 0.1195  & 0.0038  & 0.0013  & 0.1682  & 0.0034  & 0.0026  & 0.1695  & 0.0036  & 0.0019 \\
1.0300  & 5.6605  & 3.6086   & 0.1065  & 0.0035  & 0.0016  & 0.1077  & 0.0036  & 0.0012  & 0.1492  & 0.0032  & 0.0023  & 0.1500  & 0.0034  & 0.0017 \\
1.0404  & 5.7174  & 3.5322   & 0.0923  & 0.0032  & 0.0014  & 0.0934  & 0.0034  & 0.0010  & 0.1374  & 0.0031  & 0.0021  & 0.1378  & 0.0032  & 0.0016 \\
1.0507  & 5.7743  & 3.4559   & 0.0928  & 0.0033  & 0.0014  & 0.0933  & 0.0034  & 0.0010  & 0.1372  & 0.0031  & 0.0021  & 0.1365  & 0.0032  & 0.0016 \\
1.0611  & 5.8311  & 3.3796   & 0.0829  & 0.0030  & 0.0013  & 0.0835  & 0.0031  & 0.0009  & 0.1264  & 0.0029  & 0.0019  & 0.1251  & 0.0031  & 0.0014 \\
1.0714  & 5.8880  & 3.3033   & 0.0799  & 0.0030  & 0.0012  & 0.0802  & 0.0031  & 0.0009  & 0.1172  & 0.0028  & 0.0018  & 0.1156  & 0.0030  & 0.0013 \\
1.0818  & 5.9449  & 3.2270   & 0.0773  & 0.0029  & 0.0012  & 0.0774  & 0.0030  & 0.0009  & 0.1145  & 0.0027  & 0.0017  & 0.1120  & 0.0029  & 0.0013 \\
1.0921  & 6.0018  & 3.1507   & 0.0711  & 0.0027  & 0.0011  & 0.0709  & 0.0028  & 0.0008  & 0.1101  & 0.0027  & 0.0017  & 0.1071  & 0.0029  & 0.0012 \\
1.1025  & 6.0587  & 3.0743   & 0.0732  & 0.0028  & 0.0011  & 0.0720  & 0.0029  & 0.0008  & 0.0973  & 0.0025  & 0.0015  & 0.0942  & 0.0027  & 0.0011 \\
1.1128  & 6.1156  & 2.9980   & 0.0696  & 0.0026  & 0.0011  & 0.0670  & 0.0028  & 0.0008  & 0.0926  & 0.0024  & 0.0014  & 0.0890  & 0.0026  & 0.0010 \\
1.1043  & 6.0685  & 3.0612   & 0.0684  & 0.0017  & 0.0011  & 0.0672  & 0.0018  & 0.0008  & 0.0996  & 0.0012  & 0.0015  & 0.0961  & 0.0013  & 0.0011 \\
1.1162  & 6.1341  & 2.9732   & 0.0723  & 0.0017  & 0.0011  & 0.0690  & 0.0019  & 0.0008  & 0.0941  & 0.0012  & 0.0014  & 0.0901  & 0.0013  & 0.0010 \\
1.1281  & 6.1997  & 2.8852   & 0.0678  & 0.0016  & 0.0010  & 0.0629  & 0.0017  & 0.0007  & 0.0835  & 0.0011  & 0.0013  & 0.0796  & 0.0011  & 0.0009 \\
1.1401  & 6.2653  & 2.7972   & 0.0577  & 0.0014  & 0.0009  & 0.0529  & 0.0016  & 0.0006  & 0.0759  & 0.0010  & 0.0012  & 0.0720  & 0.0011  & 0.0008 \\
1.1520  & 6.3309  & 2.7092   & 0.0468  & 0.0012  & 0.0007  & 0.0440  & 0.0014  & 0.0005  & 0.0695  & 0.0009  & 0.0011  & 0.0657  & 0.0010  & 0.0008 \\
1.1640  & 6.3965  & 2.6212   & 0.0377  & 0.0011  & 0.0006  & 0.0364  & 0.0012  & 0.0004  & 0.0622  & 0.0009  & 0.0010  & 0.0586  & 0.0010  & 0.0007 \\
1.1759  & 6.4621  & 2.5332   & 0.0328  & 0.0011  & 0.0005  & 0.0321  & 0.0012  & 0.0004  & 0.0555  & 0.0008  & 0.0008  & 0.0521  & 0.0009  & 0.0006 \\
1.1878  & 6.5277  & 2.4452   & 0.0325  & 0.0011  & 0.0005  & 0.0316  & 0.0012  & 0.0004  & 0.0503  & 0.0008  & 0.0008  & 0.0471  & 0.0009  & 0.0005 \\
1.1998  & 6.5933  & 2.3572   & 0.0324  & 0.0011  & 0.0005  & 0.0303  & 0.0012  & 0.0003  & 0.0452  & 0.0008  & 0.0007  & 0.0422  & 0.0008  & 0.0005 \\
1.2117  & 6.6589  & 2.2692   & 0.0342  & 0.0011  & 0.0005  & 0.0301  & 0.0013  & 0.0003  & 0.0422  & 0.0007  & 0.0007  & 0.0390  & 0.0008  & 0.0005 \\
1.2236  & 6.7245  & 2.1812   & 0.0272  & 0.0009  & 0.0004  & 0.0236  & 0.0011  & 0.0003  & 0.0367  & 0.0007  & 0.0006  & 0.0336  & 0.0008  & 0.0004 \\
1.2356  & 6.7901  & 2.0932   & 0.0190  & 0.0008  & 0.0003  & 0.0170  & 0.0009  & 0.0002  & 0.0309  & 0.0006  & 0.0005  & 0.0286  & 0.0007  & 0.0003 \\
1.2475  & 6.8557  & 2.0052   & 0.0179  & 0.0007  & 0.0003  & 0.0161  & 0.0009  & 0.0002  & 0.0273  & 0.0006  & 0.0004  & 0.0253  & 0.0007  & 0.0003 \\
1.2595  & 6.9213  & 1.9172   & 0.0140  & 0.0007  & 0.0002  & 0.0127  & 0.0008  & 0.0001  & 0.0242  & 0.0005  & 0.0004  & 0.0225  & 0.0006  & 0.0003 \\
1.2714  & 6.9869  & 1.8291   & 0.0125  & 0.0006  & 0.0002  & 0.0113  & 0.0007  & 0.0001  & 0.0213  & 0.0005  & 0.0003  & 0.0199  & 0.0006  & 0.0002 \\
1.2833  & 7.0525  & 1.7411   & 0.0115  & 0.0006  & 0.0002  & 0.0105  & 0.0007  & 0.0001  & 0.0186  & 0.0005  & 0.0003  & 0.0175  & 0.0005  & 0.0002 \\
1.2128  & 6.6637  & 2.2624   & 0.0368  & 0.0006  & 0.0006  & 0.0324  & 0.0007  & 0.0004  & 0.0420  & 0.0005  & 0.0007  & 0.0387  & 0.0006  & 0.0005 \\
1.2259  & 6.7358  & 2.1657   & 0.0284  & 0.0005  & 0.0004  & 0.0245  & 0.0006  & 0.0003  & 0.0366  & 0.0005  & 0.0006  & 0.0336  & 0.0005  & 0.0004 \\
1.2390  & 6.8078  & 2.0691   & 0.0198  & 0.0004  & 0.0003  & 0.0178  & 0.0005  & 0.0002  & 0.0300  & 0.0004  & 0.0005  & 0.0278  & 0.0005  & 0.0003 \\
1.2521  & 6.8799  & 1.9724   & 0.0165  & 0.0004  & 0.0003  & 0.0149  & 0.0005  & 0.0002  & 0.0259  & 0.0004  & 0.0004  & 0.0240  & 0.0005  & 0.0003 \\
1.2652  & 6.9519  & 1.8758   & 0.0128  & 0.0003  & 0.0002  & 0.0116  & 0.0004  & 0.0001  & 0.0221  & 0.0004  & 0.0004  & 0.0206  & 0.0004  & 0.0003 \\
1.2783  & 7.0239  & 1.7792   & 0.0106  & 0.0003  & 0.0002  & 0.0097  & 0.0004  & 0.0001  & 0.0191  & 0.0003  & 0.0003  & 0.0180  & 0.0004  & 0.0002 \\
1.2914  & 7.0960  & 1.6825   & 0.0083  & 0.0003  & 0.0001  & 0.0079  & 0.0003  & 0.0001  & 0.0158  & 0.0003  & 0.0003  & 0.0151  & 0.0004  & 0.0002 \\
1.3045  & 7.1680  & 1.5859   & 0.0079  & 0.0003  & 0.0002  & 0.0073  & 0.0003  & 0.0001  & 0.0136  & 0.0003  & 0.0002  & 0.0133  & 0.0003  & 0.0002 \\
1.3177  & 7.2401  & 1.4892   & 0.0062  & 0.0003  & 0.0002  & 0.0057  & 0.0004  & 0.0001  & 0.0105  & 0.0002  & 0.0002  & 0.0107  & 0.0003  & 0.0001 \\
1.3308  & 7.3121  & 1.3926   & 0.0048  & 0.0002  & 0.0002  & 0.0042  & 0.0003  & 0.0001  & 0.0125  & 0.0003  & 0.0003  & 0.0122  & 0.0004  & 0.0002 \\
1.3439  & 7.3841  & 1.2959   & 0.0024  & 0.0002  & 0.0001  & 0.0027  & 0.0003  & 0.0001  & 0.0118  & 0.0003  & 0.0003  & 0.0113  & 0.0004  & 0.0002 \\
0.7445  & 5.3831  & 4.4216   & 0.1287  & 0.0037  & 0.0020  & 0.1402  & 0.0036  & 0.0016  & 0.1842  & 0.0052  & 0.0028  & 0.2018  & 0.0051  & 0.0023 \\
0.7526  & 5.4413  & 4.3483   & 0.1127  & 0.0034  & 0.0017  & 0.1232  & 0.0034  & 0.0014  & 0.1691  & 0.0048  & 0.0026  & 0.1849  & 0.0047  & 0.0021 \\
0.7606  & 5.4995  & 4.2750   & 0.1085  & 0.0033  & 0.0017  & 0.1179  & 0.0033  & 0.0013  & 0.1688  & 0.0047  & 0.0026  & 0.1830  & 0.0047  & 0.0021 \\
0.7687  & 5.5577  & 4.2017   & 0.1074  & 0.0032  & 0.0017  & 0.1158  & 0.0032  & 0.0013  & 0.1534  & 0.0044  & 0.0023  & 0.1658  & 0.0043  & 0.0019 \\
0.7767  & 5.6159  & 4.1284   & 0.0952  & 0.0030  & 0.0015  & 0.1027  & 0.0030  & 0.0012  & 0.1517  & 0.0043  & 0.0023  & 0.1627  & 0.0043  & 0.0019 \\
0.7848  & 5.6741  & 4.0551   & 0.0951  & 0.0029  & 0.0015  & 0.1017  & 0.0029  & 0.0011  & 0.1396  & 0.0040  & 0.0021  & 0.1492  & 0.0041  & 0.0017 \\
0.7928  & 5.7323  & 3.9818   & 0.0843  & 0.0028  & 0.0013  & 0.0902  & 0.0028  & 0.0010  & 0.1270  & 0.0039  & 0.0019  & 0.1353  & 0.0039  & 0.0016 \\
	    \hline
    	\hline
  	\end{tabular}
	\end{table*}

\begin{table*}
  \label{cross_sections}	
  \caption{Differential cross sections extracted from the measurements of E00-116. The normalization uncertainty is 1.75\%.}
  \begin{tabular}{ccccccccccccccc}
    \hline
    \hline
    $E^{'}$ & $Q^{2}$ & $W^{2}$ & $\frac{d \sigma^{Born}}{dE^{'}d \Omega}$ (H) & stat & syst & $\frac{d\sigma^{rad}}{dE^{'}d \Omega}$ (H) & stat & syst &
    $\frac{d \sigma^{Born}}{dE^{'}d \Omega}$ (D) & stat & syst & $\frac{d \sigma^{rad}}{dE^{'}d \Omega}$ (D) & stat & syst \\
    \hline
0.8009  & 5.7905  & 3.9085   & 0.0817  & 0.0027  & 0.0013  & 0.0869  & 0.0027  & 0.0010  & 0.1181  & 0.0037  & 0.0018  & 0.1253  & 0.0037  & 0.0014 \\
0.8089  & 5.8487  & 3.8352   & 0.0797  & 0.0027  & 0.0012  & 0.0842  & 0.0027  & 0.0009  & 0.1093  & 0.0035  & 0.0017  & 0.1156  & 0.0036  & 0.0013 \\
0.8170  & 5.9069  & 3.7619   & 0.0736  & 0.0025  & 0.0011  & 0.0776  & 0.0025  & 0.0009  & 0.1127  & 0.0035  & 0.0017  & 0.1179  & 0.0036  & 0.0014 \\
0.8250  & 5.9651  & 3.6886   & 0.0686  & 0.0025  & 0.0011  & 0.0720  & 0.0025  & 0.0008  & 0.0994  & 0.0033  & 0.0015  & 0.1039  & 0.0033  & 0.0012 \\
0.8331  & 6.0233  & 3.6153   & 0.0700  & 0.0024  & 0.0011  & 0.0729  & 0.0025  & 0.0008  & 0.0985  & 0.0032  & 0.0015  & 0.1022  & 0.0033  & 0.0012 \\
0.8411  & 6.0815  & 3.5420   & 0.0563  & 0.0022  & 0.0009  & 0.0590  & 0.0023  & 0.0007  & 0.0896  & 0.0031  & 0.0014  & 0.0927  & 0.0032  & 0.0011 \\
0.8492  & 6.1397  & 3.4687   & 0.0546  & 0.0021  & 0.0008  & 0.0570  & 0.0022  & 0.0006  & 0.0869  & 0.0029  & 0.0013  & 0.0893  & 0.0030  & 0.0010 \\
0.8572  & 6.1979  & 3.3954   & 0.0537  & 0.0021  & 0.0008  & 0.0558  & 0.0022  & 0.0006  & 0.0879  & 0.0029  & 0.0013  & 0.0895  & 0.0030  & 0.0010 \\
0.8653  & 6.2561  & 3.3221   & 0.0484  & 0.0020  & 0.0007  & 0.0503  & 0.0021  & 0.0006  & 0.0784  & 0.0027  & 0.0012  & 0.0796  & 0.0028  & 0.0009 \\
0.8461  & 6.1175  & 3.4967   & 0.0552  & 0.0013  & 0.0008  & 0.0576  & 0.0014  & 0.0006  & 0.0844  & 0.0021  & 0.0013  & 0.0871  & 0.0022  & 0.0010 \\
0.8552  & 6.1836  & 3.4134   & 0.0512  & 0.0013  & 0.0008  & 0.0535  & 0.0013  & 0.0006  & 0.0806  & 0.0020  & 0.0012  & 0.0826  & 0.0021  & 0.0009 \\
0.8644  & 6.2498  & 3.3301   & 0.0478  & 0.0012  & 0.0007  & 0.0497  & 0.0013  & 0.0006  & 0.0765  & 0.0019  & 0.0012  & 0.0779  & 0.0020  & 0.0009 \\
0.8735  & 6.3159  & 3.2468   & 0.0453  & 0.0012  & 0.0007  & 0.0470  & 0.0012  & 0.0005  & 0.0715  & 0.0018  & 0.0011  & 0.0722  & 0.0019  & 0.0008 \\
0.8827  & 6.3820  & 3.1635   & 0.0414  & 0.0011  & 0.0006  & 0.0427  & 0.0011  & 0.0005  & 0.0620  & 0.0016  & 0.0009  & 0.0624  & 0.0017  & 0.0007 \\
0.8918  & 6.4482  & 3.0802   & 0.0392  & 0.0010  & 0.0006  & 0.0399  & 0.0011  & 0.0004  & 0.0594  & 0.0016  & 0.0009  & 0.0593  & 0.0017  & 0.0007 \\
0.9010  & 6.5143  & 2.9969   & 0.0377  & 0.0010  & 0.0006  & 0.0375  & 0.0011  & 0.0004  & 0.0526  & 0.0015  & 0.0008  & 0.0522  & 0.0016  & 0.0006 \\
0.9101  & 6.5804  & 2.9136   & 0.0377  & 0.0010  & 0.0006  & 0.0363  & 0.0011  & 0.0004  & 0.0476  & 0.0014  & 0.0007  & 0.0470  & 0.0015  & 0.0005 \\
0.9193  & 6.6466  & 2.8303   & 0.0360  & 0.0009  & 0.0006  & 0.0338  & 0.0010  & 0.0004  & 0.0440  & 0.0013  & 0.0007  & 0.0431  & 0.0015  & 0.0005 \\
0.9284  & 6.7127  & 2.7470   & 0.0276  & 0.0008  & 0.0004  & 0.0265  & 0.0009  & 0.0003  & 0.0403  & 0.0013  & 0.0006  & 0.0394  & 0.0014  & 0.0005 \\
0.9376  & 6.7788  & 2.6637   & 0.0230  & 0.0008  & 0.0004  & 0.0226  & 0.0008  & 0.0003  & 0.0374  & 0.0012  & 0.0006  & 0.0363  & 0.0013  & 0.0004 \\
0.9467  & 6.8450  & 2.5804   & 0.0170  & 0.0007  & 0.0003  & 0.0174  & 0.0007  & 0.0002  & 0.0347  & 0.0012  & 0.0005  & 0.0335  & 0.0013  & 0.0004 \\
0.9559  & 6.9111  & 2.4971   & 0.0187  & 0.0007  & 0.0003  & 0.0189  & 0.0007  & 0.0002  & 0.0329  & 0.0011  & 0.0005  & 0.0317  & 0.0013  & 0.0004 \\
0.9650  & 6.9772  & 2.4138   & 0.0184  & 0.0007  & 0.0003  & 0.0182  & 0.0007  & 0.0002  & 0.0294  & 0.0010  & 0.0005  & 0.0281  & 0.0012  & 0.0003 \\
0.9742  & 7.0434  & 2.3305   & 0.0191  & 0.0007  & 0.0003  & 0.0180  & 0.0008  & 0.0002  & 0.0291  & 0.0011  & 0.0005  & 0.0277  & 0.0012  & 0.0003 \\
0.9833  & 7.1095  & 2.2472   & 0.0189  & 0.0007  & 0.0003  & 0.0170  & 0.0008  & 0.0002  & 0.0271  & 0.0011  & 0.0004  & 0.0254  & 0.0012  & 0.0003 \\
	    \hline
    	\hline
  	\end{tabular}
	\end{table*}


\begin{references}
\bibitem{BG1} E.D. Bloom, F.J. Gilman, Phys. Rev. Lett. {\bf 25}, 1140 (1970).
\bibitem{BG2} E.D. Bloom, F.J. Gilman, Phys. Rev. D {\bf 4}, 2901 (1971).
\bibitem{gp1} A. De Rujula, H. Georgi, H.D. Politzer, Phys. Lett. B {\bf 64}, 428 (1976).
\bibitem{gp2} A. De Rujula, H. Georgi, H.D. Politzer, Ann. Phys. {\bf 103}, 315 (1975).
\bibitem{wally_duality} W. Melnitchouk, R. Ent, C.E. Keppel, Physics Reports {\bf 406}, 127 (2005).
\bibitem{F2JL1} I. Niculescu {\it et al.}, Phys. Rev. Lett. {\bf 85}, 1186, (2000).
\bibitem{F2JL2} Y. Liang {\it et al.}, nucl-ex/0410027, submitted to Phys. Rev. Lett.
\bibitem{St03} D. Stump, J. Huston, J. Pumplin, W-K. Tung, H.L. Lai, S. Kuhlmann, J. Owens, JHEP 0310, 046 (2003).
\bibitem{LATTICE_MOM} D. Dolgov {\it et al.}, Phys. Rev. D {\bf 66}, 034506 (2002); \\
  M. Gockeler {\it et al.}, Nucl. Phys. Proc. Suppl. {\bf 119}, 32 (2003); \\
  W. Detmold, W. Melnitchouk, A.W. Thomas, Phys. Rev. D {\bf 66}, 054501 (2002).
\bibitem{allm97} H. Abramowicz, A. Levy, RR DESY 97-251;HEP-PH/9712415 (1997).
\bibitem{cteq} J. Pumplin, D.R. Stump, J. Huston, H.L. Lai, P. Nadolsky, W.K. Tung, JHEP 0207:012 (2002). 
\bibitem{mrst} A.D. Martin, R.G. Roberts, W.J. Stirling, R.S. Thorne, Phys. Lett. B {\bf 604}, 61 (2004).
\bibitem{alekhin_05} S.I. Alekhin, JETP Lett. {\bf 82}, 628 (2005).
\bibitem{alekhin_03} S.I. Alekhin, Phys. Rev D {\bf 63}, 094022 (2001).
\bibitem{bourrely} C. Bourrely, J. Soffer, F. Buccella, Eur. Phys. J. C {\bf 23}, 487 (2002).
\bibitem{allm} H. Abramowicz, E. Levin, A. Levy, U. Maor, Phys. Lett. B {\bf 269}, 465 (1991).
\bibitem{eric_param} M.E. Christy, Private Communication.
\bibitem{handbook_QCD} G. Sterman {\it et al}, Rev. Mod. Phys. {\bf 67}, 157 (1995).
\bibitem{f_recon} More information about the F$_{2}$ structure function reconstruction from parton distribution functions from 
CTEQ6M and MRST2004 can be provided upon request.
\bibitem{roberts} R.G. Roberts, {\it The Structure of the Proton}, Cambridge University Press (1990).
\bibitem{tzanov_nutev} M. Tzanov {\it at al.}, Phys. Rev. D {\bf 74}, 012008 (2006).
\bibitem{virchaux} M. Virchaux, A. Milzstajn, Phys. Lett. B {\bf 74}, 221 (1992).
\bibitem{yang} U.K. Yang, A. Bodek, Phys. Rev. Lett. {\bf 82}, 2467 (1999); U.K. Yang, A. Bodek, Eur. Phys. J. C {\bf 13}, 241 (2000).
\bibitem{simonetta} S. Liuti, R. Ent, C.E. Keppel, I. Niculescu, Phys. Rev. Lett. {\bf 89}, 162001 (2001).
\bibitem{liuti_ioana} I. Niculescu, C. Keppel, S. Liuti, G. Niculescu Phys. Rev. D {\bf 60}, 094001 (1999).
\bibitem{mrst_2001} A.D. Martin, R.G. Roberts, W.J. Stirling, R.S. Thorne, Eur. Phys. J C {\bf 23}, 73 (2002).
\bibitem{georgi_politzer_tmc} H. Georgi, H.D. Politzer, Phys. Rev. D {\bf 14}, 1829 (1976).
\bibitem{Bodek:1983qn} A. Bodek {\it et al.}, Phys. Rev. Lett. {\bf 50}, 1431 (1983).
\bibitem{Melnitchouk:1995fc} W. Melnitchouk and A.W. Thomas, Phys. Lett. B {\bf 377}, 11 (1996).
\bibitem{Kulagin:2004ie} S.A. Kulagin and R. Petti, Nucl. Phys. A {\bf 765}, 126 (2006).
\bibitem{Bosted:2007xd} P.E. Bosted and M.E. Christy, Phys. Rev. C {\bf 77}, 065206 (2008).
\bibitem{Arrington:2008zh} J. Arrington, F. Coester, R.J. Holt and T.S. Lee, J. Phys. G {\bf 36}, 025005 (2009).
\bibitem{d_p_antje} A. Bruell, Private Communication.
\bibitem{vladas_thesis} V. Tvaskis, Ph.D. thesis, University of Vrije (2004).
\bibitem{Hirai:2007sx} M. Hirai, S. Kumano and T.H. Nagai, Phys. Rev. C {\bf 76}, 065207 (2007).
\bibitem{Schienbein:2007fs} I. Schienbein, J.Y. Yu, C. Keppel, J.G. Morfin, F. Olness and J.F. Owens, Phys. Rev. D {\bf 77}, 054013 (2008).
\bibitem{deFlorian:2003qf} D. de Florian and R. Sassot, Phys. Rev. D {\bf 69}, 074028 (2004).
\bibitem{Eskola:2007my} K.J. Eskola, V.J. Kolhinen, H. Paukkunen and C.A. Salgado, JHEP {\bf 0705}, 002 (2007).
\bibitem{accardi_tmc1} A. Accardi, J.W. Qiu, arXiv:0805.1496v3.
\bibitem{accardi_tmc} A. Accardi, W. Melnitchouk, arXiv:0808.2397v2.
\bibitem{ingo_tm} I. Schienbein {\it at al.}, J. Phys. G {\bf 35}, 053101 (2008).
\bibitem{bpm_ref} G. Kraft, A. Hofler, {\sl How the Linac Beam Position Monitors Work}, CEBAF-TN-93-004 (1993).
\bibitem{armstr_thesis} C.S. Armstrong, Ph.D. thesis, College of William and Mary (1998).
\bibitem{m_dalton_baryon} M.M. Dalton {\it et al.}, e-Print: arXiv:0804.3509 [hep-ex], submitted to Phys. Rev. C.
\bibitem{raster_guy} C. Yan {\it et al.}, Nucl. Inst. and Meth. A {\bf 365}, 46 (1999).
\bibitem{simona_thesis} S. Malace, Ph.D. thesis, Hampton University (2006).
\bibitem{donprl} D. Dutta {\it et al.}, Phys. Rev. C {\bf 68}, 064603 (2003).
\bibitem{blok} H.P. Blok {\it et al.}, Phys. Rev. C {\bf 78}, 045202 (2008).
\bibitem{edwin_exp} Jefferson Lab E00-002, C.E. Keppel and I. Niculescu spokespersons.
\bibitem{meekins} Hall C target survey, D. Meekins, 2003.
\bibitem{pos_peter} P. Bosted, CLAS-NOTE-2004-005 (2004).
\bibitem{pion_data_slac} D.E. Wiser, Ph.D. thesis, University of Wisconsin (1977).
\bibitem{eric_elastic} M.E. Christy {\it et al.}, Phys. Rev. C {\bf 70}, 015206 (2004).
\bibitem{mo_tsai} L.W. Mo, Y.S. Tsai, Rev. Mod. Phys. {\bf 41}, 205 (1969).
\bibitem{bardin} A.A. Akhundov, D.Y. Bardin, N.M. Shumeiko, Sov. J. Nucl. Phys., 26 (1977); 
D.Y. Bardin, N.M. Shumeiko, Sov. J. Nucl. Phys., 29 (1979); A.A. Akhundov {\it at al.}, Sov. J. Nucl. 
Phys., 44 (1986).
\bibitem{liang_thesis} Y. Liang, Ph.D. thesis, The American University (2003).
\bibitem{eric_fit} M.E. Christy, P.E. Bosted, arXiv:0712.3731 [hep-ph] (2007).
\bibitem{h2_model} C.E. Keppel, Ph.D. thesis, The American University (1994).
\bibitem{bodek_fit} A. Bodek {\it et al.}, Phys. Rev. D {\bf 20}, 7 (1979).
\bibitem{r1990} L.W. Whitlow {\it et al.}, Phys. Lett. B {\bf 250}, 193 (1990).
\bibitem{r1998} K. Abe {\it et al.}, Phys. Lett. B {\bf 452}, 194 (1999).
\bibitem{ioana_thesis} I. Niculescu, Ph.D. thesis, Hampton University (1999).
\bibitem{F2JL3} I. Niculescu {\it et al.}, Phys. Rev. Lett. {\bf 85}, 1182, (2000).
\bibitem{NMC_fit} M. Arneodo {\it et al.}, Phys. Rev. B {\bf 364}, 107 (1995).
\bibitem{yong_mrst} A.D. Martin, R.G. Roberts, W.J. Stirling, R.S. Thorne, Eur. Phys. J C {\bf 4}, 463 (1998).
\bibitem{yong_cteq} H.L. Lai {\it et al.}, Eur. Phys. J. C {\bf 12}, 375 (2000).
\bibitem{thia_priv} C.E. Keppel, Private Communication.
\bibitem{whitlow_f2} L.W. Whitlow, E.M. Riordan, S. Dasu, S. Rock, A. Bodek, Phys. Lett. B {\bf 282}, 475 (1992).
\bibitem{riordan_thesis} E.M. Riordan, Ph.D. thesis, Massachusetts Institute of Technology (1973).
\bibitem{poucher_thesis} J.S. Poucher, Ph.D. thesis, Massachusetts Institute of Technology (1971).
\bibitem{bodek_thesis} A. Bodek, Ph.D. thesis, Massachusetts Institute of Technology (1972). 
\bibitem{bodek_paper} A. Bodek {\it at al.}, Phys. Rev. D {\bf 20}, 1471 (1979).
\bibitem{mestayer_paper} M.D. Mestayer {\it et al.}, Phys. Rev. D {\bf 27}, 285 (1983).
\bibitem{poucher_paper} J.S. Poucher {\it et al.}, Phys. Rev. Lett. {\bf 32}, 118 (1974).
\bibitem{atwood_paper} W.B. Atwood {\it et al.}, Phys. Lett. B {\bf 64}, 479 (1976).
\bibitem{isgur_close} F.E. Close, N. Isgur, Phys. Lett. B {\bf 509}, 81 (2001).
\bibitem{alberto_hallc_talk} A. Accardi, Private Communication.
\end{references}
\end{document}